\documentclass[12pt]{article}
\usepackage[paper=a4paper,margin=1in]{geometry}
\usepackage{t1enc}      
\usepackage{amsmath}
\usepackage{amsfonts}
\usepackage{amssymb}
\usepackage{amsthm}
\usepackage{graphicx}
\usepackage{mathrsfs}  

\newcommand{\ua}{\underline a \,}
\newcommand{\ub}{\underline b \,}
\newcommand{\uc}{\underline c \,}

\newcommand{\um}{\underline m \,}
\newcommand{\un}{\underline n \,}

\newcommand{\uA}{\underline A \,}
\newcommand{\uB}{\underline B \,}
\newcommand{\uC}{\underline C \,}
\newcommand{\uD}{\underline D \,}

\newcommand{\bi}{\bf i}
\newcommand{\bj}{\bf j}
\newcommand{\bk}{\bf k}
\newcommand{\bl}{\bf l}
\newcommand{\bm}{\bf m}

\newcommand{\bA}{\bf A}
\newcommand{\bB}{\bf B}

\newcommand{\nabbla}{{\nabla \!\!\!\!/}}

\numberwithin{equation}{section}

\begin{document}
\bibliographystyle{unsrt}

\title{Gravity, as a classical regulator for the Higgs field, and the 
origin of rest masses and electric charge
}

\author{L\'aszl\'o B. Szabados \\
  Wigner Research Centre for Physics, \\
  H--1525 Budapest 114, P. O. Box 49, European Union \\
  E-mail: lbszab@rmki.kfki.hu}

\maketitle

\begin{abstract}
The classical Einstein--Standard Model system with conformally invariant 
coupling of the Higgs field to gravity is investigated. We show that the 
energy-momentum tensor is \emph{not} polynomial in the Higgs field, and hence 
it may have two singularities: In cosmological spacetimes the usual Big Bang 
type singularity with diverging matter field variables, and a second, less 
violent one (`Small Bang'), in which it is only the geometry that is singular 
but the matter field variables remain finite. In generic spacetimes, the 
latter provides a \emph{finite, universal upper bound} for the pointwise 
norm of the Higgs field in terms of Newton's gravitational constant. 

As a consequence of this structure of the energy-momentum tensor, we also 
show that, in the presence of Friedman--Robertson--Walker or Kantowski--Sachs 
symmetries, the energy density can have finite local minimum \emph{only if 
the transitivity hypersurfaces of the spacetime symmetries are locally 
hyperboloidal and their mean curvature is less than a finite critical value}. 
In particular, in the very early era of an expanding universe or in a nearly 
spherically symmetric black hole near the central singularity, \emph{the 
Higgs sector does not have any instantaneous (symmetric or symmetry breaking) 
vacuum state}, and hence \emph{its rest mass is not defined, and}, via the 
Brout--Englert--Higgs (BEH) mechanism, \emph{the gauge and spinor fields do 
not get non-zero rest mass}. For smaller mean curvature instantaneous 
symmetry breaking vacuum states of the Higgs sector emerge, yielding non-zero 
rest mass and electric charge for some of the gauge and spinor fields via the 
BEH mechanism. These rest masses are decreasing with decreasing mean 
curvature, but the charge remains constant. It is also shown that 
\emph{globally defined} instantaneous vacuum states that are invariant with 
respect to the spacetime symmetries do not exist at all in the $k=1,0$ 
cosmological models and in Kantowski--Sachs spacetimes (e.g. inside 
spherically symmetric black holes). 
\end{abstract}


\section{Introduction}
\label{sec:0}

In classical (and quantum) mechanics the rest masses are \emph{a priori} 
given and, as an attribute, associated with the point particles; which masses 
can be determined from the small oscillations of the particles around their 
stable equilibrium states in some potential field. These stable equilibrium 
states are defined to be those configurations that are \emph{constant 
solutions} of the equations of motion and \emph{local minima} of the potential 
energy. In field theory the \emph{definition} of the rest mass of the fields 
is based just on this idea: To keep the special relativistic 
energy-momentum-rest mass relation to be valid pointwise, the rest mass of 
the fields should be \emph{defined} to be the second derivative of the 
potential energy with respect to the field variables at its stable critical 
point(s). 

At the end of the 19th century Mach raised the question why do the inertial 
frames of reference play so distinguished role in mechanics, and what is the 
origin of inertia of bodies. His (quite speculative) answer was that these 
frames are associated with the large scale distribution of the matter in the 
Universe: These are those frames from which its average mass distribution 
appears to be in uniform motion. Later, as is well known, this idea lead 
Einstein to formulate general relativity. Thus, according to Mach, the 
distinguished role of inertial frames, and also the root of inertia of 
bodies, have \emph{gravitational}, or perhaps \emph{cosmological}, origin. 

As is well known, the spacetime metric $g_{ab}$ splits in a natural way into 
the conformal class of the metric (represented by \emph{some} Lorentzian 
metric $\tilde g_{ab}$ conformal to $g_{ab}$) and a conformal factor. The 
significance of this decomposition in field theory is that all the field 
equations for the zero rest mass fields with any spin are \emph{conformally 
invariant}. In particular the Weyl neutrino fields, the massless scalar 
fields with the conformally invariant coupling to the scalar curvature of 
the spacetime (even with fourth order self-interaction) and the Yang--Mills 
fields are conformally invariant; and it is only the non-zero rest mass 
\emph{parameter} of the Higgs field that violates this invariance. Also, the 
field equations for the intrinsic degrees of freedom of the gravitational 
`field' in general relativity in \emph{vacuum} (i.e. the vacuum Bianchi 
identities) are also conformally invariant; and it is only \emph{the manner} 
in which gravitation is coupled to the matter source that violates this 
invariance. These facts motivate the idea that at the fundamental level 
elementary particles have zero rest mass, and their observed mass is a 
consequence of their interactions (see e.g. \cite{PeMc,Pe}). 

In fact, according to the electro-weak sector of the Standard Model of 
particle physics the mass of the leptons and of the vector bosons $W^\pm$ and 
$Z$ is due to their \emph{interaction} with the Higgs field. They get rest 
mass via the Brout--Englert--Higgs (or, shortly, BEH) mechanism \cite{H,EB}, 
and also the electric charge is recovered as a mixture of the two coupling 
constants of the electroweak sector of the Standard Model if the Higgs field 
has symmetry breaking vacuum states. To have such states the Higgs field 
should have an \emph{a priori} non-zero rest mass \emph{parameter} and 
self-interaction, and this rest mass parameter is the only dimensional 
parameter of the model \cite{AL73,ChLi}. 
(For the sake of completeness, it should be noted that in the QCD sector of 
the Standard Model there is another source of the rest mass, the chiral 
symmetry breaking, yielding most of the masses of the quarks, without any 
breaking of the gauge symmetry. In the present paper, however, we concentrate 
only on the origin of masses via the BEH-type mechanism.) However, it should 
be stressed that the BEH mechanism in its standard form is a \emph{purely 
kinematical} phenomenon (in the sense that in the derivation of this 
mechanism no evolution equation should be used). It is \emph{not} a dynamical 
process in which the massive leptons and vector bosons get their rest mass. 
The \emph{mere existence} of symmetry breaking vacuum states of the Higgs 
field in itself is already enough to yield rest masses. Thus, according to 
the Standard Model as it is in its present form, the a priori massless 
particles that get rest mass in the BEH mechanism are never realized in 
Nature. They are only some form of `Ideas' of the completely symmetric 
`Platonic world'. The rest masses are still inherent attributions of the 
particles rather than properties depending on the state of them (and/or, 
perhaps, the rest of the Universe). 

This state of affairs motivates the question whether or not the rest masses 
are really given once and for all, or rather the present day Standard Model 
is only an extremely good approximation of a `phase' of a more general model 
in which there could exist another phase with no symmetry breaking vacuum 
states, or with no vacuum states at all. In the latter case the rest masses 
would \emph{emerge} during such a `phase transition'. In fact, in the 
conformal cyclic cosmological (or CCC) model of Penrose \cite{PeCCC} all the 
particles on the crossover hypersurface must be massless. They should get 
rest mass \emph{after} the Big Bang in our aeon, and lose their rest mass 
\emph{before} the crossover in the previous aeon in some mechanism. Thus, in 
particle physics compatible with the CCC model, the rest masses should be 
expected to appear/disappear in some \emph{dynamical} process. But what kind 
of mechanism could yield such a phase transition? We believe that the 
ultimate explanation of the origin of the rest masses in Nature can not be 
formulated without incorporating gravity (and, perhaps, without 
incorporating the history of the Universe), just according to the ideas of 
Mach. 

In the present paper we consider the classical field theory of the coupled 
Einstein--Standard Model system, in which the Higgs field is coupled to 
gravity in a conformally invariant way (`Einstein--conformally coupled 
Standard Model', or, shortly, EccSM system). (The idea of coupling the 
fields of the Standard Model to gravity is not new. Its literature is 
enormously large, but most of the recent such investigations are motivated 
by cosmological problems. For the references to classical results see e.g. 
\cite{PR1,PR2}.) The conformal invariance is the mathematical realization 
of the idea that the fundamental particles of the model are basically 
\emph{massless}. No new field or parameter is introduced into the model. 
Our primary aim is to clarify whether or not the rest mass of elementary 
particles via the BEH mechanism could have a non-trivial `genesis', i.e. 
whether or not there could be a (very early) period in the history of the 
Universe in which the fundamental fields were massless, or perhaps their 
rest mass could not be defined at all, and they got rest mass (in some 
`phase transition') later. Also, we would like to see if in some `reverse 
BEH mechanism' these rest masses can `evanesce', e.g. in black holes. Thus, 
the present investigations can be considered as an extension of the 
\emph{classical field theoretical} investigations in the classic paper by 
Higgs \cite{H} to the case when gravitation is taken into account: We would 
like to see how the rest masses can be assigned to the matter fields, in 
particular how the BEH mechanism works in the EccSM system. 

We show that while the \emph{structure} of the field equations does not 
change, the conformally invariant coupling yields an energy-momentum tensor 
which is polynomial only in the metric and the gauge and spinor fields, but 
\emph{not} in the Higgs field. Consequently, this (and hence, via Einstein's 
equation, the spacetime geometry) may have a singularity in which \emph{all 
the matter field variables are finite}. In particular, in generic spacetimes 
we obtain \emph{a finite, universal upper bound} for the pointwise norm of 
the Higgs field in terms of Newton's gravitational constant. This provides 
a natural, non-perturbative cut-off in the field theoretic calculations. In 
fact, in a separate paper \cite{SzW} we show that, in the presence of 
Friedman--Robertson--Walker (FRW) symmetries, the field equations of the 
conformally coupled Einstein--Higgs (EccH) system do have solutions with 
scalar polynomial curvature singularity (Small Bang) when the Higgs field 
takes this finite value. 

We discuss the question of vacuum states of gravitating systems, and we find 
that the usual notion of vacuum states cannot be applied directly in such 
systems: The minimal energy density states of the matter fields in 
spacetimes with maximal number of isometries are \emph{not} solutions of the 
field equations. Hence, we are forced to generalize the notion of `spacetime 
vacuum states' to `instantaneous vacuum states', labeled by spacelike 
hypersurfaces, as certain critical configurations of the (total or 
quasi-local) energy-momentum functional. In particular, in the presence of 
FRW symmetries the energy density of the matter (in fact, the Higgs) field 
has a non-trivial dependence on the mean curvature of the $t={\rm const}$ 
hypersurface, and hence has a \emph{time dependence}. It turns out that 
there is a large, but finite \emph{critical value} of the mean curvature 
such that, with the known parameters of the Standard Model of particle 
physics, the energy density has stable (gauge symmetry breaking) minima, in 
particular has the `wine bottle' (rather than the familiar `Mexican hat') 
shape, precisely on hypersurfaces whose mean curvature is \emph{smaller} 
than the critical value above. (The Hubble time corresponding to this 
critical value is about ten Planck times.) The system does not have any 
\emph{symmetric} vacuum state. We have an analogous result in 
Kantowski--Sachs spacetimes, e.g. inside a spherically symmetric black hole. 
It is also shown that field configurations that are \emph{globally defined} 
on the $t={\rm const}$ hypersurfaces, admit the spacetime symmetries, solve 
the constraints and minimize the energy functional (i.e. globally defined 
instantaneous vacuum states) exist neither in the $k=1,0$ FRW nor in 
Kantowski--Sachs spacetimes. If the vacuum states are not required to admit 
the same symmetries that the spacetime does, i.e. if they are allowed to be 
$SO(1,3)$-symmetric even in the $k=1,0$ FRW or Kantowski--Sachs spacetimes, 
then the gauge symmetry breaking instantaneous vacuum states in these 
spacetimes can be defined at least on \emph{open subsets} of the $t={\rm 
const}$ hypersurfaces, i.e. they can exist \emph{quasi-locally}. 

Finally, we calculate the rest mass of the gauge, spinor and Higgs fields in 
the EccSM system. We find that on constant mean curvature hypersurfaces with 
mean curvature \emph{higher} than the critical value in a nearly FRW or 
Kantowski--Sachs spacetime \emph{the rest mass of the Higgs field is not 
defined, and the BEH mechanism does not work}. The instant of the 
genesis/evanescence of the rest masses is the hypersurface with the critical 
mean curvature. For smaller mean curvature we obtain \emph{time dependent} 
rest masses (though this time dependence is significant only in the very 
close vicinity of genesis/evanescence). On hypersurfaces in FRW spacetimes 
with the Hubble time equals to the characteristic time scale defined by the 
Higgs rest mass parameter $\mu$, i.e. to $5.5\times10^{-27}sec$, the rest 
mass of the electrons, the $Z$ and $W^\pm$ gauge bosons and the Higgs boson 
is still roughly \emph{twice} of their present rest mass, though the time 
dependence of the rest mass of the Higgs and the other fields is slightly 
different. Since electrodynamics is a result of the breaking of the $U(2)$ 
symmetry in the Weinberg--Salam model, the hypersurface with the critical 
mean curvature is the instant of the genesis/evanescence of the 
electromagnetism and the electric charge, too. Therefore, in the presence 
of extreme gravitational situations (e.g. in a vicinity of the initial 
singularity of the Universe or in spacetimes described e.g. by a general 
Kantowski--Sachs metric), \emph{certain concepts of field theory become 
ill-defined and particle mechanical notions (actually the rest mass) cannot 
be implemented in field theory}. The fields do not have particle 
interpretation. 

The paper is organized according to the logic of the results above: Section 
\ref{sec:1} is devoted to the definition of the EccSM system, the discussion 
of the structure of the field equations and the energy-momentum tensor. Here 
we discuss the problem of the vacuum states, too. This section is more 
pedagogical than the remaining ones, making the particle physics ideas more 
accessible for the wider readership, and, in particular, fixing the notations. 
In Section \ref{sec:2}, the critical points of the energy-momentum functional 
are determined, and, in the presence of FRW and Kantowski--Sachs symmetries, 
a detailed discussion of the qualitative properties of the energy density is 
given. The rest mass of the fields in the matter sector is calculated in 
Section \ref{sec:3}. 

Our sign conventions are those of \cite{PR1}. In particular, the signature 
of the spacetime metric is $(+,-,-,-)$ and the curvature tensor of the 
linear connection $\nabla_a$ is defined by $-R^a{}_{bcd}X^bV^cW^d:=V^c\nabla_c
(W^d\nabla_dX^a)-W^c\nabla_c(V^d\nabla_dX^a)-[V,W]^c\nabla_cX^a$ for any vector 
fields $X^a$, $V^a$ and $W^a$. Hence, Einstein's equations take the form 
$R_{ab}-\frac{1}{2}Rg_{ab}=-\kappa T_{ab}-\Lambda g_{ab}$ with $\kappa:=8\pi G$ 
(or $8\pi G/c^4$ in traditional units), where $G$ is Newton's gravitational 
constant and $\Lambda$ is the cosmological constant.


\section{The Einstein--conformally coupled Standard Model 
(EccSM) system} \label{sec:1}


\subsection{The basic fields}
\label{sub-1.1}

In the Standard Model of particle physics (even to its extension allowing 
non-zero rest-mass for the neutrinos), there are three types of matter fields: 
A Yang--Mills gauge field, a multiplet of scalar fields (the Higgs field) and 
a multiplet of Weyl spinor fields. However, primarily here we are interested 
in the \emph{general structure} of the model, in particular, in its conformal 
properties and in a potential reformulation of the classical 
Einstein--Standard Model systems. Thus, in what follows, by Standard Model we 
mean such a general Yang--Mills--Higgs--Weyl system (whose Higgs sector will 
ultimately be coupled in the conformally invariant way to gravity described 
by Einstein's theory), which mimics all the characteristic feature of the 
specific Standard Model of particle physics.

\subsubsection{The gauge fields}
\label{sub-1.1.1}

Let $P(M,G)$ be a principal fiber bundle over the spacetime manifold $(M,
g_{ab})$ with the connected Lie group $G$ as its structure group, and let 
${\cal G}$ denote the Lie algebra of $G$. If $\{g_\alpha\}=\{g_1,\cdots,g_k\}$, 
$\alpha=1,\cdots,k:=\dim{\cal G}$, is a basis in ${\cal G}$, then the 
structure constants are defined by $[g_\alpha,g_\beta]=:c^\gamma_{\alpha\beta}
g_\gamma$, and we define the metric $G_{\alpha\beta}$ to be proportional with the 
Cartan-Killing metric: $G_{\alpha\beta}:=\frac{1}{g^2}c^\mu_{\alpha\nu}c^\nu_{\beta\mu}
=G_{(\alpha\beta)}$, where $g$ is some positive constant (the `coupling 
constant'). Thus the Greek indices are referring to the basis $\{g_\alpha\}$ in 
the Lie algebra. $G_{\alpha\beta}$ is known to be \emph{negative} definite for 
compact, semisimple Lie algebras. If $G$ is the direct product of groups, 
then $G_{\alpha\beta}$ is the direct sum of the corresponding metrics, in which 
the coupling constants may be different. 

Let a connection be given on $P(M,G)$ in the form of a connection 1-form 
$\omega=\omega^\alpha g_\alpha$, which is a ${\cal G}$-valued, ${\rm Ad}(
G)$-invariant 1-form. Its pull back to open domains in $M$ along the local 
cross sections of $P(M,G)$ is denoted by $\omega_a=\omega^\alpha_ag_\alpha$. 
Then, since the linear connection $\nabla_e$ on the spacetime tensor bundles is 
torsion free, the corresponding curvature 2-form can be written\footnote{The 
particle physicists' convention is slightly different. Their gauge potential 
is $A^\alpha_e=g^{-1}\omega^\alpha_e$, where $g$ is the coupling constant. 
Hence their field strength is not simply the curvature of $A^\alpha_e$, but 
contains the coupling constant explicitly. It is $\nabla_aA^\alpha_b-\nabla
_bA^\alpha_a+gc^\alpha_{\mu\nu}A^\mu_aA^\nu_b$, which is just $g^{-1}F^\alpha
_{ab}$. This gauge potential $A^\alpha_e$ should not be confused with the 
\emph{spatial} vector potential of sections \ref{sec:2} and \ref{sec:3}.}
in the form $F^\alpha_{ab}=\nabla_a\omega^\alpha_b-\nabla_b\omega^\alpha_a+c^\alpha
_{\mu\nu}\omega^\mu_a\omega^\nu_b$. In terms of these objects the Bianchi 
identities take the form $\nabla_{[a}F^\alpha_{bc]}+c^\alpha_{\mu\beta}\omega^\mu
_{[a}F^\beta_{bc]}=0$. 

In the adjoint representation of $G$ the representation space is its own Lie 
algebra ${\cal G}$ as a $k$ dimensional vector space, and the representation 
matrices of the Lie algebra in the basis $\{g_\alpha\}$ are well known to be 
just the structure constants, i.e. $X=X^\alpha g_\alpha\in{\cal G}$ is 
represented by the $k\times k$ matrix $X^\mu c^\alpha_{\mu\beta}$. Let $A(M)$ 
denote the associated vector bundle based on the adjoint representation of 
$G$. Then the connection 1-form on $A(M)$ is $\omega^\alpha_{a\beta}:=\omega
^\mu_ac^\alpha_{\mu\beta}$, and the covariant derivative of any cross section 
$\varphi^\alpha$ of $A(M)$ is $\nabbla_e\varphi^\alpha:=\nabla_e\varphi^\alpha+
\omega^\mu_e c^\alpha_{\mu\beta}\varphi^\beta$. Considering the structure 
constants to be the components of an (1,2) type tensor \emph{field}, by the 
Jacobi identity it is constant with respect to $\nabbla_e$, too. This implies 
that the metric $G_{\alpha\beta}$ is also annihilated by $\nabbla_e$, and the 
Bianchi identities in $A(M)$ take the form $\nabbla_{[a}F^\alpha_{bc]}=0$. It is 
straightforward to rewrite the curvature 2-form of the connection $\nabbla_e$: 
It is $F^\alpha{}_{\beta cd}=F^\mu_{cd}c^\alpha_{\mu\beta}=\nabla_c\omega^\alpha_{d\beta}
-\nabla_d\omega^\alpha_{c\beta}+\omega^\alpha_{c\mu}\omega^\mu_{d\beta}-\omega^\alpha
_{d\mu}\omega^\mu_{c\beta}$.

\subsubsection{The Higgs fields}
\label{sub-1.1.2}

Let $H(M)$ be a (in general, complex) vector bundle, associated with $P(M,G)$ 
via a finite dimensional linear representation $G\rightarrow GL(N,
\mathbb{C})$, and let $T^{\bi}_{\alpha{\bj}}$, ${\bi},{\bj}=1,...,N$, denote the 
corresponding representation matrices of the Lie algebra ${\cal G}$ in the 
basis $\{g_\alpha\}$; i.e. which satisfy the matrix commutation relation 
$[T_\alpha,T_\beta]^{\bi}{}_{\bj}=c^\mu_{\alpha\beta}T^{\bi}_{\mu{\bj}}$. Thus the Lie 
algebra ${\cal G}$ is represented by the $gl(N,\mathbb{C})$ matrices given 
by $\lambda=\lambda^\alpha g_\alpha\mapsto\lambda^{\bi}{}_{\bj}:=\lambda^\alpha 
T^{\bi}_{\alpha{\bj}}$. Then the gauge group $G$ is represented by the matrices 
of the form $\Lambda^{\bi}{}_{\bj}:=\exp(\lambda)^{\bi}{}_{\bj}:=\delta^{\bi}_{\bj}
+\lambda^\alpha T^{\bi}_{\alpha{\bj}}+\frac{1}{2!}(\lambda^\alpha T^{\bi}_{\alpha{\bk}})
(\lambda^\beta T^{\bk}_{\beta{\bj}})+\cdots$. We assume that $H(M)$ admits a 
positive definite invariant Hermitian fiber metric $G_{\bi{\bj}'}$ in the sense 
that $G_{\bi{\bj}'}=G_{\bk{\bl}'}\Lambda^{\bk}{}_{\bi}\bar\Lambda^{{\bl}'}{}_{{\bj}'}$ 
holds for any $\Lambda^{\bi}{}_{\bj}$ above. Thus, over-bar denotes complex 
conjugation and the primed indices are referring to the complex conjugate 
representation. A typical cross section of $H(M)$ will be denoted by $\Phi
^{\bi}$. Locally this can be thought of as a multiplet of scalar fields on 
$M$ which, under the action of the gauge group, transforms as $\Phi^{\bi}
\mapsto\Phi^{\bj}\Lambda_{\bj}{}^{\bi}$, where $\Lambda_{\bj}{}^{\bi}$ is defined 
by $\Lambda_{\bj}{}^{\bi}\Lambda^{\bk}{}_{\bi}=\delta^{\bk}_{\bj}$. 

The connection on $P(M,G)$ defines a covariant derivative operator on $H(M)$ 
by $\nabbla_e\Phi^{\bi}:=\nabla_e\Phi^{\bi}+\omega^\alpha_e T^{\bi}_{\alpha{\bj}}
\Phi^{\bj}$. The invariance of $G_{\bi{\bj}'}$ implies that $G_{\bi{\bk}'}\bar T
^{{\bk}'}_{\alpha{\bj}'}+G_{\bk{\bj}'}T^{\bk}_{\alpha{\bi}}=0$, and hence that it is 
annihilated by $\nabbla_e$. The curvature 2-form on $H(M)$, defined by 
$F^{\bi}{}_{{\bj}cd}\Phi^{\bj}V^cW^d:=V^c\nabbla_c(W^d\nabbla_d\Phi^{\bi})-W^c
\nabbla_c(V^d\nabbla_d\Phi^{\bi})-[V,W]^c\nabbla_c\Phi^{\bi}$, is just $F^{\bi}
{}_{{\bj}cd}=F^\alpha_{cd}T^{\bi}_{\alpha{\bj}}$.

\subsubsection{The Weyl spinor fields}
\label{sub-1.1.3}

Let $G\rightarrow GL(R,\mathbb{C})$ be a linear representation of the gauge 
group, let $T^r_{\alpha s}$, $r,s=1,\cdots,R$, denote the representation 
matrices of the Lie algebra ${\cal G}$, and the corresponding associated 
vector bundle be denoted by $B(M)$. Thus the elements of $G$ are represented 
by the matrices $\Lambda^r{}_s:=\exp(\lambda)^r{}_s:=\delta^r_s+\lambda^\alpha 
T^r_{\alpha s}+\frac{1}{2!}(\lambda^\alpha T^r_{\alpha p})(\lambda^\beta T^p_{\beta s})
+\cdots$, where $\lambda^\alpha g_\alpha\in{\cal G}$. Using the complex 
conjugate representation, we can form the conjugate bundle $\bar B(M)$, in 
which the indices will be primed, e.g. $r'$, $s'$, ... etc. We assume that the 
bundle $B(M)$ admits a \emph{non-degenerate} invariant \emph{Hermitian} fiber 
metric $G_{rr'}$, i.e. $G_{rr'}=G_{ss'}\Lambda^s{}_r\bar\Lambda^{s'}{}_{r'}$. 

Let us fix a spinor structure on $M$, and let $\mathbb{S}_A(M)$ denote the 
bundle of Weyl spinors. Then let us form the tensor product bundle $F(M):=B(M)
\otimes\mathbb{S}_A(M)$. We call it the fermion bundle. Its cross sections, 
$\psi^r_A$, can be interpreted locally as multiplets of Weyl spinor fields 
transforming under the action of the gauge group as $\psi^r_A\mapsto\psi^s_A
\Lambda_s{}^r$, where $\Lambda_s{}^r$ is defined by $\Lambda^r{}_p\Lambda_s{}^p
=\delta^r_s$. 

The connection on $P(M,G)$ defines a connection on $F(M)$ in a natural way: 
The corresponding connection 1-form is $\omega^\alpha_e T^r_{\alpha s}$, and hence 
the (spacetime and gauge) covariant derivative is $\nabbla_e\psi^r_A:=\nabla_e
\psi^r_A+\omega^\alpha_e T^r_{\alpha s}\psi^s_A$. The connection 1-form on the 
complex conjugate bundle $\bar B(M)$ is $\omega^\alpha_e\bar T^{r'}_{\alpha{s'}}$; 
and $G_{rr'}T^r_{\alpha s}+G_{ss'}\bar T^{s'}_{\alpha r'}=0$ holds.


\subsection{The Lagrangians and the couplings}
\label{sub-1.2}

The Lagrangian of the gauge and Higgs fields, respectively, are chosen to be 

\begin{eqnarray}
{\cal L}_{YM}\!\!\!\!&:=\!\!\!\!&\frac{1}{4}G_{\alpha\beta}F^\alpha_{ab}F^\beta
  _{cd}g^{ac}g^{bd},\label{eq:1.2.1} \\
{\cal L}_H\!\!\!\!&:=&\!\!\!\!\frac{1}{2}G_{\bi{\bj}'}g^{ab}\bigl(\nabbla_a
  \Phi^{\bi}\bigr)\bigl(\nabbla_b\bar\Phi^{{\bj}'}\bigr)-\frac{1}{2}\alpha R
  G_{\bi{\bj}'}\Phi^{\bi}\bar\Phi^{{\bj}'}-\frac{1}{4}G_{\bi\bj{\bk}'{\bl}'}\Phi^{\bi}
  \Phi^{\bj}\bar\Phi^{{\bk}'}\bar\Phi^{{\bl}'}. \label{eq:1.2.2}
\end{eqnarray}
To ensure the positive definiteness of the `kinetic energy term' both for the 
gauge and the Higgs fields, the fiber metric $G_{\alpha\beta}$ is assumed to be 
\emph{negative}, while $G_{\bi{\bj}'}$ to be \emph{positive} definite. $\alpha$ 
is some real constant, $R$ is the curvature scalar of the spacetime, and 
$G_{\bi\bj{\bk}'{\bl}'}$ is a constant coefficient being symmetric in $\bi\bj$ and 
in $\bk'\bl'$, e.g. $\lambda G_{\bi({\bk}'}G_{{\bl}')\bj}$ for some constant 
$\lambda$, such that the last term in (\ref{eq:1.2.2}), describing the 
self-interaction of the Higgs field, be gauge invariant and positive definite. 
In particular, $T^{\bm}_{\alpha({\bi}}G_{{\bj})\bm{\bk'\bl'}}+\bar T^{{\bm}'}
_{\alpha({\bk}'}G_{{\bl}'){\bm}'\bi\bj}=0$ must hold. For the Lagrangian of the 
multiplet of the Weyl spinor fields we choose 

\begin{equation}
{\cal L}_W:=\frac{\rm i}{2}G_{rr'}g^{ab}\Bigl(\bar\psi^{r'}_{A'}\nabbla_b\psi^r_A
-\psi^r_A\nabbla_b\bar\psi^{r'}_{A'}\Bigr).
\label{eq:1.2.3}
\end{equation}
Then the Lagrangian of the total Yang--Mills--Higgs--Weyl system is ${\cal L}
:={\cal L}_{YM}+{\cal L}_H+{\cal L}_W+{\cal L}_I-V$, where ${\cal L}_I={\cal L}
_I(\Phi^{\bi},\bar\Phi^{\bi'},\psi^r_A,\bar\psi^{r'}_{A'})$ is a real, purely 
\emph{algebraic} and \emph{gauge invariant} expression of the Higgs and Weyl 
spinor fields, describing their interaction, and $V=V(\Phi^{\bi},\bar\Phi
^{{\bi}'})$ is a purely algebraic potential term. Motivated by the specific 
Standard Model, we choose them, respectively, to have the structure 

\begin{eqnarray}
{\cal L}_I\!\!\!\!&=\!\!\!\!&\frac{\rm i}{2}\Bigl(\varepsilon^{A'B'}\bar
  \psi^{r'}_{A'}\bar\psi^{s'}_{B'}\bar Y_{r's'{\bi}}\Phi^{\bi}-\varepsilon^{AB}
  \psi^r_A\psi^s_BY_{rs{\bi}'}\bar\Phi^{\bi'}\Bigr), \label{eq:1.2.5a} \\
V\!\!\!\!&=\!\!\!\!&\frac{1}{2}\mu^2G_{\bi\bj'}\Phi^{\bi}\bar\Phi^{\bj'}. 
 \label{eq:1.2.5b}
\end{eqnarray}
Here $Y_{rs{\bi}'}=-Y_{sr{\bi}'}$ are the so-called Yukawa coupling constants such 
that ${\cal L}_I$ be real and gauge invariant, i.e. $Y_{ps{\bi}'}T^p_{\alpha r}+
Y_{rp{\bi}'}T^p_{\alpha s}+Y_{rs{\bj}'}\bar T^{\bj'}_{\alpha{\bi}'}=0$, and $\mu^2$ is 
the (maybe \emph{negative}) rest mass parameter of the Higgs field. For 
$\alpha=0$ in (\ref{eq:1.2.2}) ${\cal L}_H-V$ is the standard flat-spacetime 
Lagrangian of the Higgs field, while ${\cal L}_H$ for $\alpha=1/6$ has the 
form of the Lagrangian of (a multiplet of) \emph{conformally invariant} 
self-interacting scalar fields.


\subsection{The field equations and the energy-momentum tensor}
\label{sub-1.3}

The basic spacetime covariant matter field variables are $\omega^\alpha_a$, 
$\Phi^{\bi}$ and $\psi^r_A$. Using the fact that the fiber metrics 
$G_{\alpha\beta}$, $G_{\bi\bj'}$ and $G_{rr'}$ are invariant with respect to gauge 
transformations (and hence they are annihilated by the covariant derivative 
$\nabbla_e$), a routine calculation yields that the field equations are 

\begin{eqnarray}
\bigl(\nabbla^aF^\beta_{ab}\bigr)G_{\beta\alpha}\!\!\!\!&=\!\!\!\!&\frac{1}{2}
  \Bigl((\nabbla_b\bar\Phi^{\bi'})G_{\bi'\bj}T^{\bj}_{\alpha{\bk}}\Phi^{\bk}+(\nabbla
  _b\Phi^{\bi})G_{\bi\bj'}\bar T^{\bj'}_{\alpha{\bk}'}\bar\Phi^{\bk'}\Bigr)+ 
  \nonumber \\
\!\!\!\!&{}\!\!\!\!&+\frac{\rm i}{2}G_{rr'}\Bigl(T^r_{\alpha s}\psi^s_B\bar
  \psi^{r'}_{B'}-\bar T^{r'}_{\alpha s'}\bar\psi^{s'}_{B'}\psi^r_B\Bigr), 
  \label{eq:1.3.1a} \\
\nabbla_a\bigl(\nabbla{}^a\Phi^{\bj}\bigr)G_{\bj\bi'}\!\!\!\!&=\!\!\!\!&-\bigl(
  \mu^2+\alpha R\bigr)\Phi^{\bj}G_{\bj\bi'}-G_{\bk\bl\bj'\bi'}\Phi^{\bk}\Phi^{\bl}
  \bar\Phi^{\bj'}-{\rm i}\varepsilon^{AB}\psi^r_A\psi^s_BY_{rs{\bi}'}, 
  \label{eq:1.3.1b} \\
\bigl(\nabbla^A_{A'}\psi^r_A\bigr)G_{rr'}\!\!\!\!&=\!\!\!\!&-\bar\psi^{s'}_{A'}
  \bar Y_{r's'{\bi}}\Phi^{\bi}. \label{eq:1.3.1c}
\end{eqnarray}
The two currents on the right of (\ref{eq:1.3.1a}), built from the Higgs 
scalar and the Weyl spinor multiplets, respectively, will be denoted by 
$4\pi\,{}_HJ^\beta_bG_{\beta\alpha}$ and $4\pi\,{}_WJ^\beta_bG_{\beta\alpha}$.  

Since our choice for the signature of the spacetime metric is $-2$, we should 
define the symmetric energy-momentum tensor by $T_{ab}:=2\delta I_D/\delta g
^{ab}$, where $I_D:=\int_D{\cal L}\sqrt{\vert g\vert}{\rm d}^4x$ is the action 
functional and $D\subset M$ is an open subset with compact closure. Then a 
straightforward calculation yields the energy-momentum tensor of the 
Yang--Mills--Higgs sector (governed by the Lagrangian ${\cal L}_{YM}+{\cal 
L}_H-V$). It is 

\begin{eqnarray}
T^{(YMH)}_{ab}\!\!\!\!&=\!\!\!\!&G_{\alpha\beta}F^\alpha_{ac}F^\beta_{bd}g^{cd}-
  \frac{1}{4}g_{ab}G_{\alpha\beta}F^\alpha_{ce}F^\beta_{df}g^{cd}g^{ef}+\frac{1}{2}
  g_{ab}\mu^2G_{\bi\bj'}\Phi^{\bi}\bar\Phi^{\bj'}+ \label{eq:1.3.2} \\
\!\!\!\!&+\!\!\!\!&G_{\bi\bj'}\bigl(\nabbla_{(a}\Phi^{\bi}\bigr)\bigl(\nabbla_{b)}
  \bar\Phi^{\bj'}\bigr)-\frac{1}{2}g_{ab}G_{\bi\bj'}\bigl(\nabbla_c\Phi^{\bi}\bigr)
  \bigl(\nabbla^c\bar\Phi^{\bj'}\bigr)+\frac{1}{4}g_{ab}G_{\bi\bj\bk'\bl'}
  \Phi^{\bi}\Phi^{\bj}\bar\Phi^{\bk'}\bar\Phi^{\bl'}- \nonumber \\
\!\!\!\!&-\!\!\!\!&\alpha\bigl(R_{ab}-\frac{1}{2}Rg_{ab}\bigr)G_{\bi\bj'}
  \Phi^{\bi}\bar\Phi^{\bj'}-\alpha\nabla_a\nabla_b\bigl(G_{\bi\bj'}\Phi^{\bi}\bar
  \Phi^{\bj'}\bigr)+\alpha g_{ab}\nabla_c\nabla^c\bigl(G_{\bi\bj'}\Phi^{\bi}\bar
  \Phi^{\bj'}\bigr). \nonumber
\end{eqnarray}
The apparently different energy-momentum tensor $\tilde T_{ab}$ for a single, 
conformally invariant scalar field satisfying $\nabla_c\nabla^c\Phi+\frac{1}
{6}R\Phi=0$ given in \cite{NP68} (and also appearing in \cite{PR2}) coincides 
with that given by (\ref{eq:1.3.2}) with $\alpha=1/6$ up to a numerical 
factor and the field equation: $T^{(H)}_{ab}=\frac{2}{3}\tilde T_{ab}+\frac{1}
{3}g_{ab}\Phi(\nabla_c\nabla^c\Phi+\frac{1}{6}R\Phi)$. 

Clearly, in the derivation of the field equations the variation of the spinor 
field $\psi^r_A$ and the variation of its \emph{components} $\psi^r_{\uA}$ in 
some \emph{fixed} normalized dual spinor basis $\{\varepsilon^A_{\uA},
\varepsilon^{\uA}_A\}$ are equivalent. Thus, even if we had chosen the 
components $\psi^r_{\uA}$ of the spinor fields to be the basic variable in 
the previous paragraphs, we would have obtained the same field equation 
(\ref{eq:1.3.1c}). However, the spinor bundle is linked to the orthonormal 
frame bundle, and hence the notion of spinors itself depends on the metric 
(in fact, the conformal structure) of the spacetime. Therefore, under a 
general variation of the spacetime metric quantities in \emph{different} 
orthonormal frame bundles (as different subbundles of the linear frame 
bundle) must be compared. Hence, in the calculation of the energy-momentum 
tensor of the Weyl sector of the Standard Model, it would appear to be 
natural to consider the components of the spinor fields and the tetrad 
field to be the independent variables. Nevertheless, a pure Lorentz 
transformation of the orthonormal frame field, and hence a pure $SL(2,
\mathbb{C})$ transformation of the normalized spinor basis, must yield only a 
pure (Lorentz or $SL(2,\mathbb{C})$) gauge transformation, even though the 
tensor/spinor components in these bases do change. Hence, it seems even more 
natural to choose the \emph{components} of the spinor field \emph{up to 
$SL(2,\mathbb{C})$ transformations} and the spacetime metric (rather than the 
spinor components and the orthonormal vector basis) to be the independent 
variables. 

Following this idea, in the Appendix we calculate the total variation of the 
Lagrangian of a multiplet of Weyl spinor fields in terms of the variation of 
these independent variables. Using the resulting expression (\ref{eq:A.4}) 
for this variation, it is straightforward to derive the energy-momentum 
tensor of the spinorial sector of the Standard Model (governed by the 
Lagrangian ${\cal L}_W+{\cal L}_I$ above). It is 

\begin{eqnarray}
T^{(WI)}_{ab}\!\!\!\!&=\!\!\!\!&\frac{\rm i}{4}G_{rr'}\Bigl(\bar\psi^{r'}_{A'}
  \nabbla_{BB'}\psi^r_A+\bar\psi^{r'}_{B'}\nabbla_{AA'}\psi^r_B-\psi^r_A\nabbla
  _{BB'}\bar\psi^{r'}_{A'}-\psi^r_B\nabbla_{AA'}\bar\psi^{r'}_{B'}\Bigr)- 
  \label{eq:1.3.3} \\
\!\!\!\!&-\!\!\!\!&\frac{\rm i}{2}g_{ab}\Bigl(\bar\psi^{r'}_{C'}\bigl(G_{r'r}
 \nabbla^{C'C}\psi^r_C+\varepsilon^{C'D'}\bar\psi^{s'}_{D'}\bar Y_{r's'{\bi}}
 \Phi^{\bi}\bigr)-\psi^r_C\bigl(G_{rr'}\nabbla^{CC'}\bar\psi^{r'}_{C'}+\varepsilon
 ^{CD}\psi^s_DY_{rs{\bi}'}\bar\Phi^{{\bi}'}\bigr)\Bigr). \nonumber
\end{eqnarray}
If the field equation (\ref{eq:1.3.1c}) is satisfied, then the second line 
is vanishing; and, apart from a numerical factor, for a single Weyl spinor 
field satisfying the neutrino equation the expression (\ref{eq:1.3.3}) 
reproduces the energy-momentum tensor postulated in \cite{PR1}. 

By (\ref{eq:1.3.1b}) and (\ref{eq:1.3.1c}), the trace of the energy-momentum 
tensor $T_{ab}=T^{(YMH)}_{ab}+T^{(WI)}_{ab}$ of the whole Standard Model is 

\begin{equation}
T_{ab}g^{ab}=\frac{1}{2}\bigl(6\alpha-1)\nabla_a\nabla^a\bigl(G_{\bi\bj'}\Phi^{\bi}
\bar\Phi^{\bj'}\bigr)+\mu^2G_{\bi\bj'}\Phi^{\bi}\bar\Phi^{\bj'}. \label{eq:1.3.4}
\end{equation}
Thus, even if $\alpha$ is chosen to be $1/6$ and the field equations are 
satisfied, $T_{ab}$ is not trace-free unless $\mu^2=0$. 

It is a straightforward calculation to check that (see e.g. \cite{PR1}), if 
the spacetime metric is conformally rescaled according to $g_{ab}\mapsto
\Omega^2g_{ab}$ and, at the same time, the basic matter fields are also 
rescaled according to $\omega^\alpha_e\mapsto\omega^\alpha_e$, $\Phi^{\bi}
\mapsto\Omega^{-1}\Phi^{\bi}$ and $\psi^r_A\mapsto\Omega^{-1}\psi^r_A$, then 

\begin{eqnarray*}
{\cal L}_{YM}+{\cal L}_H+{\cal L}_W+{\cal L}_I\mapsto\!\!\!\!&{}\!\!\!\!&\Omega
  ^{-4}\Bigl({\cal L}_{YM}+{\cal L}_H+{\cal L}_W+{\cal L}_I-\frac{1}{2}\nabla_e
  \bigl(\Upsilon^eG_{\bi\bj'}\Phi^{\bi}\bar\Phi^{\bj'}\bigr)\Bigr) \nonumber \\
+\!\!\!\!&{}\!\!\!\!&\Omega^{-4}\frac{1}{2}(1-6\alpha)\bigl(\nabla_e\Upsilon^e
  +\Upsilon_e\Upsilon^e\bigr)G_{\bi\bj'}\Phi^{\bi}\bar\Phi^{\bj'}, \\
V\mapsto\!\!\!\!&{}\!\!\!\!&\Omega^{-2}V; 
\end{eqnarray*}
while the spacetime volume element changes as ${\rm d}v\mapsto\Omega^4
{\rm d}v$. Therefore, if $\alpha=1/6$, then it is only the potential $V$ in 
the action $I_D$ that violates the conformal invariance. Note also that any 
higher (e.g. 6th) order self-interaction term in ${\cal L}_H$ also would 
violate the conformal invariance of the action.


\subsection{Example: The Weinberg--Salam model with a single lepton 
generation}
\label{sub-1.5}

In subsection \ref{sub-3.3} we will calculate the rest masses of the fields 
of the Einstein-conformally coupled Weinberg--Salam model. Thus (also to 
justify the above general model and to motivate the questions in the rest of 
the paper), here we review the key points of the classical theory behind 
this model. Here all the bundles are assumed to globally trivializable. For 
detailed and readable classical presentations of the model, see 
\cite{AL73,ChLi}. 

The gauge group is $U(2)=U(1)\times SU(2)$, and we choose $(\frac{\rm i}{2}
\tau_0,\frac{\rm i}{2}\tau_i)$ to be the basis $(g_0,g_i)$ in its Lie algebra 
$u(2)=u(1)\oplus su(2)$, where $\tau_0:=\delta^{\bA}_{\bB}$ and $\tau_i:=\tau
^{\bA}_{i{\bB}}$ (with $i=1,2,3$ and ${\bA,\bB}=0,1$) are the standard $SU(2)$ 
Pauli matrices (without the factor $1/\sqrt{2}$). Since the Cartan--Killing 
metric of $SU(2)$ in this basis is $-\delta_{ij}$, the metric $G_{\alpha\beta}$, 
$\alpha,\beta=0,i$, will be the direct sum of $-(1/g^2)\delta_{ij}$ and of 
$-1/{g'}^2$ for some positive coupling constants $g$ and $g'$. The 
corresponding connection 1-forms and field strengths are denoted, 
respectively, by $\omega^i_e$ and $\omega^0_e$, and by $F^i_{ab}$ and $F^0_{ab}$. 

Following \cite{AL73}, for the sake of simplicity we assume to have only the 
first lepton generation consisting of the electron and the corresponding (for 
the sake of simplicity, still massless) neutrino, represented traditionally 
by the Dirac spinor $\Psi=(\psi^A,\bar\chi_{A'})$ (as a column vector) and 
the Weyl spinor $\nu^A$, respectively. In the Weinberg--Salam model they are 
re-arranged such that the multiplets $\psi^{\bA}_A:=(\nu_A,\psi_A)$, ${\bA}=
0,1$, (as a column vector) span the representation space of the defining 
representation of $SU(2)$ (an `$SU(2)$-doublet') with the representation 
matrices $\frac{\rm i}{2}\tau^{\bA}_{i{\bB}}$, while $\bar\chi_{A'}$ is invariant 
with respect to $SU(2)$ transformations (`$SU(2)$-singlet'). 
However, since $U(1)$ is commutative, there is a freedom to choose different 
charges (actually: `hypercharges') in its different representations. In the 
Weinberg--Salam model $-1$ and $-2$ hypercharges are associated with 
$\psi^{\bA}_A$ and $\bar\chi_{A'}$, respectively, i.e. in the two cases the 
$u(1)$ algebra is represented by $-\frac{\rm i}{2}\delta^{\bA}_{\bB}$ and $-2
\frac{\rm i}{2}$, respectively. Thus, with the notation $\psi^2_A:=\chi_A$, 
the spinor representation splits to the direct sum of two and one dimensional 
irreducible representations of $u(2)=u(1)\oplus su(2)$. Therefore, the 
representation matrices $T^r_{\alpha s}$, $r,s=0,1,2$, of subsection 
\ref{sub-1.1} are 

\begin{equation*}
T^r_{0s}=\frac{\rm i}{2}\left(\begin{array}{cc}
                     -\delta^{\bA}_{\bB} & 0 \\
                     0 & 2 \\
                  \end{array}\right), 
\hskip 20pt
T^r_{is}=\frac{\rm i}{2}\left(\begin{array}{cc}
                     \tau^{\bA}_{i{\bB}} & 0\\
                     0 & 0 \\
                  \end{array}\right), 
\end{equation*}
and the (spacetime and gauge) covariant derivative of the spinor fields are 

\begin{equation*}
\nabbla_e\psi^{\bA}_A=\nabla_e\psi^{\bA}_A+\frac{\rm i}{2}\omega^i_e\tau^{\bA}
_{i{\bB}}\psi^{\bB}_A-\frac{\rm i}{2}\omega^0_e\psi^{\bA}_A, \hskip 20pt
\nabbla_e\chi_A=\nabla_e\chi_A+{\rm i}\omega^0_e\chi_A. 
\end{equation*}
To form the Lagrangian ${\cal L}_W$, we need the Hermitian metric $G_{rr'}$. 
By its invariance requirement, $G_{r's}T^s_{\alpha r}+G_{rs'}\bar T^{s'}_{\alpha r'}
=0$, it is necessarily the direct sum of two metrics: Some real constant 
times $\delta_{{\bA}{\bA}'}$ and another constant times the $1\times1$ unit 
matrix. However, by an appropriate re-definition of the spinor fields these 
constants can be chosen to be $1$ or $-1$. In the Weinberg--Salam model 
both are chosen to be 1. 

The representation for the Higgs field will also be chosen to be the defining 
representation of $u(2)$, in which the hypercharge in the representation of 
$u(1)$ is chosen to be $+1$. Thus, the representation matrices 
$T^{\bi}_{{\alpha}{\bj}}$ are 

\begin{equation*}
T^{\bi}_{0{\bj}}=\frac{\rm i}{2}\delta^{\bi}_{\bj},\hskip 20pt
T^{\bi}_{i{\bj}}=\frac{\rm i}{2}\tau^{\bi}_{i{\bj}}, 
\end{equation*}
where ${\bi},{\bj}=1,2$. Hence the covariant derivative of the Higgs field is 

\begin{equation*}
\nabbla_e\Phi^{\bi}=\nabla_e\Phi^{\bi}+\frac{\rm i}{2}\omega^i_e\tau^{\bi}_{i{\bj}}
\Phi^{\bj}+\frac{\rm i}{2}\omega^0_e\Phi^{\bi}. 
\end{equation*}
By the gauge invariance of the fiber metric $G_{\bi\bj'}$, it is necessarily 
proportional to the Euclidean metric, which should be positive definite if 
we want the kinetic term in the energy density of the Higgs field to be 
positive (rather than negative) definite. Then, however, by an appropriate 
redefinition of the Higgs field, $G_{\bi\bj'}=\delta_{\bi\bj'}$ can always be 
achieved. 

In the Weinberg--Salam model $G_{\bi\bj\bk'\bl'}=\lambda G_{\bi({\bk}'}G_{{\bl}')\bj}$ 
for some real constant $\lambda$, and by the positivity requirement of 
$G_{\bi\bj\bk'\bl'}$ this $\lambda$ must be positive. The Yukawa coupling 
constants have only one non-trivial algebraically independent component such 
that the interaction term (\ref{eq:1.2.5a}) of the Lagrangian takes the form 

\begin{equation}
-\frac{G_e}{\sqrt{2}}\Bigl(\nu_A\chi^A\bar\Phi^{{\bf 1}'}+\psi_A\chi^A\bar
\Phi^{{\bf 2}'}+\bar\nu_{A'}\bar\chi^{A'}\Phi^{\bf 1}+\bar\psi_{A'}\bar\chi^{A'}
\Phi^{\bf 2}\Bigr) \label{eq:1.5.5}
\end{equation}
for some real constant $G_e$. The above conditions with $\alpha=0$ in the 
Lagrangian specify the classical field theory behind the Weinberg--Salam 
model (with only one lepton generation) even on any curved spacetime. This 
theory depends on five parameters: $g$, $g'$, $\lambda$, $G_e$ and $\mu^2$. 
The first four are dimensionless but, in the $\hbar=c=1$ units, the physical 
dimension of the fifth is $cm^{-2}$. 

In Minkowski spacetime on a $t={\rm const}$ spacelike hyperplane with unit 
timelike normal $t^a$ (which is, in fact, a constant timelike Killing vector) 
the energy density, defined to be $T_{ab}t^at^b$, is 

\begin{eqnarray*}
\varepsilon\!\!\!\!&=\!\!\!\!&-\frac{1}{g^2}\delta_{ij}\Bigl((t^aF^i_{ac})
 (t^bF^j_{bd})-\frac{1}{4}g^{ab}F^i_{ac}F^j_{bd}\Bigr)g^{cd}-\frac{1}{g^{\prime2}}
 \Bigl((t^aF^0_{ac})(t^bF^0_{bd})-\frac{1}{4}g^{ab}F^0_{ac}F^0_{bd}\Bigr)g^{cd}+ \\
\!\!\!\!&{}\!\!\!\!&+\delta_{\bi\bj'}(t^a\nabbla_a\Phi^{\bi})(t^b\nabbla_b\bar
  \Phi^{\bj'})-\frac{1}{2}\delta_{\bi\bj'}(\nabbla_a\Phi^{\bi})(\nabbla^a\bar
  \Phi^{\bj'})+\frac{1}{4}\lambda(\delta_{\bi\bj'}\Phi^{\bi}\bar\Phi^{\bj'})^2+
  \frac{1}{2}\mu^2(\delta_{\bi\bj'}\Phi^{\bi}\bar\Phi^{\bj'})+ \\
\!\!\!\!&{}\!\!\!\!&+\frac{\rm i}{2}\delta_{{\bA}{\bA}'}t^{AA'}\bigl(\bar
  \psi^{{\bA}'}_{A'}t^b\nabbla_b\psi^{\bA}_A-\psi^{\bA}_At^b\nabbla_b\bar\psi
  ^{{\bA}'}_{A'}\bigr)+\frac{\rm i}{2}t^{AA'}\bigl(\bar\chi_{A'}t^b\nabbla_b\chi_A
  -\chi_At^b\nabbla_b\bar\chi_{A'}\bigr). 
\end{eqnarray*}
Recall that the `vacuum states' of a field theory are usually defined to be 
those configurations that are invariant under the action of the group of 
spacetime isometries, i.e. the Poincar\'e group in flat spacetime, and 
pointwise minimize the energy density. In particular, every quantity with 
spacetime (co-)vector or spinor index must be vanishing in these states. 
Hence, in such vacuum states, only the Higgs field can be non-zero, but it 
must be constant. Clearly, such configurations solve the field equations: 
They provide static, constant solutions of them. In these configurations the 
energy density above reduces to $\frac{1}{2}\mu^2\delta_{\bi\bj'}\Phi^{\bi}\bar
\Phi^{\bj'}+\frac{1}{4}\lambda(\delta_{\bi\bj'}\Phi^{\bi}\bar\Phi^{\bj'})^2$, whose 
minimum is at $\Phi^{\bi}=0$ if $\mu^2\geq0$; and at the states $\Phi^{\bi}_v$ 
for which $v^2:=\delta_{\bi\bj'}\Phi^{\bi}_v\bar\Phi^{\bj'}_v=-\mu^2/\lambda$ if 
$\mu^2<0$. In the latter case the energy-momentum tensor is proportional to 
the spacetime metric, $T_{ab}=-\frac{1}{4}\lambda v^4g_{ab}$, i.e. it is a 
pure trace. 

The significance of these vacuum states $\Phi^{\bi}_v$ is that it is only the 
\emph{set} of these vacuum states (but not the individual states) that is 
invariant under the action of the gauge group, and, in the so-called unitary 
gauge via the BEH mechanism, their non-trivial `vacuum value' $v$ yields 
mass to certain \emph{a priori massless} fields (for the details see e.g. 
\cite{AL73,ChLi}, or subsection \ref{sub-3.3} below). In particular, the 
mass of the $W^\pm$ and $Z$ gauge bosons and the electron, respectively, are 
$m_W=\frac{1}{2}gv$, $m_Z=\frac{1}{2}\sqrt{g^2+g^{\prime 2}}v$ and $m_e=
\frac{1}{\sqrt{2}}G_ev$, while the mass of the observed Higgs field is $m_H
=\sqrt{2\lambda}v$. Moreover, for the charge $e$ of the electron we obtain 
$\vert e\vert=gg'(g^2+g^{\prime 2})^{-1/2}$. Note that all these masses are 
proportional to the vacuum value $v$ of the field $\Phi^{\bi}$ at the 
symmetry breaking vacuum state. Measuring these masses, the parameters of 
the model can be determined in terms of them and the charge: $g=0.652$, 
$g'=0.357$, $G_e=2.87\times 10^{-6}$, $\lambda=1/8$ and $\vert\mu\vert\simeq 
87.2 GeV/c^2\simeq 4.4\times 10^{15}cm^{-1}$ (in the $\hbar=c=1$ units). 


\subsection{The Einstein--conformally coupled Standard Model system}
\label{sub-1.6}

The structure and the logic, and also the particular successes and results 
of the Weinberg--Salam model motivate its generalization to non-flat (in 
particular, to non-stationary) spacetime in a way such that (i.) it reduces 
to the original theory in flat spacetime (i.e. be compatible with the 
present day particle physics), and (ii.) it behaves as simply as possible 
under conformal rescalings of the spacetime metric (i.e. be compatible with 
the overall picture that basically the fundamental particles are massless). 
These requirements suggest to choose the constant $\alpha$ in the Lagrangian 
of the Higgs field to be $1/6$, but, apart from this extra coupling term, 
both the Standard Model and Einstein's General Relativity are kept as they 
are. This yields a non-trivial generalization of the flat-spacetime 
Weinberg--Salam model (or, more generally, the extended Standard Model of 
particle physics allowing even massive or sterile neutrinos), the 
Einstein--conformally coupled Standard Model (EccSM) system. Thus, in this 
model, the coupling of the matter to gravity is not only the so-called 
`minimal coupling' dictated by the principle of general covariance, but 
there is the extra term which improves the conformal properties of the 
model: It is only the Higgs potential term $V$ that violates the complete 
conformal invariance of the matter sector. Any further non-trivial change 
of the theory could easily yield effects that contradict the highly precise 
experimental tests of the Standard Model of particle physics or of 
Einstein's theory. 

With the sign conventions of the introduction, Einstein's equations take the 
form 

\begin{equation}
R_{ab}-\frac{1}{2}Rg_{ab}=-\kappa T_{ab}-\Lambda g_{ab}. \label{eq:1.6.1}
\end{equation}
Taking its trace (and assuming that $\alpha=1/6$, as in the rest of the 
paper) by (\ref{eq:1.3.4}) we obtain that $R=4\Lambda+\kappa\mu^2G_{\bi\bj'}
\Phi^{\bi}\bar\Phi^{\bj'}$. With this substitution in the field equation 
(\ref{eq:1.3.1b}) for the Higgs field we obtain that 

\begin{equation}
\nabbla_a\bigl(\nabbla{}^a\Phi^{\bj}\bigr)G_{\bj\bi'}=-\bigl(\mu^2+\frac{2}{3}
\Lambda\bigr)\Phi^{\bj}G_{\bj\bi'}-\Bigl(G_{\bi'\bj'\bk\bl}+\frac{1}{6}\kappa\mu^2
G_{\bi'\bk}G_{\bj'\bl}\Bigr)\bar\Phi^{\bj'}\Phi^{\bk}\Phi^{\bl}-{\rm i}\varepsilon
^{AB}\psi^r_A\psi^s_BY_{rs{\bi}'}. \label{eq:1.6.2}
\end{equation}
Comparing the field equations (\ref{eq:1.3.1a})-(\ref{eq:1.3.1c}) for 
$\alpha=0$ with those given by (\ref{eq:1.3.1a}), (\ref{eq:1.3.1c}) for 
$\alpha=1/6$ and (\ref{eq:1.6.2}), we see that they have \emph{exactly the 
same structure}, and hence there is no difference in the structure of their 
solution \emph{on a given spacetime geometry} (i.e. when we neglect the 
gravitational back-reaction). Thus, the conformal coupling term in ${\cal L}
_H$ does \emph{not} yield any qualitative change in the low energy particle 
physics. The only change is that in the field equations for the Higgs field 
the cosmological constant modifies the mass parameter of the Higgs field 
according to $\mu^2\mapsto\mu^2+\frac{2}{3}\Lambda$, and the self-interaction 
term is also modified by a term proportional to Newton's gravitational 
constant. In particular, in the Weinberg--Salam model the latter is $\lambda
\mapsto\lambda+\frac{1}{6}\kappa\mu^2$. Therefore, in particular, in the 
conformally coupled model in the presence of a nonzero cosmological constant 
the Higgs field would have a non-zero effective mass parameter even if 
$\mu^2$ were vanishing; i.e. even if the Higgs field were a multiplet of 
strictly \emph{conformally invariant} scalar fields. However, the 
contribution of the cosmological constant to the Higgs rest mass parameter is 
extremely tiny: The present day estimated value of the cosmological constant 
is $\Lambda\simeq 10^{-58}cm^{-2}$, while, as we noted in the 
previous subsection, $-\mu^2\simeq 1.8\times 10^{31}cm^{-2}$ (in the $\hbar=c
=1$ units). The shift of the self-interaction parameter is also very small: 
While $\lambda=1/8$, its correction term is only $\frac{1}{6}\kappa\mu^2
\simeq-2.1\times 10^{-34}$. Hence, at the present accelerator energies, both 
the Standard Model of particle physics and the EccSM system give essentially 
the \emph{same quantitative predictions}. 

Next, let us take into account Einstein's equations in the expression of 
the energy-momentum tensor, too, and introduce the notations $\vert\Phi\vert^2
:=G_{\bi\bj'}\Phi^{\bi}\bar\Phi^{\bj'}$ and $t_{ab}:=T_{ab}+\frac{1}{6}G_{ab}\vert
\Phi\vert^2$, where $G_{ab}$ is the Einstein tensor (see equation 
(\ref{eq:1.3.2})). Then by the definitions and Einstein's equations we can 
write 

\begin{eqnarray}
\kappa T_{ab}\!\!\!\!&{}\!\!\!\!&=\kappa t_{ab}-\frac{\kappa}{6}\vert\Phi
  \vert^2G_{ab}=\kappa t_{ab}\Bigl(1+\frac{\kappa}{6}\vert\Phi\vert^2\Bigr)+
  \frac{\kappa}{6}\vert\Phi\vert^2\Lambda g_{ab}-\bigl(\frac{\kappa}{6}\vert
  \Phi\vert^2\bigr)^2G_{ab}=\cdots= \nonumber \\
\!\!\!\!&{}\!\!\!\!&=\kappa\bigl(t_{ab}+\frac{\Lambda}{6}\vert\Phi\vert^2
  \bigr)\sum^\infty_{n=0}(\frac{\kappa}{6}\vert\Phi\vert^2)^n+\kappa T_{ab}
  \lim_{n\rightarrow\infty}(\frac{\kappa}{6}\vert\Phi\vert^2)^n. \label{eq:1.6.3a}
\end{eqnarray}
However, on the right the limit is finite (in fact, zero) and the geometric 
series converges precisely when $\vert\Phi\vert^2<6/\kappa$. If $\vert\Phi
\vert^2>6/\kappa$, then the analogous argumentation yields 

\begin{equation}
-\kappa T_{ab}-\Lambda g_{ab}=G_{ab}=\frac{6}{\kappa\vert\Phi\vert^2}\bigl(
\kappa t_{ab}+\Lambda g_{ab}\bigr)\sum^\infty_{n=0}\bigl(\frac{6}{\kappa\vert
\Phi\vert^2}\bigr)^n+G_{ab}\lim_{n\rightarrow\infty}\bigl(\frac{6}{\kappa\vert
\Phi\vert^2}\bigr)^n. \label{eq:1.6.3b}
\end{equation}
In both cases the energy-momentum tensor takes the form 

\begin{eqnarray}
T_{ab}\!\!\!\!&=\!\!\!\!&\bigl(1-\frac{1}{6}\kappa\vert\Phi\vert^2\bigr)^{-1}
  \Bigl\{G_{\alpha\beta}F^\alpha_{ac}F^\beta_{bd}g^{cd}-\frac{1}{4}g_{ab}G_{\alpha\beta}
  F^\alpha_{ce}F^\beta_{df}g^{cd}g^{ef}+ \label{eq:1.6.3} \\
\!\!\!\!&+\!\!\!\!&G_{\bi\bj'}\bigl(\nabbla_{(a}\Phi^{\bi}\bigr)\bigl(\nabbla_{b)}
  \bar\Phi^{\bj'}\bigr)-\frac{1}{2}g_{ab}G_{\bi\bj'}\bigl(\nabbla_c\Phi^{\bi}\bigr)
  \bigl(\nabbla^c\bar\Phi^{\bj'}\bigr)+\frac{1}{4}g_{ab}G_{\bi\bj\bk'\bl'}\Phi^{\bi}
  \Phi^{\bj}\bar\Phi^{\bk'}\bar\Phi^{\bl'}- \nonumber \\
\!\!\!\!&-\!\!\!\!&\frac{1}{6}\nabla_a\nabla_b\bigl(\vert\Phi\vert^2\bigr)+
  \frac{1}{6}g_{ab}\nabla_c\nabla^c\bigl(\vert\Phi\vert^2\bigr)+\frac{1}{2}
  g_{ab}\bigl(\mu^2+\frac{1}{3}\Lambda\bigr)\vert\Phi\vert^2+  \nonumber \\
\!\!\!\!&+\!\!\!\!&\frac{\rm i}{4}G_{rr'}\Bigl(\bar\psi^{r'}_{A'}
  \nabbla_{BB'}\psi^r_A+\bar\psi^{r'}_{B'}\nabbla_{AA'}\psi^r_B-\psi^r_A\nabbla
  _{BB'}\bar\psi^{r'}_{A'}-\psi^r_B\nabbla_{AA'}\bar\psi^{r'}_{B'}\Bigr)- 
  \nonumber \\
\!\!\!\!&-\!\!\!\!&\frac{\rm i}{2}g_{ab}\Bigl(\bar\psi^{r'}_{C'}\bigl(G_{r'r}
 \nabbla^{C'C}\psi^r_C+\varepsilon^{C'D'}\bar\psi^{s'}_{D'}\bar Y_{r's'{\bi}}
 \Phi^{\bi}\bigr)-\psi^r_C\bigl(G_{rr'}\nabbla^{CC'}\bar\psi^{r'}_{C'}+\varepsilon
 ^{CD}\psi^s_DY_{rs{\bi}'}\bar\Phi^{{\bi}'}\bigr)\Bigr)\Bigr\}. \nonumber
\end{eqnarray}
In fact, this expression of the energy-momentum tensor can be obtained 
independently of the power series arguments above simply by expressing the 
Einstein tensor by $T_{ab}$ and $\Lambda$ already from the first equality of 
(\ref{eq:1.6.3a}). However, if $\vert\Phi\vert^2=6/\kappa$, then the limit 
on the right of (\ref{eq:1.6.3a}) is just $\kappa T_{ab}$ and the geometric 
series \emph{diverges}. The breakdown of the expression (\ref{eq:1.6.3a}) 
(or of (\ref{eq:1.6.3b})) of $T_{ab}$ when $\vert\Phi\vert^2=6/\kappa$ 
indicates that this configuration corresponds to a potential singularity of 
the EccSM system. 

Thus, what changed dramatically is the structure of the energy-momentum 
tensor of the matter fields: Although this expression for $T_{ab}$ is 
polynomial in the spinor and the gauge fields, it is \emph{not} polynomial 
in the Higgs field. It, and via Einstein's equations the spacetime geometry, 
may have two different kinds of singularities. First, when the matter field 
variables (or at least some of them) are singular; and, second, when all 
these field variables are finite but the square of the norm of the Higgs 
field takes the special value $6/\kappa=3c^4/4\pi G$. (In the $\hbar=c=1$ 
units, this value is $6/\kappa\simeq 8.6\times10^{64}cm^{-2}$.) Therefore, by 
Einstein's equations we obtain the remarkable fact that \emph{the conformal 
coupling of the Higgs sector to gravity yields that the spacetime geometry 
can be singular even if all the matter fields are finite}. In particular, 
$T_{ab}$ could be diverging even if $\Phi^{\bi}$ is bounded and spatially 
constant, and the gauge and the spinor fields are identically vanishing 
(e.g. in the presence of FRW or Kantowski--Sachs symmetries). 

In the absence of the gauge and spinor fields the energy-momentum tensor near 
the first singularity diverges like $\sim\vert\Phi\vert^2$, and in the second 
like $\sim(6/\kappa-\vert\Phi\vert^2)^{-1}\sim(\sqrt{6/\kappa}-\vert\Phi
\vert)^{-1}$. In addition, since $R=4\Lambda+\kappa\mu^2\vert\Phi\vert^2$, at 
the Big Bang $R\rightarrow-\infty$, but the curvature scalar remains bounded 
when $\vert\Phi\vert^2$ tends to $6/\kappa$. Thus, the second singularity 
seems less violent than the first, and hence, motivated by the cosmological 
terminology, we call the second the `Small Bang' (though such a singularity 
may appear during a gravitational collapse, deeply behind the event horizon). 
(Note, however, that the energy-momentum tensor of the Higgs field of 
the Standard Model, i.e. without the conformal coupling term $\frac{1}{12}R
\vert\Phi\vert^2$ in its Lagrangian, has a single, but much more violent Big 
Bang type singularity whose energy-momentum tensor diverges as $\sim\vert
\Phi\vert^4$. Thus, although the conformal coupling produces an extra 
singularity, but at the same time it tempers the one in the Einstein--Standard 
Model system.) The nature of the Small Bang singularity differs from that of 
the Big Bang: While the latter is like a pole, the former is an infinite 
discontinuity of the energy-momentum tensor. The energy-momentum tensor 
\emph{changes sign} at the Small Bang singularity: If the energy density is 
positive on one side of the $\vert\Phi\vert^2=6/\kappa$ hypersurface (as a 
singularity) in the phase space, then it is negative on the other. However, 
the field configurations in which $\vert\Phi\vert^2=6/\kappa$ are \emph{not 
necessarily} singularities of the energy-momentum tensor, because the 
numerator in the expression (\ref{eq:1.6.3}) could also be zero at the same 
time (see also subsection \ref{sub-2.4.2}). Moreover, the existence of 
singularities of the energy-momentum tensor \emph{in the phase space} does 
not necessarily imply the existence of singularities \emph{in the solutions} 
of the field equations. 

In fact, the detailed analysis of the field equations of the 
Einstein-conformally coupled Higgs (EccH) system in the presence of FRW 
symmetries shows that the Small Bang singularities do appear in solutions 
\cite{SzW}: The field equations have asymptotic solutions in which $\vert
\Phi\vert^2=6/\kappa$ corresponds to a \emph{physical}, scalar polynomial 
curvature singularity in which $R_{ab}R^{ab}$ diverges. Also, there are 
asymptotic solutions in which $\vert\Phi\vert^2$ takes the value $6/\kappa$ 
at \emph{regular} spacetime points with bounded energy-momentum tensor, and 
the solution can be continued to the $\vert\Phi\vert^2>6/\kappa$ side of the 
phase space of the EccH system. Since in the present paper we study the 
consequences of the \emph{kinematical} structure of the EccSM system, now we 
do not need to know the detailed properties of the solutions. They will be 
published in a separate paper \cite{SzW}. 

The phase space of the EccSM system splits into the disjoint domains where 
$\vert\Phi\vert^2<6/\kappa$ (the states of our low energy world) and where 
$\vert\Phi\vert^2>6/\kappa$, and a part of the hypersurface $\vert\Phi\vert^2
=6/\kappa$ could represent regular (non-singular) states. 
(In the FRW case these regular states form only a \emph{2-surface} in 
the 3-dimensional $\vert\Phi\vert^2=6/\kappa$ hypersurface, see subsection 
\ref{sub-2.4.2}.) Looking at this result from a different perspective, we 
see that for \emph{generic} Yang--Mills gauge and Weyl spinor field 
configurations, i.e. when the numerator between the curly bracket in 
(\ref{eq:1.6.3}) is not zero, the Higgs field cannot be arbitrarily large: 
Its pointwise norm $\vert\Phi\vert^2$ is \emph{bounded from above} by 
$6/\kappa$. Otherwise the energy-momentum tensor and, via Einstein's 
equations, the spacetime geometry would be singular (Small Bang). The role 
of this bound is analogous to that of the speed of light $c$ in relativistic 
particle mechanics, where infinite energy would be needed to speed a particle 
up to $c$. Here, infinite energy would have to be pumped into the Higgs field 
to achieve this upper bound. We stress that this natural cut-off is 
\emph{non-perturbative}, present already in the \emph{classical} theory, and 
provided by Newton's gravitational constant $G$. In the $\hbar=c=1$ units 
this bound is roughly one order of magnitude above the Planck scale.

\subsection{The problem of `vacuum states' of gravitating systems}
\label{sub-1.7}

As we mentioned in subsection \ref{sub-1.5}, the vacuum states of a field 
theory in Minkowski spacetime are usually defined to be the field 
configurations which are Poincar\'e invariant and minimizing the energy 
density. In particular, these states are both translation and boost-rotation 
invariant, and solve the field equations, too. 

However, this definition cannot be applied directly to gravitating systems. 
Indeed, the physical system is the \emph{coupled} Einstein--matter system, 
in which the matter sector is only a subsystem of the whole, and, as a 
manifestation of the principle of equivalence, there is no non-dynamical 
(e.g. flat) background metric whose isometries could be required to be the 
symmetries of the matter fields in the vacuum state, too. Moreover, we would 
need an appropriate expression, in fact a \emph{definition}, for the energy 
density of the matter+gravity system. However, as is well known, there is 
no well defined (i.e. gauge invariant, tensorial) energy-momentum 
\emph{density} of the gravitational `field': Any such \emph{local} expression 
is \emph{necessarily} $SO(1,3)$ gauge dependent or/and pseudotensorial, as a 
consequence of the equivalence principle (and, ultimately, the E\"otv\"os 
experiment). (For a review of these difficulties, and also for the possible 
resolutions of them, see e.g. \cite{Sz09}.) 

Thus, instead of the energy density, we should use some \emph{total} or 
\emph{quasi-local} energy-momentum functional. Such a functional would be 
the integral of some local (gauge dependent) expression on a \emph{spacelike 
hypersurface} $\Sigma$. Hence, in general, the notion of the `vacuum states' 
(as states that are extremal points of such a functional) depends on the 
hypersurface. Thus, such a `vacuum state' is only an \emph{instantaneous} 
state associated with the instant represented by $\Sigma$. If, however, the 
energy-momentum in question depends only on the boundary $\partial\Sigma$ of 
the hypersurface, which could be a closed spacelike 2-surface in spacetime 
or at infinity, but does \emph{not} depend on the hypersurface itself (i.e. 
the energy-momentum is `conserved'), then the `vacuum state' introduced by 
such an energy-momentum expression can be interpreted as being associated 
with the whole domain of dependence (or Cauchy development) of $\Sigma$. 
This domain of dependence could be the whole spacetime, or only an open 
subset of it. 

Indeed, in general relativity there are various notions of total 
energy-momentum (or at least total mass), depending on the global asymptotic 
structure of the spacetime and the sign of the cosmological constant 
\cite{ADM,Bondietal,AD,SzaTo,Sz12,Sz13}. After renormalizing for the 
cosmological constant term, all these have the general form 

\begin{eqnarray}
{\tt P}[\Psi]:=\int_{\Sigma}\Bigl(\!\!\!\!&{}\!\!\!\!&\frac{2}{\kappa}t^{AA'}
  t^{BB'}t^{CC'}\bigl({\cal D}_{(AB}\lambda_{C)}{\cal D}_{(A'B'}\bar\lambda_{C')}+ 
  {\cal D}_{(AB}\mu_{C)}{\cal D}_{(A'B'}\bar\mu_{C')}\bigr)+\nonumber \\
\!\!\!\!&{}\!\!\!\!&+\frac{1}{2}t^aT_{ab}\bigl(\lambda^B\bar\lambda^{B'}+\mu^B
  \bar\mu^{B'}\bigr)\Bigr){\rm d}\Sigma. \label{eq:1.7.1}
\end{eqnarray}
Here ${\cal D}_e$ is the unitary spinor form of the so-called Sen 
connection on $\Sigma$; and the Dirac spinor $\Psi=(\lambda^A,\bar\mu^{A'})$, 
representing the spinor constituents of the vector field $K^a=\frac{1}{2}
(\lambda^A\bar\lambda^{A'}+\mu^A\bar\mu^{A'})$ that defines the appropriate 
component of the energy-momentum, is subject to a certain gauge condition. 
The gauge condition is that $\Psi$ must be a solution of an appropriate 
linear \emph{elliptic} partial differential equation, e.g. some version of 
Witten's equation, on $\Sigma$. The first term in the integrand could be 
identified with the contribution of the gravitational `field' in this gauge 
to the total energy-momentum. (For the details, see e.g. \cite{SzaTo}.) 

The significance of all these expressions in general relativity is that, 
provided the energy-momentum tensor satisfies the \emph{dominant energy 
condition} \cite{HE}, they yield non-negative total energy/mass, and have 
the so-called rigidity property: The zero (i.e. minimal) total energy 
matter+gravity configurations are the (locally) Minkowski, de Sitter or 
anti-de Sitter spacetimes (depending on the asymptotic structure of the 
spacetime and the sign of the cosmological constant) with \emph{vanishing} 
matter fields. Thus, these configurations can be interpreted to be the 
\emph{global, spacetime vacuum states} of Einstein's theory with matter 
fields satisfying the dominant energy condition. Although, in contrast to 
the total energy-momenta, there is no generally accepted and completely 
satisfactory notion of \emph{quasi-local} energy-momentum (for a 
comprehensive review of the various suggestions, see \cite{Sz09}), certain 
expressions (e.g. that of Dougan and Mason \cite{DM}) have analogous 
positivity and rigidity properties \cite{Sz93}, and hence can yield a well 
defined \emph{quasi-local spacetime vacuum state}. 

Unfortunately, however, the energy-momentum tensor (\ref{eq:1.6.3}) does 
\emph{not} satisfy even the weak energy condition. Thus, strictly speaking, 
the positivity and rigidity results for the existing total or quasi-local 
energy-momentum functionals in their present form cannot be used to define 
the total or quasi-local `vacuum states' of the EccSM system; and it is 
still not clear whether or not the above energy positivity and rigidity 
proofs could be generalized appropriately. Moreover, if the typical (partial 
Cauchy) hypersurface $\Sigma$ is not compact, then the energy-momentum 
functional is not finite unless appropriate fall-off conditions for the 
matter and geometry are imposed. Clearly, in a (more-or-less homogeneous) 
cosmological spacetime no such fall-off condition can be required to hold. 
A further potential difficulty is that while the minimal value of the total 
mass on a single hypersurface in \emph{closed universes} (with non-negative 
$\Lambda$) characterizes the locally flat/de Sitter spacetimes (i.e. the 
rigidity property can be proven if the matter fields satisfy the dominant 
energy condition), but in general this mass \emph{does} depend on the 
spacelike hypersurface \cite{Sz12,Sz13}. 

Nevertheless, although mathematically we could not \emph{derive} the (global 
or quasi-local) `spacetime vacuum states' of the EccSM system from the 
results above, \emph{on physical grounds} it seems natural to 
\emph{postulate} that these states are certain \emph{locally maximally 
symmetric spacetime+matter configurations}. All these are stationary 
configurations, but, as we will see, they do \emph{not} solve the field 
equations and extremize the energy functional at the same time. In the 
presence of gravity the familiar notion of the spacetime vacuum states is 
lost. We discuss this problem in subsection \ref{sub-1.7.1}. 

Motivated by the negative results with the spacetime vacuum states and the 
fact that in closed universes the total mass is not conserved, we should 
consider a weaker notion of vacuum states, the \emph{instantaneous} ones. 
These are defined to be the stationary points of the energy-momentum 
functional, and they depend on the hypersurface $\Sigma$. This notion is 
certainly legitimate in a cosmological context, and could provide a basis of 
the realization of Mach's idea on the origin of inertia and the rest masses. 
In fact, this notion yields the time dependence of rest masses, and, in 
particular, their non-trivial genesis. This notion will be discussed in 
subsections \ref{sub-1.7.2} and \ref{sub-3.2} in detail.

\subsubsection{The spacetime vacuum states}
\label{sub-1.7.1}

Without further mathematical justification, let us \emph{postulate} that the 
(global) spacetime vacuum states of the EccSM system correspond to certain 
locally maximally symmetric spacetimes and matter fields admitting the same 
geometric symmetries. Thus, the spacetime is assumed to be locally de Sitter, 
Minkowski or anti de Sitter. Hence the Einstein tensor is $R_{ab}-\frac{1}{2}
Rg_{ab}=-\frac{1}{4}Rg_{ab}$, and the energy-momentum tensor is a pure trace: 
$T_{ab}=\frac{1}{4}Tg_{ab}$. Hence, the energy density, seen by any observer, 
is $\varepsilon=\frac{1}{4}T$. 

Since the matter fields are required to be invariant under the action of the 
(local) isometry group of the spacetime, the matter fields at each point $p
\in M$ must be invariant under the action of the stabilizer group of $p$ in 
the isometry group, i.e. $SO(1,3)$. Therefore, all the physical fields 
specifying the spacetime vacuum state and have a spacetime tensor or spinor 
index, viz. $F^\alpha_{ab}$, $\nabbla_e\Phi^{\bi}$ and $\psi^r_A$, must be 
vanishing everywhere. In particular, the vacuum value of the Higgs field must 
be \emph{gauge covariantly} constant, and hence, by $\nabla_e\vert\Phi\vert^2=
(\nabbla_e\Phi^{\bi})G_{\bi\bj'}\bar\Phi^{\bj'}+\Phi^{\bi}G_{\bi\bj'}(\nabbla_e\bar
\Phi^{\bj'})=0$, its Hermitian pointwise norm is constant on $M$. For the sake 
of simplicity, we assume that by an appropriate globally defined gauge 
transformation the locally flat gauge field can be transformed to be 
vanishing. Hence, the Higgs field is constant on $M$, too: $\nabla_e\Phi^{\bi}
=0$. 

In these configurations gravity does not contribute explicitly to 
(\ref{eq:1.7.1}), and (\ref{eq:1.7.1}) reduces to the integral of the 
energy-momentum tensor of the matter fields. By (\ref{eq:1.6.3}) the energy 
density is 

\begin{equation}
\varepsilon=\frac{1}{4}\frac{1}{1-\frac{1}{6}\kappa\vert\Phi\vert^2}\Bigl(
2\bigl(\mu^2+\frac{\Lambda}{3}\bigr)\vert\Phi\vert^2+G_{\bi\bj\bk'\bl'}\Phi^{\bi}
\Phi^{\bj}\bar\Phi^{\bk'}\bar\Phi^{\bl'}\Bigr). \label{eq:1.7.2}
\end{equation}
For the sake of simplicity, we also assume in this subsection that the 
self-interaction coefficient is $G_{\bi\bj\bk'\bl'}=\lambda G_{\bk'(\bi}
G_{\bj)\bl'}$ with $\lambda>0$. 

The field configurations $F^\alpha_{ab}=0$, $\psi^r_A=0$ with $\nabla_e\Phi
^{\bi}=0$ solve (\ref{eq:1.3.1a}) and (\ref{eq:1.3.1c}), but (\ref{eq:1.6.2}) 
gives the additional condition 

\begin{equation}
\mu^2+\frac{2}{3}\Lambda+\lambda\vert\Phi\vert^2+\frac{1}{6}\kappa\mu^2
\vert\Phi\vert^2=0. \label{eq:1.7.5}
\end{equation}
Its solutions, denoted by $\Phi^{\bi}_g$ (`ground states'), have the norm 

\begin{equation}
\vert\Phi_g\vert^2=-\frac{\mu^2+\frac{2}{3}\Lambda}{\lambda+\frac{1}{6}\kappa
\mu^2}=-\frac{\mu^2}{\lambda}+\frac{\kappa}{6}\frac{\mu^4}{\lambda^2}-
\frac{2}{3}\frac{\Lambda}{\lambda}+...  \label{eq:1.7.6}
\end{equation}
The first term in the expansion is the well known vacuum value in the 
Standard Model in Minkowski spacetime (see subsection \ref{sub-1.5}), the 
second, being proportional to Newton's gravitational constant, is of proper 
gravitational origin, while the third has cosmological origin. 

However, the ground states $\Phi^{\bi}_g$ do \emph{not} minimize the energy 
density. In fact, by (\ref{eq:1.7.2}), the critical points of $\varepsilon$ 
are at $\Phi=0$ and at the solutions of 

\begin{equation}
-\frac{1}{12}\kappa\lambda\vert\Phi\vert^4+\lambda\vert\Phi\vert^2+(\mu^2+
\frac{1}{3}\Lambda)=0. \label{eq:1.7.3}
\end{equation}
Its solutions, representing the minima of $\varepsilon$ and denoted by 
$\Phi^{\bi}_v$ (`vacuum states'), have the norm 

\begin{equation}
\vert\Phi_v\vert^2=\frac{6}{\kappa}\Bigl(1-\sqrt{1+\frac{\kappa}{3\lambda}
(\mu^2+\frac{1}{3}\Lambda)}\Bigr)=-\frac{\mu^2}{\lambda}+\frac{\kappa}{12}
\frac{\mu^4}{\lambda^2}-\frac{1}{3}\frac{\Lambda}{\lambda}+.... 
\label{eq:1.7.4}
\end{equation}
($\Phi=0$ and the solution with the $+$ sign in front of the square root 
are local \emph{maxima} rather than minima of the energy density.) The 
structure of its expansion and the meaning of the corrections are similar 
to those of $\vert\Phi_g\vert^2$. The minimum value of $\varepsilon$ is 
$-\frac{1}{4}\lambda\vert\Phi_v\vert^4$; i.e. in the vacuum states the 
spacetime is anti-de Sitter (rather than Minkowski). 

Comparing $\vert\Phi_g\vert^2$ and $\vert\Phi_v\vert^2$ we find that these 
do not coincide. (\ref{eq:1.7.6}) would be a solution of (\ref{eq:1.7.3}) 
precisely when $(\Lambda-\kappa\mu^4/4\lambda)(\Lambda+3\mu^2+9\lambda/
\kappa)=0$, i.e. if $\Lambda=\kappa\mu^4/4\lambda$ or $\Lambda=-3\mu^2-9
\lambda/\kappa$ held. However, their left hand side is $\simeq10^{-58}cm^{-2}$, 
but the right hand sides are $\simeq1.1\times10^{-2}cm^{-2}$ and $\simeq-1.6
\times10^{64}cm^{-2}$, respectively. 

Next, let us calculate the rest masses (see e.g. \cite{AL73}). Let $\Phi^{\bi}
_0$ denote a constant Higgs field on $M$ (which could be $\Phi^{\bi}_v$ or 
$\Phi^{\bi}_g$), and choose the basis $\{g_\alpha\}$ of the Lie algebra of the 
gauge group such that $T^{\bi}_{\alpha{\bj}}\Phi^{\bj}_0=0$ for $\alpha=1,...,k_0$, 
and $T^{\bi}_{\alpha{\bj}}\Phi^{\bj}_0\not=0$ for $\alpha=k_0+1,...,k$. (Thus 
$\{g_1,...,g_{k_0}\}$ is a basis in the Lie algebra of the stabilizer subgroup 
of $\Phi^{\bi}_0$ in $G$.) Then, as Weinberg showed \cite{We73}, for any 
compact gauge group and Higgs field there is a gauge, the so-called unitary 
gauge, in which the Higgs field is $\Phi^{\bi}=\Phi^{\bi}_0+H^{\bi}$, where 
$H^{\bi}$ is the sum of a field proportional to $\Phi^{\bi}_0$, say $H\vert
\Phi_0\vert^{-1}\Phi^{\bi}_0$ for some \emph{real} function $H$, and another one 
orthogonal to all the vectors $T^{\bi}_{\alpha{\bj}}\Phi^{\bj}_0$ for $\alpha=
k_0+1,...,k$. (N.B.: For \emph{real} Higgs fields $\Phi^{\bi}_0$ is always 
$G_{\bi\bj}$-orthogonal to all the vectors $T^{\bi}_{\alpha{\bj}}\Phi^{\bj}_0$ for 
$\alpha=k_0+1,...,k$, but for \emph{complex} Higgs fields $\bar\Phi^{\bi'}_0
G_{\bi'\bj}T^{\bj}_{\alpha{\bk}}\Phi^{\bk}_0$ is not zero, it is only purely 
imaginary.) In terms of these 

\begin{eqnarray}
{\cal L}_{H}\!\!\!\!&{}\!\!\!\!&-V=\frac{1}{2}G_{\bi\bj'}g^{ab}\bigl(\nabla_a
  H^{\bi}\bigr)\bigl(\nabla_b\bar H^{\bj'}\bigr)+\frac{1}{2}g^{ab}\omega^\alpha_a
  \omega^\beta_b\bigl(T^{\bi}_{\alpha{\bk}}\Phi^{\bk}_0\bigr)G_{\bi\bj'}\bigl(\bar 
  T^{\bj'}_{\beta{\bl'}}\bar\Phi^{\bl'}_0\bigr)- \nonumber \\
\!\!\!\!&{}\!\!\!\!&-\frac{1}{2}\bigl(\mu^2+\frac{1}{2}\lambda\vert\Phi_0\vert
  ^2+\frac{1}{6}R\bigr)\vert\Phi_0\vert^2-\frac{1}{2}\bigl(\mu^2+\lambda
  \vert\Phi_0\vert^2+\frac{1}{6}R\bigr)\Bigl(G_{\bi\bj'}\Phi^{\bi}_0\bar H^{\bj'}
  +G_{\bi\bj'}H^{\bi}\bar\Phi^{\bj'}_0\Bigr)- \nonumber \\
\!\!\!\!&{}\!\!\!\!&-\frac{1}{2}\bigl(\mu^2+\lambda\vert\Phi_0\vert^2+
  \frac{1}{6}R\bigr)\vert H^{\bi}\vert^2-\frac{1}{4}\lambda\Bigl(G_{\bi\bj'}
  \Phi^{\bi}_0\bar H^{\bj'}+G_{\bi\bj'}H^{\bi}\bar\Phi^{\bj'}_0\Bigr)^2+{\cal O}(3), 
  \label{eq:1.7.7}
\end{eqnarray}
and 

\begin{equation*}
{\cal L}_I=\frac{\rm i}{2}\Bigl(\varepsilon^{A'B'}\bar\psi^{r'}_{A'}\bar
\psi^{s'}_{B'}\bar Y_{r's'{\bi}}\Phi^{\bi}_0-\varepsilon^{AB}\psi^r_A\psi^s_B
Y_{rs{\bi}'}\bar\Phi^{\bi'}_0\Bigr)+{\cal O}(3). 
\end{equation*}
Here $\vert H^{\bi}\vert^2:=G_{\bi\bj'}H^{\bi}\bar H^{\bj'}$ and ${\cal O}(3)$ 
stands for all the terms cubic or higher order in the field variables 
$H^{\bi}$, $\omega^\alpha_a$ and $\psi^r_A$. The rest mass of the gauge and the 
spinor fields can be read-off from these expressions. 

In particular, in the Einstein--conformally coupled Weinberg--Salam model, 
$\Phi^{\bi}_0$ can be chosen to have the form $(0,\vert\Phi_0\vert)$, and also 
$H^{\bi}=(0,H)$ (as column vectors). Then we find $m_e=\frac{1}{\sqrt{2}}G_e
\vert\Phi_0\vert$, $m_W=\frac{1}{2}g\vert\Phi_0\vert$, $m_Z=\frac{1}{2}
\sqrt{g^2+g'{}^2}\vert\Phi_0\vert$ and the photon is massless (for the details 
see e.g. \cite{AL73,ChLi}, or subsection \ref{sub-3.3} below). Thus, all 
these masses are given by their expression in the Weinberg--Salam model 
except that the vacuum value $v=\sqrt{-\mu^2/\lambda}$ of the Higgs field in 
that model should be replaced by $\vert\Phi_0\vert$ (i.e. by $\vert\Phi_v
\vert$ or $\vert\Phi_g\vert$). To determine the rest mass of the Higgs field, 
too, we should calculate the derivatives of ${\cal L}_H-V$ with respect to 
$H$ at the state $\Phi^{\bi}_0$ (i.e. at $H=0$). They are 

\begin{eqnarray}
&{}&\bigl(\frac{\partial({\cal L}_H-V)}{\partial H}\bigr)_0=-\Bigl(\mu^2+
  \frac{2}{3}\Lambda+\lambda\vert\Phi_0\vert^2+\frac{1}{6}\kappa\mu^2\vert
  \Phi_0\vert^2\Bigr)\vert\Phi_0\vert, \label{eq:1.7.8a} \\
&{}&\bigl(\frac{\partial^2({\cal L}_H-V)}{\partial H^2}\bigr)_0=-\Bigl(\mu^2+
  \frac{2}{3}\Lambda+3\lambda\vert\Phi_0\vert^2+\frac{1}{6}\kappa\mu^2\vert
  \Phi_0\vert^2\Bigr). \label{eq:1.7.8b}
\end{eqnarray}
Therefore, comparing these with (\ref{eq:1.7.5}), we see that \emph{the 
critical point of ${\cal L}_H-V$ is the solution $\Phi^{\bi}_g$ of the field 
equations (the `ground state'), rather than the `vacuum state' $\Phi^{\bi}_v$}. 
Hence, the rest mass of $H$ cannot be read-off from (\ref{eq:1.7.8b}) if 
$\Phi^{\bi}_0$ is chosen to be the minimal energy density state $\Phi^{\bi}_v$. 
That would have to be the solution $\Phi^{\bi}_g$ of the field equations, 
which does \emph{not} minimize the energy density.

\subsubsection{Preliminary remarks on the instantaneous  vacuum states}
\label{sub-1.7.2}

What we learnt in subsection \ref{sub-1.7.1} is that \emph{the uniqueness of 
the notion of the usual `spacetime vacuum states' is lost}: The two key 
properties of the usual vacuum states, viz. that they minimize the energy 
density and solve the field equations, split. The uniquely determined global 
spacetime vacuum states of the Standard Model in Minkowski spacetime seem to 
be analogous to absolute parallelism, i.e. the existence of globally defined 
Cartesian coordinate frames, in differential geometry. The conformally 
invariant coupling to gravity rules out the very existence of such uniquely 
defined vacuum states. Thus, to find the appropriate notion of the `vacuum 
states' we should rethink this concept and the mathematical realization of 
these states. 

Let us recall that the states in classical field theory, represented by 
certain spinor and tensor fields, are specified on a 3-manifold, which will 
be the typical Cauchy hypersurface in the spacetime. These fields form the 
correct initial data set for the evolution equations. What we want to 
identify as the instantaneous vacuum states are certain special 
\emph{physical states}, defined e.g. as the extremal points of some energy 
functional. Thus, in particular, in a constrained system, these states must 
solve the \emph{constraint parts} of the field equations.

In a (gauge symmetry breaking) local classical field theory the role of the 
vacuum states is to provide a \emph{non-trivial reference configuration}, 
whose `vacuum value' is present in the outcome of certain \emph{local 
experiments}, e.g. in the measurement of the vector boson masses. However, 
by the principle of locality, it is hard to imagine how the result of such 
a local measurement could depend on the state of the world in the remote 
future. The outcome of a local experiment should be determined by the 
instantaneous state of the system in which the experiment was carried out. 
Therefore, the notion of the `vacuum states' should also be instantaneous, 
and the `vacuum state' at one instant is not \emph{a priori} required to be 
the time evolution of the `vacuum state' at an earlier instant. The evolution 
may take these `instantaneous vacuum states' into non-vacuum states in the 
next instant. 

Nevertheless, since the `vacuum states' are \emph{special} states, they may 
have some \emph{spatially non-local} character, like the spinor field $\Psi$ 
in the total energy-momentum expression (\ref{eq:1.7.1}) that satisfies an 
\emph{elliptic} partial differential equation on $\Sigma$. Thus, the `vacuum 
states' are local in time, but could be non-local in space. On the other 
hand, there might be (and, as we will see, there are) situations in which 
the field configurations that are to be the instantaneous vacuum states are 
well defined only on \emph{open subsets} of the hypersurface $\Sigma$ 
defining the instant. If these field configurations are well defined on the 
whole $\Sigma$, then the instantaneous vacuum state will be called 
\emph{global}, otherwise only \emph{quasi-local}. To formulate these states 
mathematically, we should investigate the energy-momentum functional and 
split the spacetime in a 3+1 way with respect to the hypersurface $\Sigma$.


\section{The energy-momentum functional}
\label{sec:2}

\subsection{The 3+1 form of the field equations}
\label{sub-2.1}

Let $\Sigma$ be a smooth spacelike hypersurface, $t^a$ its future pointing unit 
timelike normal and define $P^a_b:=\delta^a_b-t^at_b$, the $g_{ab}$-orthogonal 
projection to $\Sigma$. Then the induced metric and the extrinsic curvature of 
$\Sigma$ are defined, respectively, by $h_{ab}:=P^c_aP^d_bg_{cd}$ and $\chi_{ab}
:=P^c_aP^d_b\nabla_ct_d$. The intrinsic Levi-Civita derivative operator will 
be denoted by $D_e$. The induced volume 3-form (and the orientation) on 
$\Sigma$ is defined by the convention $\varepsilon_{abc}:=t^e\varepsilon_{eabc}$, 
where $\varepsilon_{abcd}$ is the spacetime volume 4-form. 

Next we decompose the Yang--Mills connection 1-form into its scalar and 
spatial vector potential according to $\omega^\alpha_a=t_a\phi^\alpha+A^\alpha_a$, 
where $A^\alpha_a:=P^b_a\omega^\alpha_b$; and define the electric and magnetic 
field strengths, respectively, by $E^\alpha_a:=F^\alpha_{ab}t^b$ and $B^\alpha_{ab}
:=F^\alpha_{cd}P^c_aP^d_b$. The latter is just the field strength of the spatial 
vector potential: $B^\alpha_{ab}=D_aA^\alpha_b-D_bA^\alpha_a+c^\alpha_{\mu\nu}A^\mu_a
A^\nu_b$. 

Let us define $\mathbb{D}_e\varphi^\alpha:=D_e\varphi^\alpha+A^\mu_ec^\alpha
_{\mu\nu}\varphi^\nu$ for any $\varphi^\alpha$, the \emph{spatial} gauge 
covariant derivative in the pull back of the adjoint vector bundle $A(M)$ to 
$\Sigma$. The spatial gauge covariant derivative on the pull back of the 
Higgs and fermion bundles to $\Sigma$ will also be denoted by $\mathbb{D}_e$. 
However, while on the Higgs bundle it is the pull back of $\nabbla_e$, i.e. 
$P^b_e\nabbla_b\Phi^{\bi}=\mathbb{D}_e\Phi^{\bi}:=D_e\Phi^{\bi}+A^\alpha_eT^{\bi}
_{\alpha{\bj}}\Phi^{\bj}$ for any cross section $\Phi^{\bi}$ of the pulled back 
Higgs bundle, on the pulled back fermion bundle $P^b_e\nabbla_b$ deviates 
from $\mathbb{D}_e$. The latter is given by $\mathbb{D}_e\psi^r_A:=D_e\psi^r_A
+A^\alpha_eT^r_{\alpha s}\psi^s_A$, where $D_e$ is the intrinsic Levi-Civita 
covariant derivative operator on the pulled back spinor bundle. The 
difference of $P^b_e\nabbla_b$ (the `Sen type' connection) and $\mathbb{D}_e$ 
is the extrinsic curvature of the hypersurface: $P^b_e\nabbla_b\psi^r_A=
\mathbb{D}_e\psi^r_A+\chi_{eAA'}t^{A'B}\psi^r_B$, where we converted the second 
index of the extrinsic curvature tensor $\chi_{ea}$ into the pair $AA'$ of 
Weyl spinor indices. 

Since the field equations for the Yang--Mills and the Higgs fields are 
\emph{second} order, in the Cauchy problem for them we can choose $(A^\alpha_a,
E^\alpha_a,\Phi^{\bi},t^a(\nabbla_a\Phi^{\bi}))$ as the initial data set. On the 
other hand, since the field equations for the Weyl spinor fields are 
\emph{first order}, the initial data set consists only of the spinor field 
$\psi^r_A$ itself. Therefore, we choose $(A^\alpha_a,E^\alpha_a,\Phi^{\bi},t^a
(\nabbla_a\Phi^{\bi}),\psi^r_A)$ to represent the field configuration of the 
Standard Model at the instant defined by $\Sigma$. $E^\alpha_a$ is essentially 
the momentum canonically conjugate to $A^\alpha_a$. 

In terms of these variables the projection to $\Sigma$ of the Bianchi identity 
for $F^\alpha_{cd}$, and the contraction of the Yang--Mills equation 
(\ref{eq:1.3.1a}) with $t^b$, respectively, yield 

\begin{eqnarray}
\mathbb{D}_{[a}B^\alpha_{bc]}\!\!\!\!&=\!\!\!\!&0,  \label{eq:2.1.1a} \\
\mathbb{D}^aE^\alpha_a\!\!\!\!&=\!\!\!\!&4\pi \bigl({}_HJ^\alpha_a+{}_WJ^\alpha_a
  \bigr)t^a. \label{eq:2.1.1b}
\end{eqnarray}
Here ${}_HJ^\alpha_a$ and ${}_WJ^\alpha_a$ are the currents introduced in 
connection with equation (\ref{eq:1.3.1a}). (\ref{eq:2.1.1a}) is just the 
Bianchi identity for the spatial vector potential $A^\alpha_a$, while 
(\ref{eq:2.1.1b}) is the Gauss equation, a constraint, in which ${}_H
J^\alpha_at^a$ and ${}_WJ^\alpha_at^a$ are the charge densities. Similarly, 
the 3+1 form of the field equation (\ref{eq:1.3.1c}) for the Weyl spinor 
fields is 

\begin{equation}
t^e\bigl(\nabbla_e\psi^r_A\bigr)=2t_A{}^{B'}\bigl(\mathbb{D}_{B'}{}^B\psi^r_B
\bigr)-\frac{1}{2}\chi\psi^r_A+2G^{rr'}t_A{}^{A'}\bar\psi^{s'}_{A'}\bar Y
_{r's'{\bi}}\Phi^{\bi}. \label{eq:2.1.1c}
\end{equation}
However, to find the 3+1 form of the Higgs and of the remaining part of the 
Yang--Mills field equations, we should have a \emph{foliation} of the 
spacetime by a family $\Sigma_t$ of smooth spacelike hypersurfaces (rather 
than to have only a single $\Sigma$). Thus, let $N$ denote the lapse function 
of the foliation, defined by $1=:Nt^a\nabla_at$, by means of which the 
acceleration of the leaves is $a_e:=t^b\nabla_bt_e=-D_e\ln N$. Then the 
expression of the electric field strength in terms of the scalar and spatial 
vector potentials yields 

\begin{equation}
t^e\bigl(\nabla_eA^\alpha_b\bigr)P^b_a+\phi^\mu c^\alpha_{\mu\nu}A^\nu_a=\frac{1}{N}
D_a\bigl(N\phi^\alpha\bigr)-A^\alpha_b\chi^b{}_a-E^\alpha_a; \label{eq:2.1.2a}
\end{equation}
but there is no evolution equation for $\phi^\alpha$. The 3+1 form of the 
remaining part of the Bianchi identity and of the Yang--Mills field equation, 
respectively, are 

\begin{eqnarray}
t^e\bigl(\nabbla_eB^\alpha_{cd}\bigr)P^c_aP^d_b\!\!\!\!&=\!\!\!\!&B^\alpha_{ae}
  \chi^e{}_b-B^\alpha_{be}\chi^e{}_a-\frac{1}{N}\Bigl(\mathbb{D}_a\bigl(NE^\alpha_b
  \bigr)-\mathbb{D}_b\bigl(NE^\alpha_a\bigr)\Bigr),  \label{eq:2.1.2b} \\
t^e\bigl(\nabbla_eE^\alpha_b\bigr)P^b_a\!\!\!\!&=\!\!\!\!&\bigl(\chi_a{}^b-
  \chi\delta^b_a\bigr)E^\alpha_b+\frac{1}{N}\mathbb{D}^b\bigl(NB^\alpha_{ba}\bigr)
  -4\pi \bigl({}_HJ^\alpha_b+{}_WJ^\alpha_b\bigr)P^b_a. \label{eq:2.1.2c}
\end{eqnarray}
Finally, 

\begin{eqnarray}
t^a\nabbla_a\bigl(t^b\nabbla_b\Phi^{\bi}\bigr)\!\!\!\!&=\!\!\!\!&-\chi t^a
  \bigl(\nabbla_a\Phi^{\bi}\bigr)-\mathbb{D}_a\mathbb{D}^a\Phi^{\bi}-\frac{1}{N}
  \bigl(D^aN\bigr)\mathbb{D}_a\Phi^{\bi}-\bigl(\mu^2+\frac{2}{3}\Lambda\bigr)
  \Phi^{\bi}-\nonumber \\
\!\!\!\!&{}\!\!\!\!&-\Bigl(G^{\bi\bj'}G_{\bj'\bk'\bl\bm}+\frac{1}{6}\kappa\mu^2
  \delta^{\bi}_{\bl}G_{\bm\bk'}\Bigr)\bar\Phi^{\bk'}\Phi^{\bl}\Phi^{\bm}-{\rm i}
  \varepsilon^{AB}\psi^r_A\psi^s_BY_{rs{\bj}'}G^{\bi\bj'} \label{eq:2.1.2d}
\end{eqnarray}
is the 3+1 form of (\ref{eq:1.6.2}). 


\subsection{The 3+1 form of the energy-momentum tensor}
\label{sub-2.2}

From (\ref{eq:1.6.3}) it is straightforward to calculate the energy density 
$\varepsilon:=T_{ab}t^at^b$, the momentum density $\pi_a:=P^b_aT_{bc}t^c$ and 
the spatial stress $\sigma_{ab}:=P^c_aP^d_bT_{cd}$ of the matter fields, seen by 
the observers at rest with respect to the hypersurface $\Sigma$. Introducing 
the notation $\Pi^{\bi}:=t^e\nabbla_e\Phi^{\bi}+\frac{1}{3}\chi\Phi^{\bi}$, 
which, as we will see in subsection \ref{sub-3.1}, is essentially the 
momentum canonically conjugate to $\bar\Phi^{\bi'}$, for the energy density 
we obtain 

\begin{eqnarray}
\varepsilon\!\!\!\!&{}\!\!\!\!&\bigl(1-\frac{1}{6}\kappa\vert\Phi\vert^2
  \bigr)=\frac{1}{2}G_{\alpha\beta}E^\alpha_aE^\beta_bh^{ab}-\frac{1}{4}
  G_{\alpha\beta}B^\alpha_{ab}B^\beta_{cd}h^{ac}h^{bd}+ \label{eq:2.2.1a} \\
\!\!\!\!&{}\!\!\!\!&+\frac{1}{2}G_{\bi\bj'}\Pi^{\bi}\bar\Pi^{\bj'}-\frac{1}{2}
  G_{\bi\bj'}h^{ab}\bigl(\mathbb{D}_a\Phi^{\bi}\bigr)\bigl(\mathbb{D}_b\bar
  \Phi^{\bj'}\bigr)+ \nonumber \\
\!\!\!\!&{}\!\!\!\!&+\frac{1}{4}G_{\bi\bj\bk'\bl'}\Phi^{\bi}\Phi^{\bj}\bar\Phi
  ^{\bk'}\bar\Phi^{\bl'}+\frac{1}{2}\bigl(\mu^2+\frac{1}{3}\Lambda-\frac{1}{9}
  \chi^2\bigr)\vert\Phi\vert^2+\frac{1}{6}D_aD^a\bigl(\vert\Phi\vert^2\bigr)- 
  \nonumber \\
\!\!\!\!&{}\!\!\!\!&-\frac{\rm i}{2}G_{rr'}\Bigl(\bar\psi^{r'}_{A'}\mathbb{D}^{A'A}
  \psi^r_A-\psi^r_A\mathbb{D}^{AA'}\bar\psi^{r'}_{A'}\Bigr)+\frac{\rm i}{2}\Bigl(
  \varepsilon^{AB}\psi^r_A\psi^s_BY_{rs{\bi}'}\bar\Phi^{\bi'}-\varepsilon^{A'B'}
  \bar\psi^{r'}_{A'}\bar\psi^{s'}_{B'}\bar Y_{r's'{\bi}}\Phi^{\bi}\Bigr), \nonumber
\end{eqnarray}
while for the momentum density 

\begin{eqnarray}
\pi_a\!\!\!\!&{}\!\!\!\!&\bigl(1-\frac{1}{6}\kappa\vert\Phi\vert^2\bigr)=
  G_{\alpha\beta}E^\alpha_bh^{bc}B^\beta_{ca}+\frac{1}{2}G_{\bi\bj'}\Bigl(\Pi^{\bi}
  \bigl(\mathbb{D}_a\bar\Phi^{\bj'}\bigr)+\bar\Pi^{\bj'}\bigl(\mathbb{D}_a
  \Phi^{\bi}\bigr)\Bigr)-\label{eq:2.2.1b} \\
\!\!\!\!&{}\!\!\!\!&-\frac{1}{6}D_a\Bigl(G_{\bi\bj'}\Pi^{\bi}\bar\Phi^{\bj'}+
  G_{\bi\bj'}\bar\Pi^{\bj'}\Phi^{\bi}-\frac{2}{3}\chi\vert\Phi\vert^2\Bigr)+
  \frac{1}{6}\bigl(\chi_a{}^b-\chi\delta^b_a\bigr)D_b\bigl(\vert\Phi\vert^2
  \bigr)+ \nonumber\\
\!\!\!\!&{}\!\!\!\!&+\frac{\rm i}{4}G_{rr'}t^{BB'}\Bigl(\bar\psi^{r'}_{B'}
  \mathbb{D}_a\psi^r_B-\psi^r_B\mathbb{D}_a\bar\psi^{r'}_{B'}\Bigr)+\frac{\rm i}
  {2}G_{rr'}\Bigl(\bar\psi^{r'}_{B'}t_B{}^{C'}\mathbb{D}_{C'}{}^C\psi^r_C-\psi^r_B
  t_{B'}{}^C\mathbb{D}_C{}^{C'}\bar\psi^{r}_{C'}\Bigr)P^b_a. \nonumber
\end{eqnarray}
These are expressions of the initial value of the fields on $\Sigma$. Note 
that in the derivation of the above form of $\varepsilon$ we used no field 
equation except the Hamiltonian constraint part of Einstein's equations, 
i.e. the first of 

\begin{equation}
\bigl(R_{ab}-\frac{1}{2}Rg_{ab}\bigr)t^at^b=-\kappa\varepsilon-\Lambda, 
\hskip 20pt
\bigl(R_{bc}-\frac{1}{2}Rg_{bc}\bigr)t^bP^c_a=-\kappa\pi_a. \label{eq:2.2.2}
\end{equation}
On the other hand, in the derivation of $\pi_a$ above we used not only the 
momentum constraint, the second of (\ref{eq:2.2.2}), but the 3+1 form 
(\ref{eq:2.1.1c}) of the field equation for the Weyl spinor fields, too. 
Otherwise $t^e(\nabbla_e\psi^r_A)$, appearing in the 3+1 form of 
(\ref{eq:1.6.3}), could not be expressed by the initial data on $\Sigma$. 
In the present paper we need only the isotropic pressure $P:=-\frac{1}{3}
\sigma_{ab}h^{ab}=-\frac{1}{3}T_{ab}h^{ab}$ but not the spatial stress itself. 
Using the field equations (\ref{eq:2.1.1c}) and (\ref{eq:2.1.2d}), we obtain 
that $3P=\varepsilon-\mu^2\vert\Phi\vert^2$. Thus, if $\vert\Phi\vert^2
\rightarrow6/\kappa$ and $\varepsilon$ is diverging (and hence $P$ also), 
then the term $\frac{1}{3}\mu^2\vert\Phi\vert^2$ is less and less significant. 
Therefore, when $\vert\Phi\vert^2$ approaches $6/\kappa$, the 
$\varepsilon$--$P$ relation is getting to be that in the phenomenological 
equation of state of incoherent pure radiation. Nevertheless, this is still 
not a fluid, as it has the (also diverging) non-isotropic spatial stress. 
Moreover, apart from the Yang--Mills sector for compact gauge groups, 
neither the Higgs nor the spinor sector of the energy-momentum tensor 
satisfy the usual energy conditions.


\subsection{The critical points of the energy-momentum functional}
\label{sub-2.3}

We need to know the critical points of the energy-momentum functional with 
respect to the \emph{matter field} variables. As we have already seen, this 
functional is the sum of a term depending only on the gravitational field 
variables and the energy-momentum functional of the matter fields, where the 
latter is 

\begin{equation}
{\tt Q}[K]:=\int_{\Sigma} K^aT_{ab}t^b{\rm d}\Sigma=\int_{\Sigma}(\varepsilon M+
\pi_aM^a){\rm d}\Sigma. \label{eq:2.3.1}
\end{equation}
Here the $3+1$ form of the generator vector field is $K^a=Mt^a+M^a$. Since 
in the (total or quasi-local) energy-momentum of the matter+gravity system 
it is only ${\tt Q}[K]$ that depends on the matter field variables, in the 
calculation of its variational derivatives with respect to them it is only 
${\tt Q}[K]$ that matters. The fields $M$ and $M^a$ play the role only as 
`parameter fields'.

\subsubsection{The functional derivatives}
\label{sub-2.3.1}

Let $A^\alpha_a(u)$, $E^\alpha_a(u)$, $\Phi^{\bi}(u)$, $\Pi^{\bi}(u)$ and 
$\psi^r_A(u)$ be any smooth one-parameter families of field configurations on 
$\Sigma$. Denoting by $\delta$ the derivative with respect to the parameter 
$u$ at $u=0$ we obtain that 

\begin{eqnarray}
\delta{\tt Q}[K]=\int_{\Sigma}\Bigl\{\!\!\!\!&{}\!\!\!\!&D_aB^a+\frac{\delta
  {\tt Q}}{\delta A^\alpha_a}\delta A^\alpha_a+\frac{\delta{\tt Q}}{\delta 
  E^\alpha_a}\delta E^\alpha_a+ \label{eq:2.3.2} \\
+\!\!\!\!&{}\!\!\!\!&\frac{\delta{\tt Q}}{\delta \Phi^{\bi}}\delta\Phi^{\bi}+
  \frac{\delta{\tt Q}}{\delta\bar\Phi^{\bi'}}\delta\bar\Phi^{\bi'}+
  \frac{\delta{\tt Q}}{\delta \Pi^{\bi}}\delta\Pi^{\bi}+\frac{\delta{\tt Q}}
  {\delta\bar\Pi^{\bi'}}\delta\bar\Pi^{\bi'}+\frac{\delta{\tt Q}}{\delta\psi^r_A}
  \delta\psi^r_A+\frac{\delta{\tt Q}}{\delta\bar\psi^{r'}_{A'}}\delta\bar
  \psi^{r'}_{A'}\Bigr\}{\rm d}\Sigma, \nonumber
\end{eqnarray}
where the functional derivatives themselves are 

\begin{eqnarray}
\frac{\delta{\tt Q}}{\delta A^\alpha_a}=\frac{1}{1-\frac{1}{6}\kappa\vert
 \Phi\vert^2}\Bigl(\!\!\!\!&{}\!\!\!\!&M(\mathbb{D}^cB^\beta_{cb})h^{ba}-M4\pi
 \bigl({}_HJ^\beta_b+{}_WJ^\beta_b\bigr)h^{ba}+ \label{eq:2.3.3a} \\
+\!\!\!\!&{}\!\!\!\!&M^c(\mathbb{D}_cE^\beta_b)h^{ba}-M^a(\mathbb{D}^bE^\beta_b)
 +M^a4\pi({}_HJ^\beta_b+{}_WJ^\beta_b)t^b\Bigr)G_{\beta\alpha}+ \nonumber \\
+D^b\bigl(\frac{M}{1-\frac{1}{6}\kappa\vert\Phi\vert^2}\bigr)\!\!\!\!&
 {}\!\!\!\!&B^\beta_{bc}G_{\beta\alpha}h^{ca}+D_c\bigl(\frac{M^d}{1-\frac{1}{6}
 \kappa\vert\Phi\vert^2}\bigr)\bigl(\delta^c_dh^{ab}-\delta^a_dh^{cb}\bigr)
 E^\beta_bG_{\beta\alpha}, \nonumber \\
\frac{\delta{\tt Q}}{\delta E^\alpha_a}=\frac{1}{1-\frac{1}{6}\kappa\vert
 \Phi\vert^2}\bigl(\!\!\!\!&{}\!\!\!\!&ME^\beta_b-M^cB^\beta_{cb}\bigr)
 G_{\beta\alpha}h^{ba}, \label{eq:2.3.3b} 
\end{eqnarray}

\begin{eqnarray}
\frac{\delta{\tt Q}}{\delta\Phi^{\bi}}=\frac{M}{1-\frac{1}{6}\kappa\vert
 \Phi\vert^2}\frac{1}{2}\Bigl(\!\!\!\!&{}\!\!\!\!&\mathbb{D}^a\bigl(
 \mathbb{D}_a\bar\Phi^{\bj'}\bigr)G_{\bj'\bi}+\frac{1}{3}\kappa\varepsilon\bar
 \Phi^{\bj'}G_{\bj'\bi}+(\mu^2+\frac{1}{3}\Lambda-\frac{1}{9}\chi^2)\bar\Phi
 ^{\bj'}G_{\bj'\bi}+ \nonumber \\
+\!\!\!\!&{}\!\!\!\!&G_{\bi\bj\bk'\bl'}\Phi^{\bj}\bar\Phi^{\bk'}\bar\Phi^{\bl'}-
 {\rm i}\varepsilon^{A'B'}\bar\psi^{r'}_{A'}\bar\psi^{s'}_{B'}\bar Y_{r's'{\bi}}
 \Bigr)+\nonumber \\
+\frac{1}{2}D^a\bigl(\!\!\!\!&{}\!\!\!\!&\frac{M}{1-\frac{1}{6}\kappa\vert
 \Phi\vert^2}\bigr)(\mathbb{D}_a\bar\Phi^{\bj'})G_{\bj'\bi}+\frac{1}{6}D_aD^a
 \Bigl(\frac{M}{1-\frac{1}{6}\kappa\vert\Phi\vert^2}\Bigr)\bar\Phi^{\bj'}
 G_{\bj'\bi}+ \nonumber \\
+\frac{1}{2}\frac{M^a}{1-\frac{1}{6}\kappa\vert\Phi\vert^2}\Bigl(\!\!\!\!&{}
 \!\!\!\!&\frac{1}{3}\kappa\pi_a\bar\Phi^{\bj'}-\mathbb{D}_a\bar\Pi^{\bj'}
 -\frac{1}{3}D_b(\chi^b{}_a-\chi\delta^b_a)\bar\Phi^{\bj'}\Bigr)G_{\bj'\bi}- 
 \nonumber \\
-\frac{1}{3}D_a\bigl(\!\!\!\!&{}\!\!\!\!&\frac{M^b}{1-\frac{1}{6}\kappa\vert
 \Phi\vert^2}\bigr)\Bigl(\delta^a_b\bar\Pi^{\bj'}+\frac{1}{2}(\chi^a{}_b-
 \frac{1}{3}\chi\delta^a_b)\bar\Phi^{\bj'}\Bigr)G_{\bj'\bi}, \label{eq:2.3.3c} \\
\frac{\delta{\tt Q}}{\delta\Pi^{\bi}}=\frac{1}{2}\frac{1}{1-\frac{1}{6}\kappa
 \vert\Phi\vert^2}\Bigl(\!\!\!\!&{}\!\!\!\!&M\bar\Pi^{\bj'}+M^a(\mathbb{D}_a
 \bar\Phi^{\bj'})\Bigr)G_{\bj'\bi}+\frac{1}{6}D_a\bigl(\frac{M^a}{1-\frac{1}{6}
 \kappa\vert\Phi\vert^2}\bigr)\bar\Phi^{\bj'}G_{\bj'\bi},  \label{eq:2.3.3d}
 \end{eqnarray}

\begin{eqnarray}
\frac{\delta{\tt Q}}{\delta\psi^r_A}=\frac{\rm i}{2}\frac{M}{1-\frac{1}{6}
 \kappa\vert\Phi\vert^2}\Bigl(\!\!\!\!&{}\!\!\!\!&(\mathbb{D}^{AA'}\bar
 \psi^{r'}_{A'})G_{r'r}+2\varepsilon^{AB}\psi^s_BY_{rs{\bi}'}\bar\Phi^{\bi'}\Bigr)+
 \frac{\rm i}{2}D^a\bigl(
 \frac{M}{1-\frac{1}{6}\kappa\vert\Phi\vert^2}\bigr)\bar\psi^{r'}_{A'}G_{r'r}-
 \nonumber \\
-\frac{\rm i}{2}\frac{M^b}{1-\frac{1}{6}\kappa\vert\Phi\vert^2}\Bigl(
 \!\!\!\!&{}\!\!\!\!&(\mathbb{D}_b\bar\psi^{r'}_{A'})t^{A'A}-t_{B'C}(\mathbb{D}
 ^{CC'}\bar\psi^{r'}_{C'})\delta^A_B-t_{BA'}(\mathbb{D}^{A'A}\bar\psi^{r'}_{B'})
 \Bigr)G_{r'r}+ \nonumber \\
+\frac{\rm i}{2}D^a\bigl(\frac{M^b}{1-\frac{1}{6}\kappa\vert\Phi\vert^2}\bigr)
 \!\!\!\!&{}\!\!\!\!&t_{BA'}\bar\psi^{r'}_{B'}G_{r'r}-\frac{\rm i}{4}D_b\bigl(
\frac{M^b}{1-\frac{1}{6}\kappa\vert\Phi\vert^2}\bigr)\bar\psi^{r'}_{A'}t^{A'A}
G_{r'r}. \label{eq:2.3.3e}
\end{eqnarray}
The total divergence, $D_aB^a$, can also be given explicitly, and the 
vanishing of its integral is the condition of the classical functional 
differentiability of ${\tt Q}[K]$. However, in the present paper we do not 
need it. We discuss its effect and meaning elsewhere.

\subsubsection{The critical configurations}
\label{sub-2.3.2}

Since the `parameter fields' $M$ and $M^a$ are involved in the functional 
derivatives, the critical configurations depend on our choice for them. For 
example, if the spinor fields $\lambda^A$ and $\mu^A$ in (\ref{eq:1.7.1}) 
solve the Witten equation, ${\cal D}_{A'A}\lambda^A=0$ and ${\cal D}_{A'A}\mu^A
=0$, then $M\chi+D_aM^a={\cal D}_aK^a=0$ holds. An even more restrictive 
condition could be that $M$ is constant and $M^a$ divergence-free, like for 
the translation Killing fields on spacelike hyperplanes in Minkowski 
spacetime, or the Killing fields in the FRW and Kantowski--Sachs examples 
below. In this subsection we show that the critical configuration for 
\emph{arbitrary} $M$ and $M^a$ is the trivial one in the matter sector; but 
for constant $M$ and divergence-free $M^a$ we obtain non-trivial ones in 
which the Higgs field is non-zero but spatially constant. 

Let us start with equation (\ref{eq:2.3.3b}) with $M^a=0$. The vanishing of 
$\delta{\tt Q}/\delta E^\alpha_a$ for any constant $M$ yields that $E^\alpha_a
=0$. Substituting this back into (\ref{eq:2.3.3b}) and using that $M^a$ is 
(divergence-free, but otherwise) arbitrary, we obtain that the magnetic field 
strength is also vanishing, i.e. in the critical configurations 

\begin{equation}
E^\alpha_a=0, \hskip 20pt B^\alpha_{ab}=0. \label{eq:2.3.4a}
\end{equation}
Then by (\ref{eq:2.3.3a}) the vanishing of $\delta{\tt Q}/\delta A^\alpha_a$ 
gives that ${}_HJ^\alpha_a+{}_WJ^\alpha_a=0$. In the critical configurations it 
follows from (\ref{eq:2.3.3d}) with $M^a=0$ that $\Pi^{\bi}=0$. Substituting 
this back into (\ref{eq:2.3.3d}), from $\delta{\tt Q}/\delta\Pi^{\bi}=0$ it 
follows that 

\begin{equation}
M^a(\mathbb{D}_a\Phi^{\bi})+\frac{1}{3}(D_aM^a)\Phi^{\bi}+\frac{1}{18}
\kappa\Phi^{\bi}\frac{1}{1-\frac{1}{6}\kappa\vert\Phi\vert^2}M^aD_a\vert\Phi
\vert^2=0. \label{eq:2.3.4x}
\end{equation}
Since $D_aM^a=0$, the second term is vanishing. Contracting the resulting 
equation with $\bar\Phi^{\bj'}G_{\bj'\bi}$, adding to its own complex conjugate, 
and assuming that $\vert\Phi\vert^2\not=18/\kappa$, we find that $M^aD_a\vert
\Phi\vert^2=0$. Substituting this back into the above equation we obtain that 
$\Phi^{\bi}$ itself is gauge covariantly constant, i.e. 

\begin{equation}
\Pi^{\bi}=0, \hskip 20pt \mathbb{D}_a\Phi^{\bi}=0. \label{eq:2.3.4b}
\end{equation}
These two together imply that ${}_HJ^\alpha_a=0$, and hence that ${}_WJ^\alpha
_a=0$, too. Note that for \emph{arbitrary} $M^a$, whose divergence is not 
required to be vanishing, (\ref{eq:2.3.4x}) implies the vanishing of the 
coefficient of $D_aM^a$, i.e. $\Phi^{\bi}=0$, too. 

Next, from $\delta{\tt Q}/\delta\psi^r_A=0$ and (\ref{eq:2.3.3e}) with 
$M=0$ and divergence-free $M^a$ we obtain 

\begin{equation}
M^b\Bigl((\mathbb{D}_b\bar\psi^{r'}_{A'})t^{A'A}-t_{B'C}(\mathbb{D}^{CC'}\bar
\psi^{r'}_{C'})\delta^A_B-t_{BA'}(\mathbb{D}^{A'A}\bar\psi^{r'}_{B'})\Bigr)-
(D^aM^b)t_{A'B}\bar\psi^{r}_{B'}=0. \label{eq:2.3.5}
\end{equation}
Since at each point $M^{AB'}$ and $(D^{AA'}M^{BB'})t_{A'B}$ are independent, this 
implies that 

\begin{equation}
\psi^r_A=0. \label{eq:2.3.4c}
\end{equation}
With this substitution $\delta{\tt Q}/\delta\psi^r_A=0$ is already satisfied. 

From $\delta{\tt Q}/\delta\Phi^{\bi}=0$ and (\ref{eq:2.3.3c}) with $M={\rm 
const}$ and $M^a=0$ we obtain 

\begin{equation}
\frac{1}{3}\kappa\varepsilon\bar\Phi^{\bj'}G_{\bj'\bi}+\bigl(\mu^2+\frac{1}
{3}\Lambda-\frac{1}{9}\chi^2\bigr)\bar\Phi^{\bj'}G_{\bj'\bi}+G_{\bi\bj\bk'\bl'}
\Phi^{\bj}\bar\Phi^{\bk'}\bar\Phi^{\bl'}=0, \label{eq:2.3.4d}
\end{equation}
whose contraction with $\Phi^{\bi}$ (together with (\ref{eq:2.2.1a})) yields 

\begin{equation}
\Bigl(1-\frac{\kappa}{12}\vert\Phi\vert^2\Bigr)G_{\bi\bj\bk'\bl'}\Phi^{\bi}
\Phi^{\bj}\bar\Phi^{\bk'}\bar\Phi^{\bl'}+\bigl(\mu^2+\frac{\Lambda}{3}-
\frac{1}{9}\chi^2\bigr)\vert\Phi\vert^2=0. \label{eq:2.3.4dc}
\end{equation}
Using this, (\ref{eq:2.3.4d}) yields that $4\varepsilon=-G_{\bi\bj\bk'\bl'}
\Phi^{\bi}\Phi^{\bj}\bar\Phi^{\bk'}\bar\Phi^{\bl'}$. In particular, for 
$G_{\bi\bj\bk'\bl'}=\lambda G_{\bk'(\bi}G_{\bj)\bl'}$ (and $\vert\Phi\vert\not=0$) 
(\ref{eq:2.3.4dc}) reduces to 

\begin{equation}
-\frac{1}{12}\kappa\lambda\vert\Phi\vert^4+\lambda\vert\Phi\vert^2+\bigl(
\mu^2+\frac{\Lambda}{3}-\frac{1}{9}\chi^2\bigr)=0, \label{eq:2.3.4dWS}
\end{equation}
whose solution is 

\begin{equation}
\vert\Phi\vert^2=\frac{6}{\kappa}\Bigl(1\pm\sqrt{1+\frac{\kappa}{3\lambda}
(\mu^2+\frac{1}{3}\Lambda-\frac{1}{9}\chi^2)}\Bigr). \label{eq:2.3.4dWSs}
\end{equation}
Since $\mu^2+\Lambda/3<0$, by (\ref{eq:2.3.4dWSs}) $0<\vert\Phi\vert^2<12/
\kappa$ holds. However, the requirement of the reality of $\vert\Phi\vert^2$ 
gives a non-trivial condition on the extrinsic curvature: That cannot be 
arbitrarily large: $\chi^2\leq\chi^2_c:=9(3\lambda/\kappa+\mu^2+\Lambda/3)$ 
must hold. This \emph{critical value} $\chi_c$ of the mean curvature will 
play fundamental role in what follows. We discuss this issue in detail in 
subsection \ref{sub-2.4.2}, and its consequences in subsections \ref{sub-3.2} 
and \ref{sub-3.3}. 

Therefore, the critical points of the energy-momentum functional (with 
constant $M$ and divergence-free $M^a$) are those matter field configurations 
which satisfy (\ref{eq:2.3.4a}), (\ref{eq:2.3.4b}), (\ref{eq:2.3.4c}) and  
(\ref{eq:2.3.4d}); i.e. this is a state of the conformally coupled 
Einstein--Higgs system with spatially gauge-covariantly constant Higgs field. 
By $E^\alpha_a=0$ and ${}_HJ^\alpha_a+{}_WJ^\alpha_a=0$ the only constraint for 
the matter fields, equation (\ref{eq:2.1.1b}), is already satisfied. Since 
we assumed that the locally flat gauge potentials can be transformed to zero 
even globally, the Higgs field is spatially constant: $D_e\Phi^{\bi}=0$. Note 
that, apart from the mean curvature term and the $\pm$ sign, 
(\ref{eq:2.3.4dWSs}) is exactly the equation (\ref{eq:1.7.4}) for the 
spacetime vacuum states of subsection \ref{sub-1.7.1}. 

Since $D_a\Phi^{\bi}=0$, by (\ref{eq:2.3.4dc}) \emph{the mean curvature $\chi$ 
must be constant} on $\Sigma$. Hence, the momentum density is vanishing: 
$\pi_a=0$. Then by (\ref{eq:2.3.3c}) from $\delta{\tt Q}/\delta\Phi^{\bi}=0$ 
with $M=0$ it follows that 

\begin{equation*}
D_a\bigl(\chi^a{}_b-\chi\delta^a_b\bigr)M^b+(D_aM^b)\bigl(\chi^a{}_b-
\frac{1}{3}\chi\delta^a_b\bigr)=0.
\end{equation*}
However, this yields that \emph{the extrinsic curvature is a constant pure 
trace}, i.e. 

\begin{equation}
\chi_{ab}=\frac{1}{3}\chi h_{ab}, \hskip 20pt \chi=\chi(t). 
\label{eq:2.3.4e}
\end{equation}
Then the momentum constraint of general relativity, i.e. the second of 
(\ref{eq:2.2.2}) given in terms of the three-dimensional quantities by $D_b(
\chi^b{}_a-\chi\delta^b_a)=\kappa\pi_a$, is already satisfied. Finally, let 
us take into account the explicit form $\frac{1}{2}({\cal R}+\chi^2-\chi_{ab}
\chi^{ab})=\kappa\varepsilon+\Lambda$ of the Hamiltonian constraint part of 
Einstein's equations, the first of (\ref{eq:2.2.2}). (Here ${\cal R}$ is the 
curvature scalar of the intrinsic 3-geometry of $\Sigma$.) By 
(\ref{eq:2.3.4d}) this yields that 

\begin{equation}
\frac{1}{2}{\cal R}=\Lambda-\frac{1}{3}\chi^2-\frac{\kappa}{4}G_{\bi\bj\bk'\bl'}
\Phi^{\bi}\Phi^{\bj}\bar\Phi^{\bk'}\bar\Phi^{\bl'}=\frac{1}{1-\frac{\kappa}{12}
\vert\Phi\vert^2}\bigl(\Lambda+\frac{\kappa}{4}\mu^2\vert\Phi\vert^2-
\frac{1}{3}\chi^2\bigr), \label{eq:2.3.4f}
\end{equation}
where, in the second equality, we used (\ref{eq:2.3.4dc}), too. Thus, 
\emph{the spatial curvature scalar is constant on $\Sigma$}. Moreover, with 
the parameters of the Weinberg--Salam model, the first two terms together in 
the brackets on the right of (\ref{eq:2.3.4f}) is negative for $\chi^2
\leq\chi^2_c$ (see (\ref{eq:2.3.4dWSs})). Hence, with the Weinberg--Salam 
parameters, in the critical configurations \emph{the curvature scalar of the 
spatial 3-metric is negative}.

Therefore, to summarize, the critical points of the energy-momentum 
functional with respect to the matter fields (with constant $M$ and 
divergence-free $M^a$) that also solve the constraints are those states, in 
which the only non-zero matter field is a constant Higgs field satisfying 
(\ref{eq:2.3.4d}), and the state of the gravitational `field' is specified by 
a 3-metric with constant curvature scalar given by (\ref{eq:2.3.4f}) and the 
extrinsic curvature is a pure constant trace. The only freely specifiable 
fields are $\chi$, which is in fact an extrinsic time (the so-called York 
time) parameter and which fixes both $\vert\Phi\vert^2$ and the curvature 
scalar ${\cal R}$, and the 3-metric up to the constraint (\ref{eq:2.3.4f}). 
The energy density in these states is $\varepsilon=-\frac{1}{4}
G_{\bi\bj\bk'\bl'}\Phi^{\bi}\Phi^{\bj}\bar\Phi^{\bk'}\bar\Phi^{\bl'}$, and, with the 
parameters of the Weinberg--Salam model, the spatial curvature scalar is 
negative. 

Finally, considering these critical configurations as a 1-parameter family of 
physical states (parametrized by the label $t$ of the hypersurface $\Sigma
_t$), and substituting this into (\ref{eq:2.1.2d}) and the evolution part of 
Einstein's equations, 

\begin{equation}
({\pounds}_t\chi_{cd})P^c_aP^d_b=-{\cal R}_{ab}+2\chi_{ac}\chi^c{}_b-\chi\chi
_{ab}+\Lambda h_{ab}+\kappa\bigl(-\sigma_{ab}-\frac{3}{2}Ph_{ab}+\frac{1}{2}
\varepsilon h_{ab}\bigr), \label{eq:2.3.6}
\end{equation}
we obtain that $\Phi^{\bi}=0$. (Here ${\pounds}_t$ denotes Lie derivative 
along the timelike normal $t^a$ of the leaves $\Sigma_t$ of the foliation 
and ${\cal R}_{ab}$ is the Ricci tensor of the intrinsic 3-metric.) Therefore, 
the 1-parameter family of critical configurations does \emph{not} solve the 
evolution equations unless $\Phi^{\bi}=0$ even though they are physical 
states, i.e. solve the constraint equations on every $\Sigma_t$.


\subsection{Example: Configurations with FRW symmetries}
\label{sub-2.4}

The critical configurations of the energy-momentum functional is reminiscent 
of that in the Friedman--Robertson--Walker (FRW) spacetimes: The spatial 
distribution of the matter fields is homogeneous and isotropic, the extrinsic 
curvature is a spatially constant pure trace and the curvature scalar of the 
hypersurface is also constant. Thus, almost the whole FRW symmetries have 
been recovered in the critical configurations. The only difference between the 
exact FRW-symmetric and the critical configurations is that in the latter the 
spatial 3-metric is not necessarily homogeneous and isotropic. Therefore, it 
could be worth discussing the EccSM system with the FRW symmetries in detail. 
Although the extremal configurations with the FRW symmetries can be obtained 
by evaluating the general results of subsection \ref{sub-2.3.2}, it could be 
instructive to determine them directly from (\ref{eq:2.2.1a}) by elementary 
methods.

\subsubsection{The matter fields with the FRW symmetries}
\label{sub-2.4.1}

Let $\Sigma_t:=\{t={\rm const}\}$ be the foliation of the spacetime with the 
FRW symmetries by the transitivity surfaces of the isometries, where $t$ is 
the proper time coordinate along the integral curves of the future pointing 
unit normals of the hypersurfaces $\Sigma_t$ (see e.g. \cite{HE,Wa}). Thus 
the lapse is $N=1$. Let $S=S(t)$ be the (strictly positive) scale function 
for which the induced metric on $\Sigma_t$ is $h_{ab}=S^2{}_1h_{ab}$, where 
${}_1h_{ab}$ is the standard negative definite metric on the unit 3-sphere, 
the flat 3-space and the unit hyperboloidal 3-space, respectively, for $k=
1,0,-1$. The extrinsic curvature of the hypersurfaces is $\chi_{ab}={}_1
h_{ab}S\dot S$, where overdot denotes derivative with respect to $t$, and 
hence its trace is $\chi=3\dot S/S$. The curvature scalar of the intrinsic 
3-metric is ${\cal R}=6k/S^2$. In the initial value formulation of Einstein's 
theory the initial data are $h_{ab}$ and $\chi_{ab}$, and hence in the present 
case $S$ and $\dot S$, restricted by the first of the constraints 
(\ref{eq:2.2.2}). 

Let us suppose that the fields of the matter sector of the EccSM system 
admit the isometries of the spacetime as symmetries. Then, by the argument 
similar to that in subsection \ref{sub-1.7.1}, it follows that $E^\alpha_a$, 
$B^\alpha_{ab}$ and $\psi^r_A$ are all vanishing and $\Phi^{\bi}$ and $\Pi^{\bi}$ 
are constant on the hypersurfaces $\Sigma_t$. (We assume that the locally 
flat Yang--Mills connection is globally flat, too, and hence by an 
appropriate gauge transformation $\phi^\alpha=0$ and $A^\alpha_a=0$ can be 
achieved even globally.) These yield that ${}_WJ^\alpha_a=0$, $\pi_a=0$ and 
that $\sigma_{ab}$ is pure trace; and, by the Gauss equation (\ref{eq:2.1.1b}), 
also that ${}_HJ^\alpha_a=0$. Thus, the EccSM system restricted by the FRW 
symmetries reduces to the Einstein--conformally coupled Higgs (EccH) system. 
For the sake of simplicity, we also assume that the Higgs self-interaction 
coefficient is $G_{\bi\bj\bk'\bl'}=\lambda G_{\bk'(\bi}G_{\bj)\bl'}$. 

For the metric with FRW symmetries Einstein's equations are well known 
\cite{HE,Wa} to reduce to 

\begin{equation}
3\bigl(\frac{\dot S}{S}\bigr)^2=\Lambda+\kappa\varepsilon-3\frac{k}{S^2}, 
\hskip 20pt
3\frac{\ddot S}{S}=\Lambda-\frac{1}{2}\kappa\bigl(\varepsilon+3P\bigr). 
\label{eq:2.4.1}
\end{equation}
The first of these is just the Hamiltonian constraint, the first of 
(\ref{eq:2.2.2}), while the second is the evolution equation 
(\ref{eq:2.3.6}). The field equation (\ref{eq:2.1.2d}) for the Higgs field is 

\begin{equation}
\ddot\Phi^{\bi}+\chi\dot\Phi^{\bi}=-\bigl(\mu^2+\frac{2}{3}\Lambda\bigr)
\Phi^{\bi}-\bigl(\lambda+\frac{1}{6}\kappa\mu^2\bigr)\vert\Phi\vert^2\Phi^{\bi}. 
\label{eq:2.4.2}
\end{equation}
The initial data for the evolution equations is the quadruplet $(\Phi^{\bi},
S;\dot\Phi^{\bi},\dot S)$, or, equivalently, $(\Phi^{\bi},S;\Pi^{\bi},\chi)$, 
subject to the constraint part of (\ref{eq:2.4.1}). The isotropic pressure 
is $P=\frac{1}{3}\varepsilon-\frac{1}{3}\mu^2\vert\Phi\vert^2$, and hence 
in the $\vert\Phi\vert^2\rightarrow6/\kappa$ limit when $\varepsilon$ 
diverges the $\varepsilon$--$P$ relation tends to the phenomenological 
equation of state of a null fluid of incoherent pure radiation. We discuss 
the properties of the energy density $\varepsilon$ in the next subsection.

\subsubsection{The energy density}
\label{sub-2.4.2}

Introducing the notation $\vert\Pi\vert^2:=G_{\bi\bj'}\Pi^{\bi}\bar\Pi^{\bj'}$, 
in the presence of FRW symmetries the energy density (\ref{eq:2.2.1a}) reduces 
to 

\begin{equation}
\varepsilon=\frac{1}{2}\frac{1}{1-\frac{1}{6}\kappa\vert\Phi\vert^2}\Bigl(
\vert\Pi\vert^2+\bigl(\mu^2+\frac{1}{3}\Lambda-\frac{1}{9}\chi^2\bigr)\vert
\Phi\vert^2+\frac{1}{2}\lambda\vert\Phi\vert^4\Bigr). \label{eq:2.4.3}
\end{equation}
This depends on $\Phi^{\bi}$, $\dot\Phi^{\bi}$, $S$ and $\dot S$ only through 
the positive definite norms $\vert\Phi\vert^2$, $\vert\Pi\vert^2$ and $\chi^2$. 
Hence, $\varepsilon$ is, in fact, an \emph{even} function of \emph{three 
variables}. Therefore, all the properties of $\varepsilon=\varepsilon(\Phi
^{\bi},\Pi^{\bi},\chi)$ can be determined from the special case when the Higgs 
field is a single real scalar field with the gauge group $\mathbb{Z}_2$ 
acting on $\Phi$ as $\Phi\mapsto-\Phi$. Moreover, since, for given $\Phi$ 
and $\chi$, the energy density is a simple quadratic function of $\Pi$, all 
the qualitative properties of $\varepsilon$ can be determined easily from 
the special case when $\Pi$ is kept fixed, e.g. $\Pi=0$. Thus, first we 
consider $\varepsilon$ as a function of $\Phi$ and $\chi$ with $\Pi=0$, 

\begin{equation}
\varepsilon(\Phi,0,\chi)=\frac{1}{18}\bigl(\chi^2_c-\chi^2\bigr)\frac{\Phi^2}
{1-\frac{1}{6}\kappa\Phi^2}-\frac{3\lambda}{2\kappa}\Phi^2, \label{eq:2.4.3a}
\end{equation}
where $\chi^2_c:=9(\frac{3\lambda}{\kappa}+\mu^2+\frac{1}{3}\Lambda)$ is the 
\emph{critical mean curvature}; and we discuss the general case at the end 
of this subsection. 

Clearly, in addition to the trivial zero of $\varepsilon$ at $\Phi=0$, it 
has nontrivial ones at 

\begin{equation}
\Phi^2_0=-\frac{2}{\lambda}\bigl(\mu^2+\frac{1}{3}\Lambda-\frac{1}{9}\chi^2
\bigr)=\frac{6}{\kappa}-\frac{2}{9}\lambda\bigl(\chi^2_c-\chi^2\bigr). 
\label{eq:2.4.4}
\end{equation}
Since with the value of $\mu^2$ in the Standard Model and the value of the 
observed cosmological constant $\mu^2+\Lambda/3<0$ holds, the right hand 
side is positive. However, for $\chi^2=\chi^2_c$ the energy density 
$\varepsilon(\Phi,0,\chi_c)$ is not vanishing at $\Phi^2=6/\kappa$; it has 
only the trivial zero $\Phi=0$. Hence, $\varepsilon$ has non-trivial zeros 
for any $\chi^2\not=\chi^2_c$. Since 

\begin{equation}
\frac{\partial\varepsilon}{\partial\Phi}=\frac{\Phi}{(1-\frac{1}{6}\kappa
\Phi^2)^2}\Bigl(-\frac{1}{12}\kappa\lambda\Phi^4+\lambda\Phi^2+\bigl(\mu^2+
\frac{1}{3}\Lambda-\frac{1}{9}\chi^2\bigr)\Bigr), \label{eq:2.4.5}
\end{equation}
the trivial zero is always an extremal point of $\varepsilon$ for any $\chi$. 
At the non-trivial zeros this is 

\begin{equation}
\bigl(\frac{\partial\varepsilon}{\partial\Phi}\bigr)_0=\frac{\lambda}{2}
\frac{\Phi^3_0}{(1-\frac{1}{6}\kappa\Phi^2_0)}=\frac{9\lambda^2}{2\kappa}
\frac{\Phi^3_0}{\chi^2_c-\chi^2}.  \nonumber
\end{equation}
Therefore, if $\chi^2<\chi^2_c$, then $\varepsilon$ is increasing at the 
nontrivial zero $\Phi_0>0$ (and decreasing at $-\Phi_0$); and if $\chi^2>
\chi^2_c$, then $\varepsilon$ is decreasing at $\Phi_0>0$ (and increasing at 
$-\Phi_0$). In the former case $\Phi^2_0<6/\kappa$, but in the latter 
$\Phi^2_0>6/\kappa$. 

In addition to the $\Phi=0$ trivial extremal point, $\varepsilon$ also has 
non-trivial ones. By (\ref{eq:2.4.5}) they are given by 

\begin{equation}
\Phi^2_{\pm}=\frac{6}{\kappa}\Bigl(1\pm\sqrt{1+\frac{\kappa}{3\lambda}\bigl(
\mu^2+\frac{1}{3}\Lambda-\frac{1}{9}\chi^2\bigr)}\Bigr). \label{eq:2.4.6}
\end{equation}
Since $\Phi^2$ should be real, 

\begin{equation}
\frac{1}{9}\chi^2\leq\frac{1}{9}\chi^2_c:=\frac{3\lambda}{\kappa}+\mu^2+
\frac{1}{3}\Lambda \label{eq:2.4.7}
\end{equation}
must hold. Clearly, $\Phi^2_+\geq6/\kappa$, and $\Phi^2_{-}\leq6/\kappa$; and 
$\Phi^2_\pm=6/\kappa$ when the equality holds in (\ref{eq:2.4.7}). Since 
$\Phi^2_\pm$ must be positive, in the latter case $\mu^2+\Lambda/3<\frac{1}{9}
\chi^2$ must also hold. However, as we already mentioned in connection with 
the non-trivial zeros, this condition is satisfied with the known value of 
$\mu^2$ and $\Lambda$. The extremal value of the energy density is 
$\varepsilon(\Phi_\pm,0,\chi)=-\frac{1}{4}\lambda\Phi^4_\pm$. 

The second derivative of $\varepsilon$ at the trivial zero/critical point 
$\Phi=0$ is 

\begin{equation*}
\Bigl(\frac{\partial^2\varepsilon}{\partial\Phi^2}\Bigr)_0=\mu^2+\frac{1}{3}
\Lambda-\frac{1}{9}\chi^2<0;
\end{equation*}
i.e. $\Phi=0$ is \emph{always} a \emph{local maximum} of the energy density. 
(If $\mu^2+\Lambda/3>0$ held, then $\Phi=0$ could be a local minimum for 
small enough $\chi^2$.) On the other hand, at the non-trivial critical points 

\begin{equation}
\Bigl(\frac{\partial^2\varepsilon}{\partial\Phi^2}\Bigr)_{\pm}=\frac{2\lambda
\Phi^2_\pm}{1-\frac{1}{6}\kappa\Phi^2_\pm}, \label{eq:2.4.8}
\end{equation}
which is positive for $\Phi_->0$, but it is negative for $\Phi_+>0$. 
Therefore, $\Phi_-$ (and $-\Phi_-$ also) is a local \emph{minimum}, while 
$\Phi_+$ (and $-\Phi_+$, too) is a local \emph{maximum} of $\varepsilon$. 

\begin{figure}[htbp]
\centerline{\includegraphics[width=0.9\textwidth]{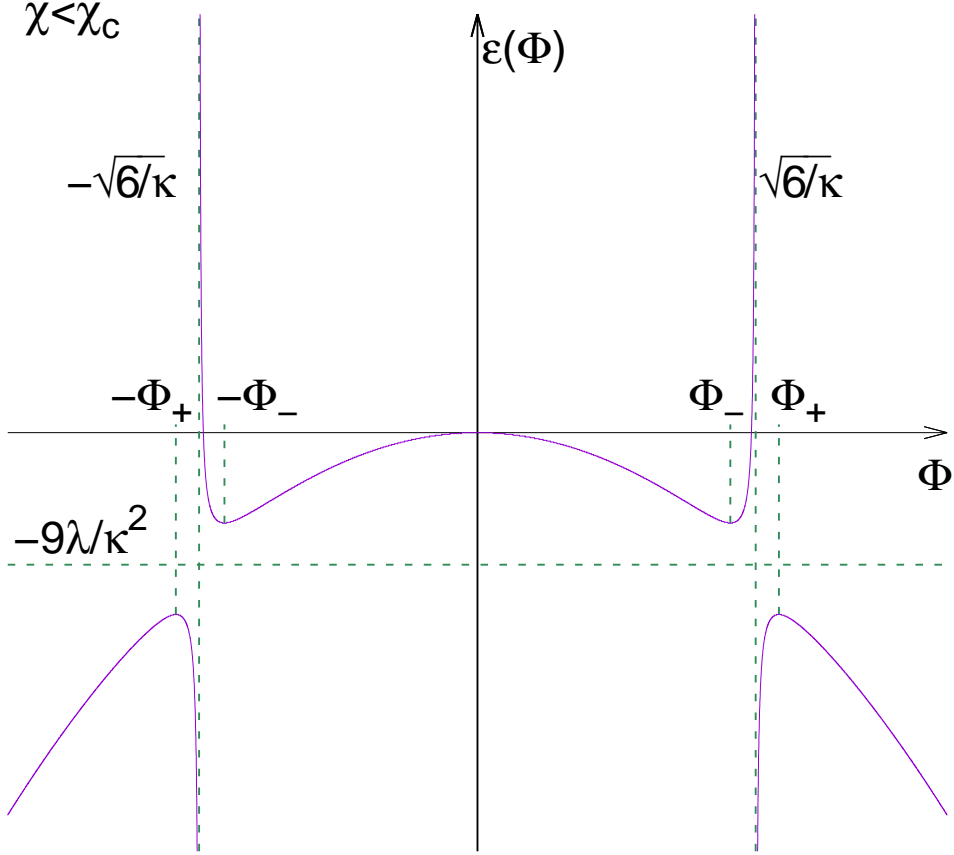}}
\caption{\label{fig:1}
The energy density $\varepsilon$ as a function of $\Phi$ with $\Pi=0$ and 
given $\chi^2<\chi^2_c$. $\varepsilon(\Phi)$ has the `wine bottle' (rather 
than the `Mexican hat') shape, in particular it has minima at $\pm\Phi_-$, 
only in the domain $\Phi^2<6/\kappa$. The critical points $\pm\Phi_+$ are 
maxima of $\varepsilon$. If $\chi\rightarrow\chi_c$, then $\Phi_0,\Phi_{\pm}
\rightarrow\sqrt{6/\kappa}$ and $\varepsilon(\Phi_{\pm})\rightarrow-9\lambda/
\kappa^2$, where $\pm\Phi_0$ are the two non-trivial zeros. }
\end{figure}

Summarizing the above results, we can give a qualitative picture on the 
behaviour of $\varepsilon$ as a function of $\Phi$ for $\Pi=0$ and for 
given $\chi$: First, suppose that $\chi^2<\chi^2_c$. Then, in the domain 
$\Phi<-\sqrt{6/\kappa}$, the energy density $\varepsilon$ is increasing 
from $-\infty$ to the local maximum $-\frac{1}{4}\lambda\Phi^4_+$ at 
$-\Phi_+$, and then it is decreasing and tends to $-\infty$ as $\Phi
\rightarrow-\sqrt{6/\kappa}$. In the domain $-\sqrt{6/\kappa}<\Phi<
\sqrt{6/\kappa}$, the energy density is decreasing from $+\infty$, at 
$-\Phi_0$ it has a zero, and it is decreasing further until $-\Phi_-$, 
where it takes its local minimum. Then it is increasing, takes its local 
maximum at $\Phi=0$, and then it is decreasing until the other local 
minimum at $\Phi_-$. Then, it is increasing again, taking the zero value 
at $\Phi_0$, and tends to $+\infty$ as $\Phi\rightarrow\sqrt{6/\kappa}$. 
Therefore, in this domain at fixed $\chi$, the graph of the function 
$\varepsilon=\varepsilon(\Phi,0,\chi)$ has the `wine bottle' shape, i.e. 
it is like a `Mexican hat', but the `brim' of this `hat' does not extend 
beyond $\Phi=\pm\sqrt{6/\kappa}$. In the domain $\sqrt{6/\kappa}<\Phi$, 
$\varepsilon$ is increasing from $-\infty$ to its local maximum at $\Phi_+$, 
and then decreasing back to $-\infty$ as $\Phi\rightarrow\infty$. The 
energy density has infinite discontinuities at $\mp\sqrt{6/\kappa}$. 

If we increase $\chi^2$ to tend to $\chi^2_c$, then the local maximum in 
the domain $\Phi<-\sqrt{6/\kappa}$ and maximal value, and the local minimum 
in the domain $-\sqrt{6/\kappa}<\Phi<0$ and the corresponding minimal value 
tend, respectively, to $-\sqrt{6/\kappa}$ and $-9\lambda/\kappa^2$. 
Similarly, the local maximum in $\Phi>\sqrt{6/\kappa}$ and maximal value, 
and the local minimum in $0<\Phi<\sqrt{6/\kappa}$ and the minimal value 
tend, respectively, to $\sqrt{6/\kappa}$ and $-9\lambda/\kappa^2$. 
(Fig.~\ref{fig:1})

\begin{figure}[htbp]
\centerline{\includegraphics[width=0.9\textwidth]{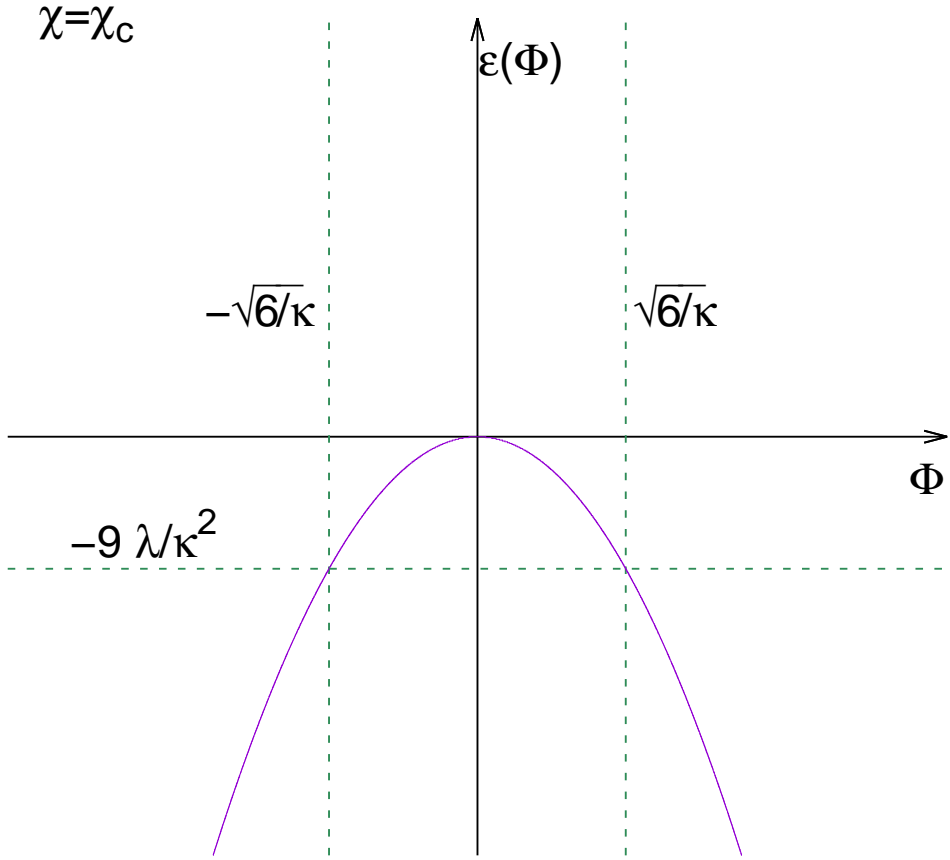}}
\caption{\label{fig:2}
The energy density $\varepsilon$ as a function of $\Phi$ with $\Pi=0$ 
and $\chi^2=\chi^2_c$. $\varepsilon(\Phi)$ is not bounded from below.}

\end{figure}

If $\chi^2=\chi^2_c$, then $\varepsilon(\Phi,0,\chi_c)=-\frac{3}{2}
\frac{\lambda}{\kappa}\Phi^2$. Hence, along the $\chi^2=\chi^2_c$ line, the 
energy density and its derivatives are finite even at $\Phi=\pm\sqrt{6/
\kappa}$ and changes smoothly and monotonically across $\Phi^2=6/\kappa$. 
(Fig.~\ref{fig:2})

\begin{figure}[htb]
\centerline{\includegraphics[width=0.9\textwidth]{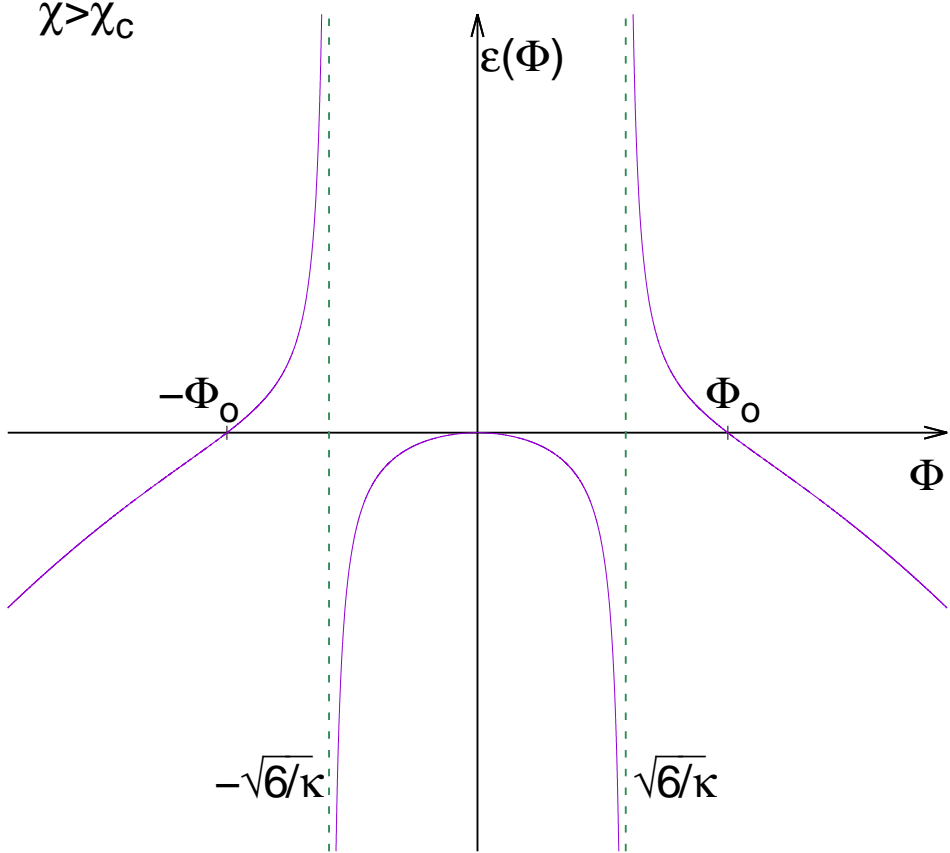}}
\caption{\label{fig:3}
The energy density $\varepsilon$ as a function of $\Phi$ with $\Pi=0$ and 
given $\chi^2>\chi^2_c$. $\varepsilon(\Phi)$ is not bounded from below. If 
$\chi\rightarrow\chi_c$, then $\Phi_0\rightarrow\sqrt{6/\kappa}$.}
\end{figure}

Finally, suppose that $\chi^2>\chi^2_c$. Then, in the domain $\Phi<-\sqrt{6/
\kappa}$, $\varepsilon$ is increasing from $-\infty$, takes the zero value 
at $-\Phi_0$, and it is increasing further and tends to $+\infty$ as $\Phi
\rightarrow-\sqrt{6/\kappa}$. In the domain $-\sqrt{6/\kappa}<\Phi<\sqrt{6/
\kappa}$, the energy density is increasing from $-\infty$, it takes its 
local maximum at $\Phi=0$, and then it is decreasing and tends to $-\infty$ 
as $\Phi\rightarrow\sqrt{6/\kappa}$. Thus, the graph of the function 
$\varepsilon=\varepsilon(\Phi,0,\chi)$ in this domain is a `simple hat', 
whose `brim' does not bend `upwards' and does not extend beyond $\Phi=\pm
\sqrt{6/\kappa}$. Finally, in the domain $\Phi>\sqrt{6/\kappa}$, 
$\varepsilon$ is decreasing from $+\infty$ to $-\infty$ and between these, 
at $\Phi_0$, it takes zero. The energy density has infinite discontinuities 
at $\mp\sqrt{6/\kappa}$. If we decrease $\chi^2$ to tend to $\chi^2_c$, then 
the zeros $\pm\Phi_0$ tend, respectively, to $\pm\sqrt{6/\kappa}$. 
(Fig.~\ref{fig:3})

The three disconnected parts of the graph of the function $\varepsilon=
\varepsilon(\Phi,0,\chi)$ in the $\chi^2<\chi^2_c$ regime join smoothly to 
the corresponding parts in the $\chi^2>\chi^2_c$ regime, and, apart from the 
exceptional points at $(\Phi,\chi^2)=(\pm\sqrt{6/\kappa},\chi^2_c)$, they 
form three disconnected leaves. However, the `asymptotic ends' of any of 
these leaves near the $\Phi=\pm\sqrt{6/\kappa}$ singularities behave 
oppositely in the two regimes. For example, the `wine bottle' of the $\chi^2
<\chi^2_c$ regime is continued in the `simple hat' of the $\chi^2>\chi^2_c$ 
regime such that near the singularity $\Phi=-\sqrt{6/\kappa}$ the resulting 
leaf tends to $-\infty$ for $\chi^2>\chi^2_c$, but it tends to $+\infty$ for 
$\chi^2<\chi^2_c$. 

Therefore, the $\Phi,\chi^2$--(half) plane is naturally split into six 
domains by the $\chi^2=\chi^2_c$ and $\Phi=\pm\sqrt{6/\kappa}$ lines, and it 
is only the domain $-\sqrt{6/\kappa}<\Phi<\sqrt{6/\kappa}$, $\chi^2<\chi^2_c$ 
where $\varepsilon$ has finite local minima. For fixed $\chi^2$ these minima 
are global, but the corresponding minimal value, $-\frac{1}{4}\lambda
\Phi^4_-$, is \emph{decreasing} and tends to the \emph{finite} value 
$-9\lambda/\kappa^2$ as $\chi^2\rightarrow\chi^2_c$ (or, equivalently, as 
$\Phi^2_-\rightarrow6/\kappa$). However, in this limit the second derivative 
$(\partial^2\varepsilon/\partial\Phi^2)_{-}$, given by (\ref{eq:2.4.8}), 
tends to $+\infty$. On the other hand, as we saw above, if we consider the 
limit $\Phi\rightarrow-\sqrt{6/\kappa}$ along the line $\chi^2=\chi^2_c$, 
then $\varepsilon$ still tends to $-9\lambda/\kappa^2$, but $(\partial^2
\varepsilon/\partial\Phi^2)\rightarrow-3\lambda/\kappa$. Hence, there are 
special configurations with $\vert\Phi\vert^2=6/\kappa$ in which the 
energy-momentum tensor (\ref{eq:1.6.3}) is finite, but its derivatives are 
not; indicating that these configurations are probably unstable. 
Nevertheless, these provide the only `bridge' between the $\Phi^2>6/\kappa$ 
and $\Phi^2<6/\kappa$ parts of the phase space. 

Since, for given $\chi$, $(\partial\varepsilon/\partial\Pi)=0$ implies $\Pi
=0$, the extremal points of $\varepsilon(\Phi,\Pi,\chi)$ with respect to 
$\Phi$ and $\Pi$ are just the extremal points of $\varepsilon(\Phi,0,\chi)$ 
with respect to $\Phi$. On the other hand, for $\Pi\not=0$ the domain of 
positivity of $\varepsilon$ changes: On the $\Phi,\chi^2$--(half) plane this 
domain is `wider' for $\Pi^2>0$ than for $\Pi^2=0$, and, in particular, 
$\varepsilon$ is positive on a strip with the $\Phi=0$ center-line. In fact, 
for given $\chi^2$, in the $\Phi,\Pi^2$-(half) plane the domain on which 
$\varepsilon$ is non-positive is 

\begin{equation*}
\Pi^2+\Bigl(\sqrt{\frac{\lambda}{2}}\Phi^2+\sqrt{\frac{1}{2\lambda}}\bigl(
\mu^2+\frac{1}{3}\Lambda-\frac{1}{9}\chi^2\bigr)\Bigr)^2\leq\frac{1}
{2\lambda}\bigl(\mu^2+\frac{1}{3}\Lambda-\frac{1}{9}\chi^2\bigr)^2. 
\end{equation*}
For given $\chi^2$ this domain is compact, but since the right hand side of 
this expression is never zero, this domain is not compact in the 
$\chi^2$--direction. Also, if $\Pi\not=0$, then the locus of the `bridge' 
between the $\Phi^2>6/\kappa$ and $\Phi^2<6/\kappa$ parts of the phase space 
changes: It is given by the two hyperbolas $\chi^2-\frac{3}{2}\kappa\Pi^2=
\chi^2_c$ in the $\Phi=\pm\sqrt{6/\kappa}$, $S={\rm const}$ 2-planes in the 
phase space. For a more detailed discussion of the behaviour of $\varepsilon$ 
see \cite{SzW}.

\subsubsection{The static solution}
\label{sub-2.4.3}

It is an easy exercise to check that the equations 
(\ref{eq:2.4.1})-(\ref{eq:2.4.3}) formally admit a static solution, which is 
just the Minkowski spacetime, but yield a condition for the numerical value 
of the constants $\lambda$, $\mu^2$, $\Lambda$ and $\kappa$ that is not 
satisfied in the observed Universe. In fact, $\Phi^{\bi}=0$ solves 
(\ref{eq:2.4.2}) and yields $\Pi^{\bi}=0$, and hence $\varepsilon=0$ and $P=0$. 
Then, by the second of (\ref{eq:2.4.1}), $\Lambda=0$ follows, which, by the 
first of (\ref{eq:2.4.1}), yields that $k=0$. Similarly, if $\Phi^{\bi}$ is a 
non-zero static solution of (\ref{eq:2.4.2}), then $\Pi^{\bi}=-\chi\Phi^{\bi}
/3$ and $\vert\Phi\vert^2=-(\mu^2+2\Lambda/3)/(\lambda+\kappa\mu^2/6)$. 
Substituting these into (\ref{eq:2.4.3}) and using the second of 
(\ref{eq:2.4.1}) we find that 

\begin{equation*}
\vert\Phi\vert^2=-\frac{\mu^2}{\lambda}, \hskip 20pt 
\varepsilon=-\frac{\Lambda}{\kappa}=-\frac{1}{4}\lambda\vert\Phi\vert^4.
\end{equation*}
For $\lambda>0$ these imply that $\mu^2<0$ and $\Lambda>0$ must hold. Finally, 
by the first of (\ref{eq:2.4.1}) $k=0$ follows. It might be worth noting 
that this non-trivial static state of the Higgs field is \emph{precisely} 
the symmetry breaking vacuum state that we saw in the Weinberg--Salam model, 
but clearly this is \emph{not} a critical point of the energy density 
(\ref{eq:2.4.3}). In this static solution it would be a large positive value 
of the cosmological constant that compensates the large negative energy 
density of the Higgs vacuum. On the other hand, although the \emph{observed} 
cosmological constant is \emph{strictly positive}, but it is much-much 
smaller than that would follow from the Weinberg--Salam model via these 
expressions. In fact, in traditional units the energy density in the vacuum 
state would be $\varepsilon_v=-2.2\times 10^{46}erg/cm^3$, while that 
corresponding to the observed cosmological constant is $\varepsilon_\Lambda=
4.7\times 10^{-11}erg/cm^3$. Thus, this static solution cannot be expected to 
give a reliable model of the observed Universe. That should be dynamical.


\subsection{Example: Configurations with Kantowski--Sachs 
symmetries}
\label{sub-2.5}

In the present subsection we determine a class of field configurations that 
admit the isometries of the Kantowski--Sachs metrics as symmetries in which 
the (global or quasi-local) instantaneous vacuum states should be searched 
for. Since the number of spacetime symmetries is less than the maximal one, 
the Kantowski--Sachs case is technically more complicated than the FRW case. 
In subsection \ref{sub-2.5.1}, we determine the matter field configurations 
that admit the Kantowski--Sachs symmetries, and then, in subsection 
\ref{sub-2.5.2}, those with minimal energy density.

\subsubsection{Matter fields with Kantowski--Sachs symmetries}
\label{sub-2.5.1}

Let us consider spacetimes with the Kantowski--Sachs line element $ds^2=dt^2-
X^2dr^2-Y^2d\Omega^2$, where $X=X(t)$ and $Y=Y(t)$ are positive functions and 
$d\Omega^2$ denotes the line element on the unit 2-sphere (see e.g. 
\cite{Coll,MacCallum}). For example the line element inside the Schwarzschild 
black hole belongs to the Kantowski--Sachs class (see e.g. appendix B in 
\cite{HE}). These metrics admit four spacelike Killing vectors, the three 
familiar ones $K^a_1$, $K^a_2$ and $K^a_3$ for the spherical symmetry with 
transitivity surfaces $t={\rm const}$, $r={\rm const}$; and the fourth is 
$K^a_4=(\partial/\partial r)^a$, which commutes with the previous three. Let 
$v^a:=X^{-1}K^a_4$, the unit normal of the transitivity surfaces of the 
spherical symmetry in the $\Sigma_t:=\{t={\rm const}\}$ hypersurfaces. The 
extrinsic curvature of these hypersurfaces is $\chi_{ab}=-\dot XX^{-1}v_av_b+
\dot YY^{-1}q_{ab}$, where $q_{ab}$ is the induced (negative definite) metric 
on the $t={\rm const}$, $r={\rm const}$ 2-spheres and over-dot denotes 
derivative with respect to $t$. Note that $\chi=\chi(t)$, i.e. the mean 
curvature of the leaves $\Sigma_t$ is spatially constant. In the coordinates 
$(r,\theta,\phi)$ the only nonzero component of the curvature tensor of the 
intrinsic geometry of the hypersurfaces is ${\cal R}^2{}_{3cd}=-\sin^2\theta
(\delta^2_c\delta^3_d-\delta^2_d\delta^3_c)$; and hence the corresponding 
curvature scalar is ${\cal R}=2/Y^2>0$. A direct calculation shows that 
$D_a(\chi^a{}_b-\chi\delta^a_b)=0$. 

Recall that in field theory the vacuum states are defined to be those field 
configurations that (i.) do not break the spacetime symmetries and (ii.) 
minimize the energy functional. In particular, by the first requirement (a.) 
no state that can be a potential (instantaneous) vacuum state can specify 
any direction different from $v^a$; and (b.) any such state must be invariant 
(in some suitable sense) under the action of the spacetime symmetries. By 
(a.) the field strengths $E^\alpha_a$ and $B^\alpha_{ab}$, the Higgs field 
$\Phi^{\bi}$ and a spinor multiplet $\psi^r_A$ in any particular vacuum state 
must be such that $E^\alpha_a=E^\alpha v_a$, $B^\alpha_{ab}=B^\alpha\varepsilon
_{ab}$, $\mathbb{D}_a\Phi^{\bi}=\varphi^{\bi}v_a$ and $\psi^r_A=\psi^r
\varepsilon_A$, where $E^\alpha$, $B^\alpha$, $\varphi^{\bi}$ and $\psi^r$ are 
gauge covariant vector and spatial scalar fields, $\varepsilon_{ab}$ is the 
area 2-form on the $t={\rm const}$, $r={\rm const}$ 2-surfaces and 
$\varepsilon_A$ is one of the vectors of the normalized spinor dyad $\{o_A,
\iota_A\}$ adapted to the $t={\rm const}$, $r={\rm const}$ 2-surfaces (in 
the sense that $\sqrt{2}t_a=o_A\bar o_{A'}+\iota_A\bar\iota_{A'}$ and 
$\sqrt{2}v_a=o_A\bar o_{A'}-\iota_A\bar\iota_{A'}$). (To see that any spinor 
field of the spinor multiplet $\psi^r_A$ in the vacuum state must be 
proportional to either $o_A$ or $\iota_A$, it is enough to recall that the 
projections $P^b_ao_B\bar o_{B'}$ and $P^b_a\iota_B\bar\iota_{B'}$ are 
proportional to $v_a$, but for any nontrivial combination $\psi_A=-\psi_0
\iota_A+\psi_1o_A$ the projection $P^b_a\psi_B\bar\psi_{B'}$ is not.) 

To formulate mathematically part (b.) of requirement (i.), let us recall how 
the spacetime symmetries are implemented in gauge theories (see \cite{FoMa}): 
In the actual system the state represented by a field configuration is said 
to be symmetric if for each Killing vector $K^a$ there exists a Lie algebra 
valued function $\lambda^\alpha g_\alpha$ such that 

\begin{eqnarray}
&{}&{\pounds}_KE^\alpha_a=\lambda^\mu c^\alpha_{\mu\nu}E^\nu_a, \hskip 10pt
    {\pounds}_KB^\alpha_{ab}=\lambda^\mu c^\alpha_{\mu\nu}B^\nu_{ab}, \nonumber \\
&{}&{\pounds}_K\Phi^{\bi}=\lambda^\mu T^{\bi}_{\mu{\bj}}\Phi^{\bj}, \hskip 10pt
    {\pounds}_K\Pi^{\bi}=\lambda^\mu T^{\bi}_{\mu{\bj}}\Pi^{\bj}, \hskip 10pt
    {\pounds}_K\psi^r_A=\lambda^\mu T^r_{\mu{s}}\psi^s_A; \label{eq:2.5.1}
\end{eqnarray}
i.e. it is required that the Lie dragging of the fields along the integral 
curves of the Killing vectors could always be compensated by an appropriate 
gauge transformation. Here ${\pounds}_K$ denotes Lie derivative along $K^a$, 
and actually the structure constants $c^\alpha_{\mu\nu}$ play the role of the 
representation matrices of the Lie algebra in the adjoint representation. 
Note that although the Lie derivative of a Weyl spinor field along a general 
vector field is not canonically defined, it is well defined if $K^a$ is a 
(conformal) Killing vector \cite{HuTo}. 

However, if a given field configuration is intended to represent a vacuum 
state, then it reduces the gauge group to the identity (symmetric vacuum) or 
to a non-trivial, smaller gauge group (symmetry breaking vacuum). In the 
latter case this subgroup is just the stabilizer subgroup $G_v\subset G$ of 
the vacuum state, and the Lie algebra valued function $\lambda^\alpha$ in the 
criteria (\ref{eq:2.5.1}) should be restricted to take its values in the Lie 
algebra of the reduced gauge group $G_v$. But then, just by the definition 
of $G_v$, the right hand sides of equations in (\ref{eq:2.5.1}) are all 
vanishing for such restricted Lie algebra valued functions. Thus 

\begin{equation}
{\pounds}_KE^\alpha_a=0, \hskip 10pt
{\pounds}_KB^\alpha_{ab}=0, \hskip 10pt
{\pounds}_K\Phi^{\bi}=0, \hskip 10pt
{\pounds}_K\Pi^{\bi}=0, \hskip 10pt
{\pounds}_K\psi^r_A=0 \label{eq:2.5.2}
\end{equation}
must hold for the fields in the vacuum states for any Killing vector $K^a$. 
Solving the first four of these equations is a straightforward exercise, 
and probably the simplest way of solving the fifth of these is based on the 
following observation: If the rotation Killing vectors are given explicitly 
by 

\begin{equation*}
K^a_1=-\sin\phi(\frac{\partial}{\partial\theta})^a-\cot\theta\cos\phi(
 \frac{\partial}{\partial\phi})^a, \hskip 8pt 
K^a_2=\cos\phi(\frac{\partial}{\partial\theta})^a-\cot\theta\sin\phi(
 \frac{\partial}{\partial\phi})^a, \hskip 8pt 
K^a_3=(\frac{\partial}{\partial\phi})^a,
\end{equation*}
then the Lie derivative of the vectors of the normalized spin frame $\{o_A,
\iota_A\}$ is 

\begin{equation*}
{\pounds}_{K_l}o_A=-\frac{\rm i}{2}Y_lo_A, \hskip 20pt
{\pounds}_{K_l}\iota_A=\frac{\rm i}{2}Y_l\iota_A, \hskip 25pt 
l=1,2,3,4;
\end{equation*}
where $Y_l=(\sin\theta\sin\phi,\sin\theta\cos\phi,\cos\theta,0)$. (For the 
explicit form of the Lie derivative of spinors in terms of their covariant 
derivative and the Killing field, see \cite{HuTo}.) Then, writing $\psi^r_A=
\psi^ro_A$ or $\psi^r\iota_A$ (just according to the second paragraph of the 
present subsection) and substituting this into the fifth of (\ref{eq:2.5.2}), 
we find that $\psi^r=0$. Thus, the vacuum states of the EccSM system should 
be among the states represented by the field configurations of the form: 

\begin{equation}
E^\alpha_a=E^\alpha(t)v_a, \hskip 10pt
B^\alpha_{ab}=B^\alpha(t)\varepsilon_{ab}, \hskip 10pt
\Phi^{\bi}=\Phi^{\bi}(t), \hskip 10pt
\Pi^{\bi}=\Pi^{\bi}(t), \hskip 10pt
\psi^r_A=0.  \label{eq:2.5.3}
\end{equation}
This structure of the magnetic field strength makes it possible to find a 
gauge in which $A^\alpha_a$ is also aligned with the spacetime symmetries. In 
fact, in the coordinate system $(t,r,\theta,\phi)$ the expression of the 
magnetic field strength $B^\alpha_{ab}=\partial_aA^\alpha_b-\partial_bA^\alpha_a+
c^\alpha_{\mu\nu}A^\mu_aA^\nu_b$, together with $\sqrt{\vert q\vert}=Y^2(t)\sin
\theta$, yields 

\begin{eqnarray*}
&{}&\partial_rA^\alpha_\theta-\partial_\theta A^\alpha_r+c^\alpha_{\mu\nu}
  A^\mu_rA^\nu_\theta=0, \\
&{}&\partial_rA^\alpha_\phi-\partial_\phi A^\alpha_r+c^\alpha_{\mu\nu}
  A^\mu_rA^\nu_\phi=0, \\
&{}&\partial_\theta A^\alpha_\phi-\partial_\phi A^\alpha_\theta+c^\alpha
  _{\mu\nu}A^\mu_\theta A^\nu_\phi=B^\alpha Y^2\sin\theta.
\end{eqnarray*}
The third of these equations decouples from the first two, and does not depend 
on $r$. Hence it can be solved for $A^\alpha_\theta$ and $A^\alpha_\phi$, and the 
solution can be chosen to be independent of $r$. (Their $t$-dependence cannot 
be chosen arbitrarily, that is determined by the evolution equations.) Then 
the first two of these equations form a system of partial differential 
equations for $A^\alpha_r$, whose integrability condition (using the third of 
these equations) is just $c^\alpha_{\mu\nu}B^\mu A^\nu_r=0$. Hence, by $\partial
_rA^\alpha_\theta=\partial_rA^\alpha_\phi=0$, there is a gauge in which $v^a
A^\alpha_a=0$ holds. 

Finally, by the result for the form of $\mathbb{D}_a\Phi^{\bi}$ (obtained in 
the second paragraph above) and the third of (\ref{eq:2.5.3}) we have that 
$v_a\varphi^{\bi}=\mathbb{D}_a\Phi^{\bi}=A^\alpha_aT^{\bi}_{\alpha{\bj}}\Phi^{\bj}$. 
This implies, first, that the components $A^\alpha_\theta$ and $A^\alpha_\phi$ 
of the spatial vector potential take their values in the Lie algebra of the 
reduced gauge group $G_v$; and, second, that $\varphi^{\bi}=0$ by $A^\alpha_r
=0$, i.e. 

\begin{equation}
\mathbb{D}_a\Phi^{\bi}=0. \label{eq:2.5.4}
\end{equation}
We search the instantaneous vacuum states among the field configurations 
satisfying (\ref{eq:2.5.3})-(\ref{eq:2.5.4}), and we specify them in the 
next sections, where we evaluate requirement (ii.) above, too. 


\subsubsection{The minimal energy configurations}
\label{sub-2.5.2}

In the special field configurations (\ref{eq:2.5.3})-(\ref{eq:2.5.4}), the 
energy density $\varepsilon$ of the matter fields (with the specific 
self-interaction term of the Weinberg--Salam model) reduces to 

\begin{equation}
\varepsilon=\frac{1}{1-\frac{1}{6}\kappa\vert\Phi\vert^2}\Bigl(-\frac{1}{2}
G_{\alpha\beta}\bigl(E^\alpha E^\beta+B^\alpha B^\beta\bigr)+\frac{1}{2}\vert\Pi
\vert^2+\frac{1}{2}\bigl(\mu^2+\frac{1}{3}\Lambda-\frac{1}{9}\chi^2\bigr)
\vert\Phi\vert^2+\frac{1}{4}\lambda\vert\Phi\vert^4\Bigr),  \label{eq:2.5.5}
\end{equation}
while the momentum density $\pi_a$ is zero. Thus, the matter sector of the 
(total or quasi-local) energy-momentum functional in these special 
configurations is simply the hypersurface integral of the (spatially 
constant) $\varepsilon$. Hence, to find the instantaneous vacuum states, we 
should determine the local minima of $\varepsilon$. 

The structure of $\varepsilon=\varepsilon(E^\alpha,B^\alpha,\Phi^{\bi},\Pi^{\bi};
\chi)$ is exactly the same that of $\varepsilon=\varepsilon(\Phi^{\bi},\Pi
^{\bi};\chi)$ in the FRW case. Thus, they have the same qualitative properties 
that $\varepsilon=\varepsilon(\Phi,0;\chi)$ has for a single real scalar 
field $\Phi$. In particular, the critical points of $\varepsilon=\varepsilon
(E^\alpha,B^\alpha,\Phi^{\bi},\Pi^{\bi};\chi)$ with respect to the matter field 
variables are at $E^\alpha=0$, $B^\alpha=0$, $\Pi^{\bi}=0$ and $\Phi^{\bi}$ 
solving (\ref{eq:2.3.4dWS}). Its solutions have the pointwise norm 
(\ref{eq:2.4.6}), just like in the FRW case, whose reality is ensured 
precisely by the bound (\ref{eq:2.4.7}) for the mean curvature. It is only 
$\pm\Phi_-$ that represents local minima of the energy density.


\section{The EccSM system with instantaneous vacuum states}
\label{sec:3}

\subsection{The 3+1 form of the Lagrangian}
\label{sub-3.1}

Let the foliation $\Sigma_t$ of the spacetime be fixed, let $N$ be its lapse, 
and consider the so-called evolution vector field $\xi^a=Nt^a+N^a$, where 
$N^a$, the shift part of $\xi^a$, is tangent to $\Sigma_t$. The 3+1 form of 
the spacetime volume element is ${\rm d}v=N{\rm d}\Sigma{\rm d}t$, where, in 
a local coordinate system $(x^1,x^2,x^3)$ on $\Sigma_t$, the hypersurface 
volume element is ${\rm d}\Sigma=\sqrt{\vert h\vert}{\rm d}^3x$. The time 
derivative of a spatial tensor field, say $T^{a...}_{b...}$, is defined by 
$\dot T^{a...}_{b...}:=({\pounds}_\xi T^{c...}_{d...})P^a_c\cdots P^d_b\cdots$. 
Thus, the Lagrangian variables in the Yang--Mills--Higgs sector are $\phi
^\alpha$, $\dot\phi^\alpha$, $A^\alpha_a$, $\dot A^\alpha_a$, $\Phi^{\bi}$ and 
$\dot\Phi^{\bi}$. Since, however, the Lie derivative of a spinor field along 
a general vector field is not canonically defined, for the Lagrangian spinor 
field variables we choose the \emph{components} $\psi^r_{\uA}$ of the spinor 
field $\psi^r_A$ and their time derivative $\dot\psi^r_{\uA}$, where the 
spinor components are defined with respect to some fixed normalized dual spinor 
basis $\{\varepsilon^A_{\uA},\varepsilon^{\uA}_A\}$ by $\psi^r_A=:\psi^r_{\uA}
\varepsilon^{\uA}_A$. 

Clearly, the form of the Lagrangian ${\cal L}_I$ and the potential $V$ remains 
the same in their 3+1 form, and it is straightforward to find the 3+1 form of 
${\cal L}_{YM}$ and ${\cal L}_W$. The former is 

\begin{equation}
{\cal L}_{YM}=\frac{1}{2}G_{\alpha\beta}E^\alpha_aE^\beta_bh^{ab}+\frac{1}{4}
G_{\alpha\beta}B^\alpha_{ac}B^\beta_{bd}h^{ab}h^{cd}, \label{eq:3.2.1}
\end{equation}
where by (\ref{eq:2.1.2a}) the electric field strength is $E^\alpha_a=\frac{1}
{N}(D_a(N\phi^\alpha)+A^\mu_ac^\alpha_{\mu\nu}N\phi^\nu-\dot A^\alpha_a+{\pounds}_N
A^\alpha_a)$ and the magnetic field strength is expressed by the spatial vector 
potential and its derivative; and the latter is 

\begin{eqnarray}
{\cal L}_{W}=\frac{\rm i}{2N}G_{rr'}t^{AA'}\Bigl(\!\!\!\!&{}\!\!\!\!&\bar\psi
 ^{r'}_{A'}\varepsilon^{\uA}_A\bigl(\dot\psi^r_{\uA}-N^eD_e\psi^r_{\uA}-Nt^e
 \Gamma^{\uB}_{e{\uA}}\psi^r_{\uB}\bigr)-\nonumber \\
-\!\!\!\!&{}\!\!\!\!&\psi^r_A\bar\varepsilon^{{\uA}'}_{A'}\bigl(\dot{\bar\psi}
 ^{r'}_{{\uA}'}-N^eD_e\bar\psi^{r'}_{{\uA}'}-Nt^e\bar\Gamma^{{\uB}'}_{e{\uA}'}\bar
 \psi^{r'}_{{\uB}'}\bigr)\Bigr)+ \label{eq:3.2.2} \\
+\frac{\rm i}{2}G_{rr'}t^{AA'}\phi^\alpha\bigl(\!\!\!\!&{}\!\!\!\!&T^r
 _{\alpha s}\psi^s_A\bar\psi^{r'}_{A'}-\bar T^{r'}_{\alpha s'}\bar\psi^{s'}
 _{A'}\psi^r_A\bigr)+\frac{\rm i}{2}G_{rr'}h^{ab}\bigl(\bar\psi^{r'}_{A'}
 \mathbb{D}_b\psi^r_A-\psi^r_A\mathbb{D}_b\bar\psi^{r'}_{A'}\bigr), \nonumber
\end{eqnarray}
where $\Gamma^{\uA}_{e{\uB}}:=\varepsilon^{\uA}_A\nabla_e\varepsilon^A_{\uB}$, and 
note that the spatial vector potential $A^\alpha_a$ is involved in $\mathbb{D}
_e$. Thus, if we introduce the `mechanical' Lagrangians $L_{YM}:=\int_\Sigma
{\cal L}_{YM}N\sqrt{\vert h\vert}{\rm d}^3x$ and $L_W:=\int_\Sigma{\cal L}_WN
\sqrt{\vert h\vert}{\rm d}^3x$, then for the canonical momenta we obtain 

\begin{equation*}
\frac{\delta L_{YM}}{\delta\dot A^\alpha_a}=-G_{\alpha\beta}E^\beta_bh^{ba}, 
\hskip 20pt
\frac{\delta L_W}{\delta\dot\psi^r_{\uA}}=\frac{\rm i}{2}G_{rr'}\bar\psi^{r'}
_{A'}t^{A'A}\varepsilon^{\uA}_A.
\end{equation*}
Thus, as we claimed, $E^\alpha_a$ is essentially the momentum conjugate to 
$A^\alpha_a$, while the momentum conjugate to $\phi^\alpha$ is zero. The momentum 
conjugate to the spinor multiplet is just its own complex conjugate, in 
accordance with the fact that the field equations for the spinor fields are 
only first order. 

On the other hand, because of the term $\frac{1}{12}R\vert\Phi\vert^2$ in 
${\cal L}_H$, finding the most convenient 3+1 form of the Higgs Lagrangian is 
slightly more complicated. It is known \cite{IsNe} that the spacetime 
curvature scalar can be decomposed as 

\begin{equation*}
R={\cal R}+\chi_{ab}\chi^{ab}-\chi^2+\frac{2}{N\sqrt{\vert h\vert}}\frac{\rm d}
{{\rm d}t}\bigl(\chi\sqrt{\vert h\vert}\bigr)+\frac{2}{N}D_a\bigl(D^aN-\chi 
N^a\bigr),
\end{equation*}
where ${\cal R}$ is the curvature scalar of the intrinsic geometry of the 
hypersurface; by means of which 

\begin{eqnarray*}
R\vert\Phi\vert^2\!\!\!\!&=\!\!\!\!&\bigl({\cal R}+\chi_{ab}\chi^{ab}-\chi^2
 \bigr)\vert\Phi\vert^2-\frac{2}{N}\chi\bigl(\dot\Phi^{\bi}G_{\bi\bj'}\bar
 \Phi^{\bj'}+\Phi^{\bi}G_{\bi\bj'}\dot{\bar\Phi}^{\bj'}\bigr)-\frac{2}{N}\bigl(
 D^aN-\chi N^a\bigr)D_a\vert\Phi\vert^2 \\
\!\!\!\!&+\!\!\!\!&\frac{2}{N\sqrt{\vert h\vert}}\frac{\rm d}{{\rm d}t}\bigl(
 \chi\vert\Phi\vert^2\sqrt{\vert h\vert}\bigr)+\frac{2}{N}D_a\bigl((D^aN-\chi 
 N^a)\vert\Phi\vert^2\bigr). 
\end{eqnarray*}
Thus, in the 3+1 form of the Higgs Lagrangian, it seems natural to drop the 
last two terms, which, after integration, would give a total time derivative 
and the integral of a total spatial divergence, respectively. Hence, we 
choose the 3+1 form of the Higgs Lagrangian to be 

\begin{eqnarray}
\hat{\cal L}_H:\!\!\!\!&=\!\!\!\!&\frac{1}{2}G_{\bi\bj'}t^a\bigl(\nabbla_a
 \Phi^{\bi}\bigr)t^b\bigl(\nabbla_b\bar\Phi^{\bj'}\bigr)+\frac{1}{2}G_{\bi
 \bj'}h^{ab}\bigl(D_a\Phi^{\bi}+A^\alpha_aT^{\bi}_{\alpha{\bk}}\Phi^{\bk}
 \bigr)\bigl(D_b\bar\Phi^{\bj'}+A^\beta_b\bar T^{\bj'}_{\beta{\bl'}}\bar
 \Phi^{\bl'}\bigr)- \nonumber \\
\!\!\!\!&-\!\!\!\!&\frac{1}{4}G_{\bi\bj\bk'\bl'}\Phi^{\bi}\Phi^{\bj}\bar
 \Phi^{\bk'}\bar\Phi^{\bl'}-\frac{1}{12}\bigl({\cal R}+\chi_{ab}\chi^{ab}-
 \chi^2\bigr)G_{\bi\bj'}\Phi^{\bi}\bar\Phi^{\bj'}+ \nonumber \\
\!\!\!\!&+\!\!\!\!&\frac{1}{6}\frac{1}{N}\chi G_{\bi\bj'}\bigl(\dot\Phi^{\bi}
 \bar\Phi^{\bj'}+\Phi^{\bi}\dot{\bar\Phi}^{\bj'}\bigr)+\frac{1}{6}\frac{1}{N}
 \bigl(D^aN-\chi N^a\bigr)D_a\vert\Phi\vert^2, \label{eq:3.2.3}
\end{eqnarray}
where 

\begin{equation}
t^a\nabbla_a\Phi^{\bi}=\frac{1}{N}\bigl(\dot\Phi^{\bi}+N\phi^\alpha T^{\bi}_{\alpha
{\bk}}\Phi^{\bj}-N^aD_a\Phi^{\bi}\bigr). \label{eq:3.2.4}
\end{equation}
The canonical momentum conjugate to $\Phi^{\bi}$, calculated from the 
`mechanical' Lagrangian $\hat L_H:=\int_\Sigma\hat{\cal L}_HN\sqrt{\vert h
\vert}{\rm d}^3x$, is 

\begin{equation}
\frac{\delta \hat L_H}{\delta\dot\Phi^{\bi}}=\frac{1}{2}G_{\bi\bj'}\bigl(t^a
\nabbla_a\bar\Phi^{\bj'}+\frac{1}{3}\chi\bar\Phi^{\bj'}\bigr)=\frac{1}{2}
G_{\bi\bj'}\bar\Pi^{\bj'}. \label{eq:3.2.5}
\end{equation}
Therefore, $\Pi^{\bi}=t^a\nabbla_a\Phi^{\bi}+\frac{1}{3}\chi\Phi^{\bi}$ is, in 
fact, essentially the canonical momentum conjugate to $\bar\Phi^{\bi'}$, as we 
claimed in subsection \ref{sub-2.2}.

\subsection{The instantaneous vacuum states}
\label{sub-3.2}

\subsubsection{The definition of the instantaneous vacuum states}
\label{sub-3.2.1}

We search for the instantaneous vacuum states among the critical points of 
${\tt Q}[K]$. In these states the gauge fields and the spinor fields are 
vanishing, and the Higgs field $\Phi^{\bi}_v$ is such that $D_e\Phi^{\bi}_v=0$, 
it solves (\ref{eq:2.3.4d}), and its canonical momentum is zero: $\Pi^{\bi}_v
=0$. By (\ref{eq:3.2.4}) the last condition is equivalent to 

\begin{equation}
\frac{1}{N}\dot\Phi^{\bi}_v=\frac{1}{N}\bigl(\dot\Phi^{\bi}_v-N^aD_a\Phi^{\bi}_v
\bigr)=t^e(\nabla_e\Phi^{\bi}_v)=t^e(\nabbla_e\Phi^{\bi}_v)=\Pi^{\bi}_v-\frac{1}
{3}\chi\Phi^{\bi}_v=-\frac{1}{3}\chi\Phi^{\bi}_v. \label{eq:3.2.4v}
\end{equation}
Here we used also that the scalar and spatial vector potentials, $\phi
^\alpha$ and $A^\alpha_a$, can be chosen to be vanishing even globally. 

At the end of subsection \ref{sub-2.3.2} we saw that the 1-parameter family 
of these states does \emph{not} solve all of the field equations. 
Nevertheless, we want these instantaneous states to be \emph{physical 
states}, thus they are required to solve the \emph{constraint} parts of the 
field equations. Thus, an instantaneous vacuum state is represented not 
only by a special configuration of the matter fields alone, but by such a 
configuration of the matter fields \emph{and} the initial data set for 
Einstein's equations. As we saw, in the gravitational part of such a data 
both the mean curvature and the intrinsic curvature scalar must be constant. 

Moreover, if the spacetime admits some geometric symmetry, then it seems 
natural to require that the instantaneous vacuum states be represented by 
field configurations with the same geometric symmetry. Or, in other words, 
these states could be expected to break only the \emph{internal gauge} 
symmetries, but \emph{not} the geometric symmetries of the spacetime. In 
particular, if the spacetime admits the FRW or Kantowski--Sachs symmetries, 
then one could expect that the instantaneous vacuum states have the same 
geometric symmetries, too. In this case, the (spatially constant) mean 
curvature in the gravitational part of the vacuum state, which plays the 
role of the extrinsic York time parameter and labels the hypersurfaces, is 
identical with the mean curvature of the actual hypersurface on which e.g. 
the rest mass of the matter fields is calculated.

\subsubsection{The existence of the instantaneous vacuum states: A 
condition on the mean curvature}
\label{sub-3.2.2}

For the usual self-interaction coefficient of the Higgs field, $G_{\bi\bj\bk'\bl'}
=\lambda G_{\bk'(\bi}G_{\bj)\bl'}$, the norm of the `vacuum value' of the Higgs 
field (see equation (\ref{eq:2.3.4dWSs})) is 

\begin{equation}
\vert\Phi_v\vert^2=\frac{6}{\kappa}\Bigl(1-\sqrt{1+\frac{\kappa}{3\lambda}
\bigl(\mu^2+\frac{1}{3}\Lambda-\frac{1}{9}\chi^2\bigr)}\Bigr)=-\frac{\mu^2}
{\lambda}+\frac{\kappa}{12}\frac{\mu^4}{\lambda^2}-\frac{1}{\lambda}\Bigl(
\frac{\Lambda}{3}-\frac{1}{9}\chi^2\Bigr)+..., \label{eq:3.2.6}
\end{equation}
just the solution $\Phi^2_-$ given by (\ref{eq:2.4.6}). (The solution with 
the $+$ sign in front of the square root is a local \emph{maximum} rather 
than a minimum of the energy density in the domain $\vert\Phi\vert^2\geq6/
\kappa$, see subsection \ref{sub-2.4.2}.) Thus the structure of $\vert\Phi_v
\vert^2$ (and the interpretation of the various corrections in $\vert\Phi_v
\vert^2$ to the flat-spacetime Standard Model expression) is very similar 
to that of (\ref{eq:1.7.4}). The only (but essential) difference between 
these two is the extrinsic curvature term in (\ref{eq:3.2.6}), which makes 
the instantaneous vacuum states \emph{time dependent}. 

However, for given parameters $\lambda$, $\mu^2$ and $\Lambda$ the inequality 
(\ref{eq:2.4.7}) is a nontrivial condition on $\chi^2$, i.e. on the mean 
curvature of the hypersurface $\Sigma_t$ on which instantaneous vacuum states 
are possible. This is a \emph{finite upper bound}, and its value is given by 
$\chi^2_c:=9(3\lambda/\kappa+\mu^2+\Lambda/3)\simeq 4.8\times10^{64}cm^{-2}$. 
On hypersurfaces with greater mean curvature \emph{there are no such vacuum 
states at all}. The corresponding characteristic (Hubble) time is $t_c:=3/
\chi_c\simeq 4.5\times 10^{-43}sec$, which is almost ten times longer than 
the Planck time $T_P\simeq 5.39\times 10^{-44} sec$. 

In particular, in a FRW spacetime the instantaneous vacuum states start to 
emerge on the hypersurface for which the Hubble time $t_H:=S/\dot S$ was 
$t_c$. The present value of the Hubble constant in our observed Universe is 
$H_0:=(\dot S/S)_{\rm now}=\frac{1}{3}\chi_{\rm now}\simeq 7\times 10^{-28}cm
^{-1}$. Hence, the condition (\ref{eq:2.4.7}) is satisfied in the present 
epoch of the evolution of the Universe. On the other hand, in \emph{all} the 
asymptotic power series solutions of the field equations of the EccH system 
with a Small Bang singularity, in a vicinity of the singularity the mean 
curvature of the $t={\rm const}$ hypersurfaces does exceed $\chi_c$ 
\cite{SzW}. Also, in \emph{all} of these solutions in which $\vert\Phi
\vert^2$ takes $6/\kappa$ at \emph{regular} spacetime points, the mean 
curvature either exceeds $\chi_c$ in a vicinity of, or takes the value 
$\chi_c$ at the $\vert\Phi\vert^2=6/\kappa$ hypersurface. Therefore, in both 
cases there is a regime in the spacetime where instantaneous vacuum states do 
\emph{not} exist. 

Similarly, since the mean curvature e.g. on the standard foliation of the 
interior Schwarzschild solution diverges as $\sim t^{-3/2}$, inequality 
(\ref{eq:2.4.7}) is a non-trivial condition in Kantowski--Sachs spacetimes, 
too: Near the spacetime singularity, deeply behind the black hole horizon, 
instantaneous vacuum states do \emph{not} exist. 

On the other hand, note that by the specific form of the energy density in 
the FRW and Kantowski--Sachs cases, given, respectively, by (\ref{eq:2.4.3}) 
and (\ref{eq:2.5.5}), the instantaneous vacuum states, when they exist, are 
\emph{gauge symmetry breaking} vacuum states. In the presence of these 
geometric symmetries the energy density does \emph{not} have any \emph{gauge 
symmetric} stable minimum.

\subsubsection{The existence of the instantaneous vacuum states: A 
condition on the geometric symmetries}
\label{sub-3.2.3}

Since the instantaneous vacuum states are expected to be physical states, 
they must solve the Hamiltonian constraint. Hence, by (\ref{eq:2.3.4f}), 
\emph{the corresponding curvature scalar ${\cal R}_v$ must be negative}. 
Therefore, \emph{globally defined} instantaneous vacuum states exist only 
when the hypersurface $\Sigma_t$ admits hyperboloidal 3-geometries and the 
group of spacetime symmetries is compatible with ${\cal R}_v<0$. In 
particular, in the presence of FRW symmetries, $k$ in (\ref{eq:2.4.1}) 
must be $-1$. Thus, to have \emph{global} instantaneous vacuum states 
\emph{the Universe must be open}. 

In Kantowski--Sachs spacetimes, by $E^\alpha_a=0$, $B^\alpha_{ab}=0$, $\psi^r_A
=0$ and $\mathbb{D}_a\Phi^{\bi}=0$ the only constraint in the matter sector, 
the Gauss constraint, is satisfied. Remarkably enough, although \emph{a 
priori} we allowed the matter field variables to have special direction 
dependence, i.e. to be aligned with the distinguished vector field $v^a$, in 
the spacetime-symmetric minimal energy states the matter field configurations 
turn out to be \emph{isotropic}. This result is compatible with the general 
structure of the critical configurations of the general energy-momentum 
functional obtained in \ref{sub-2.3.2}. Since $D_a(\chi^a{}_b-\chi\delta^a_b)
=0$ holds by the results of the first paragraph of subsection \ref{sub-2.5.1} 
and the momentum density built from the Kantowski--Sachs invariant fields of 
the EccSM model is vanishing, the momentum constraint of General Relativity 
is identically satisfied. On the other hand, the Hamiltonian constraint, 
written in the form 

\begin{equation*}
{\cal R}-(\chi_{ab}-\frac{1}{3}\chi h_{ab})(\chi^{ab}-\frac{1}{3}\chi h^{ab})=
2\Lambda+2\kappa\varepsilon-\frac{2}{3}\chi^2, 
\end{equation*}
is a non-trivial condition on the difference of the spatial curvature scalar 
and the square of the trace-free part of the extrinsic curvature. Although 
the matter field variables in the instantaneous vacuum states must be 
isotropic on $\Sigma_t$, apparently the extrinsic curvature need not. 
Nevertheless, the general analysis of the critical configurations of the 
energy-momentum functional in subsection \ref{sub-2.3.2} shows that the 
extrinsic curvature \emph{must} be a constant pure trace (see equation 
(\ref{eq:2.3.4e})). Otherwise the field configuration would not be critical 
with respect to \emph{general} variations, though they are with respect to 
variations in the Kantowski--Sachs class. 

With this additional restriction coming from the results of the general 
analysis in subsection \ref{sub-2.3.2}, in the instantaneous vacuum states, 
the mean curvature $\chi$ determines the spatial curvature scalar on 
$\Sigma_t$ completely via equation (\ref{eq:3.2.6}) and the Hamiltonian 
constraint: 

\begin{equation}
\frac{1}{2}{\cal R}_v=\Lambda-\frac{1}{4}\kappa\lambda\vert\Phi_v\vert^4-
\frac{1}{3}\chi^2=\frac{1}{1-\frac{1}{12}\kappa\vert\Phi_v\vert^2}\bigl(
\Lambda+\frac{1}{4}\kappa\mu^2\vert\Phi_v\vert^2-\frac{1}{3}\chi^2\bigr). 
\label{eq:3.2.7}
\end{equation}
Here, in the second equality, we used (\ref{eq:2.3.4dWS}). However, by 
(\ref{eq:3.2.6}), the first two terms together in the brackets on the right 
is negative for any $\chi^2\leq\chi^2_c$, and hence \emph{the curvature 
scalar ${\cal R}_v$ must be negative}, while, according to the first 
paragraph of subsection \ref{sub-2.5.1}, the curvature scalar on any 
Kantowski--Sachs symmetric 3-space is \emph{strictly positive}. Therefore, 
\emph{global} instantaneous vacuum states whose matter \emph{and} 
gravitational sectors would be Kantowski--Sachs invariant do \emph{not} exist. 

On the other hand, if the vacuum states are not required to admit the same 
symmetries that the spacetime has, i.e. if they are allowed to be 
$SO(1,3)$-symmetric even in the $k=1,0$ FRW or Kantowski--Sachs spacetimes, 
too, then the instantaneous vacuum states can be defined at least on 
\emph{open subsets} of the $t={\rm const}$ hypersurfaces, i.e. they can exist 
\emph{quasi-locally}. In particular, in the $k=0$ FRW spacetime the 
instantaneous vacuum states can exist globally, but in the $k=1$ case only 
on proper subsets of $\Sigma_t\approx S^3$. 

Similarly, in the Kantowski--Sachs case, if we require the 3-metric in the 
gravitational sector of the vacuum state to be $O(1,3)$-invariant (rather 
than to belong to the Kantowski--Sachs class with isometry group $\mathbb{R}
\times O(3)$), just like the matter fields and the extrinsic curvature, 
then that should be the hyperboloidal metric $dh_v^2=-\frac{K^2}{K^2+\rho^2}
d\rho^2-\rho^2d\Omega^2$ on some domain $r>\ln\rho_0$ of the 3-manifold 
$\Sigma_t\approx\mathbb{R}\times S^2$ with $\rho_0>0$, where the radial 
coordinate is $\rho:=\exp(r)-\rho_0$, $K^2:=-6/{\cal R}_v$, and the 
2-sphere $r=\ln\rho_0$ corresponds to the (missing) origin of the 
hyperboloidal space. Clearly, this \emph{quasi-locally} defined 
instantaneous vacuum state on the given open subset of $\Sigma_t$ exists if 
the mean curvature $\chi$ satisfies the strict inequality in 
(\ref{eq:2.4.7}), in which case it is determined completely by the mean 
curvature.

\subsection{The genesis/evanescence and time dependence of the rest 
masses}
\label{sub-3.3}

Following the general ideas in \cite{AL73}, now we calculate the rest mass 
of the fields of the EccSM system via the BEH mechanism in nearly FRW and 
Kantowski--Sachs spacetime: We assume that the spacetime satisfies those 
conditions that ensure the existence of instantaneous vacuum states. In 
particular, it should admit a foliation by constant mean curvature 
hypersurfaces $\Sigma_t$, whose mean curvature $\chi$ can be used as an 
external time variable (the so-called York time). 

However, as we saw in subsection \ref{sub-3.2} (and showed explicitly in the 
FRW case), the existence of local minima of the spatially constant energy 
density, and hence the existence of the instantaneous vacuum states, is 
guaranteed only on the spacelike hypersurfaces with mean curvature smaller 
than $\chi_c$. Moreover, if the instantaneous vacuum states exist, then they 
are \emph{necessarily} gauge symmetry breaking states. The existence of the 
bound $\chi_c$ yields a non-trivial `genesis'/`evanescence' of the rest 
masses of certain fields of the standard model, depending on whether $\chi$ 
is decreasing or increasing when $\chi$ takes the critical value $\chi_c$. 
Also, the electromagnetic interaction (together with the electric charge) 
emerges from the $U(2)$ gauge theory in the Weinberg--Salam model as an 
`effective' gauge theory \emph{only} in the presence of the (global or 
quasi-local) instantaneous gauge symmetry breaking vacuum states (see 
\cite{AL73}). Thus, the hypersurface whose mean curvature is just the bound 
$\chi_c$ is the `moment of genesis/evanescence' of the (globally or 
quasi-locally defined) electric charge, too. 

Thus let us suppose that $\chi^2<\chi^2_c$, and let us fix a particular 
(gauge symmetry breaking) vacuum state $\Phi^{\bi}_v$. Then, in the unitary 
gauge (see \cite{AL73,We73}, or subsection \ref{sub-1.7.1}), the independent 
Lagrangian field variables of the matter sector are $\phi^\alpha$, $A^\alpha_a$, 
$\psi^r_A$, $H^{\bi}:=\Phi^{\bi}-\Phi^{\bi}_v$ and the corresponding velocities; 
and the vacuum states correspond to the vanishing of these variables. The 
mass of the various fields can be read off from the second derivative of the 
Lagrangian with respect to $\phi^\alpha$, $A^\alpha_a$, $H^{\bi}$ and $\psi^r_A$ 
at the vacuum state, \emph{provided} its first derivatives are vanishing. 
Denoting the 3+1 form of the matter Lagrangian density by $\hat{\cal L}$ 
(which is just ${\cal L}$ in which ${\cal L}_H$ is replaced by $\hat{\cal L}
_H$), these first derivatives with respect to the gauge potentials at the 
vacuum states are 

\begin{eqnarray*}
&{}&\bigl(\frac{\partial\hat{\cal L}}{\partial\phi^\alpha}\bigr)_v=\bigl(
  \frac{\partial\hat{\cal L}_H}{\partial\phi^\alpha}\bigr)_v=-\frac{1}{6}
  \chi\bigl(G_{\bi\bk'}\bar T^{\bk'}_{\alpha{\bj'}}+G_{\bj'\bk}T^{\bk}_{\alpha{\bi}}
  \bigr)\Phi^{\bi}_v\bar\Phi^{\bj'}_v=0, \\
&{}&\bigl(\frac{\partial\hat{\cal L}}{\partial A^\alpha_a}\bigr)_v=\bigl(
  \frac{\partial\hat{\cal L}_H}{\partial A^\alpha_a}\bigr)_v=\frac{1}{2}
  G_{\bi\bj'}\Bigl(\bigl(D^a\bar\Phi^{\bj'}_v\bigr)T^{\bi}_{\alpha{\bk}}\Phi^{\bk}_v
  +\bigl(D^a\Phi^{\bi}_v\bigr)\bar T^{\bj'}_{\alpha{\bl'}}\bar\Phi^{\bl'}_v\Bigr)=0. 
\end{eqnarray*}
Here, in the first of these, we used (\ref{eq:3.2.4v}) and the gauge 
invariance of the fiber metric $G_{\bi\bj'}$; and, in the second, we used $D_a
\Phi^{\bi}_v=0$ and $A^\alpha_e=0$. These two can be summarized as $(\partial
\hat{\cal L}/\partial\omega^\alpha_a)_v=((\partial\hat{\cal L}/\partial\phi
^\alpha)t^a+(\partial\hat{\cal L}/\partial A^\alpha_b)P^a_b)=0$, and hence the 
mass matrix for the gauge fields, 

\begin{equation}
M^2_{\alpha\beta}:=\frac{1}{4}\bigl(g_{ab}\frac{\partial^2\hat{\cal L}}
{\partial\omega^\alpha_a\partial\omega^\beta_b}\bigr)_v=\frac{1}{2}\Bigl(
\bigl(T^{\bi}_{\alpha{\bk}}\Phi^{\bk}_v\bigr)G_{\bi\bj'}\bigl(\bar T^{\bj'}
_{\beta{\bl'}}\bar\Phi^{\bl'}_v\bigr)+\bigl(T^{\bi}_{\beta{\bk}}\Phi^{\bk}_v
\bigr)G_{\bi\bj'}\bigl(\bar T^{\bj'}_{\alpha{\bl'}}\bar\Phi^{\bl'}_v\bigr)
\Bigr), \label{eq:3.3.1}
\end{equation}
is a well defined real, symmetric and positive semi-definite matrix. The 
first derivative of $\hat{\cal L}$ with respect to the spinor field $\psi^r
_A$ at the vacuum state is also zero (by $\psi^r_A=0$), and the corresponding 
mass matrix is 

\begin{equation}
M_{rs}:=-\frac{1}{2}\bigl(\epsilon_{\uA\uB}\frac{\partial^2\hat{\cal L}}
{\partial\psi^r_{\uA}\partial\psi^s_{\uB}}\bigr)_v={\rm i}Y_{rs{\bi'}}\bar\Phi
^{\bi'}_v, \label{eq:3.3.2}
\end{equation}
just the component of the Yukawa coupling determined by the Higgs field in 
the vacuum state, $\Phi^{\bi}_v$. (Here $\epsilon_{\uA\uB}$ is the anti-symmetric 
Levi-Civita symbol.) The first derivative of $\hat{\cal L}$ with respect to 
$H^{\bi}$ is 

\begin{eqnarray*}
\bigl(\frac{\partial\hat{\cal L}}{\partial H^{\bi}}\bigr)_v\!\!\!\!&=\!\!\!\!&
  \bigl(\frac{\partial(\hat{\cal L}_H-V)}{\partial H^{\bi}}\bigr)_v=-\frac{1}
  {18}\chi^2G_{\bi\bj'}\bar\Phi^{\bj'}_v-\frac{1}{2}\Bigl(\frac{1}{6}\bigl({\cal 
  R}_v+\chi^{ab}\chi_{ab}-\chi^2\bigr)+\mu^2\Bigr)G_{\bi\bj'}\bar\Phi^{\bj'}_v\\
\!\!\!\!&-\!\!\!\!&\frac{1}{2}G_{\bi\bj\bk'\bl'}\Phi^{\bj}_v\bar\Phi^{\bk'}_v\bar
  \Phi^{\bl'}_v= \\
\!\!\!\!&=\!\!\!\!&-\frac{1}{2}\Bigl(\frac{1}{3}\kappa\varepsilon_v+\mu^2+
  \frac{1}{3}\Lambda-\frac{1}{9}\chi^2\Bigr)G_{\bi\bj'}\bar\Phi^{\bj'}_v-
  \frac{1}{2}G_{\bi\bj\bk'\bl'}\Phi^{\bj}_v\bar\Phi^{\bk'}_v\bar\Phi^{\bl'}_v=0.
\end{eqnarray*}
Here ${\cal R}_v$ and $\varepsilon_v$ denote the spatial curvature scalar and 
energy density in the vacuum state $\Phi^{\bi}_v$, respectively; and first we 
used $\phi^\alpha=0$, $A^\alpha_e=0$, $D_a\Phi^{\bi}_v=0$ and (\ref{eq:3.2.4v}), 
and then the Hamiltonian constraint, (\ref{eq:2.3.4e}) and (\ref{eq:2.3.4d}). 
Therefore, in contrast to the spacetime vacuum states of subsection 
\ref{sub-1.7.1}, \emph{the instantaneous gauge symmetry breaking vacuum 
states are critical points of the Lagrangian}. Hence, its second derivatives 
at these states, 

\begin{eqnarray}
M^2_{\bi\bj'}:=-\bigl(\frac{\partial^2\hat{\cal L}}{\partial H^{\bi}\partial
  \bar H^{\bj'}}\bigr)_v\!\!\!\!&=\!\!\!\!&\Bigl(\frac{1}{6}\kappa\varepsilon
  _v+\frac{1}{6}\Lambda+\frac{1}{2}\mu^2-\frac{1}{9}\chi^2\Bigr)G_{\bi\bj'}+
  G_{\bi\bk\bj'\bl'}\Phi^{\bk}_v\bar\Phi^{\bl'}_v, \label{eq:3.3.3a} \\
M^2_{\bi\bj}:=-\bigl(\frac{\partial^2\hat{\cal L}}{\partial H^{\bi}\partial H
  ^{\bj}}\bigr)_v\!\!\!\!&=\!\!\!\!&\frac{1}{2}G_{\bi\bj\bk'\bl'}\bar\Phi^{\bk'}
  _v\bar\Phi^{\bl'}_v, \label{eq:3.3.3b}
\end{eqnarray}
are well defined. Note that, even if the instantaneous vacuum state were 
only a \emph{quasi-local} one (in the $k=1,0$ FRW and Kantowski--Sachs 
cases), these mass matrices would still be well defined and independent of 
the parameter $\rho_0$ fixing the domain of the hyperboloidal line element 
$dh_v^2$ (see the end of subsection \ref{sub-3.2.3}). Because of the extrinsic 
curvature term in (\ref{eq:2.4.6}), the instantaneous gauge symmetry breaking 
vacuum states are in general \emph{time dependent}, and hence the mass 
matrices are also time dependent. 

In the Einstein--conformally coupled Weinberg--Salam model, $\Phi^{\bi}_v$ is 
chosen to have the form $(0,\vert\Phi_v\vert)$ (as a column vector), and 
$H^{\bi}=(0,H)$ for some real function $H$. Then by (\ref{eq:1.5.5}) and 
(\ref{eq:3.3.2}) the rest mass of the electron is $m_e=\frac{1}{\sqrt{2}}G_e
\vert\Phi_v\vert$, and the neutrino is massless. Evaluating the mass matrix 
for the gauge fields in (\ref{eq:3.3.1}) and recalling that in particle 
physics the gauge fields are defined to be the connection 1-forms divided by 
the corresponding coupling constants (see the footnote to subsection 
\ref{sub-1.1}), we find that the fields $Z_e:=\frac{1}{\sqrt{g^2+g'{}^2}}
(\omega^0_e-\omega^3_e)$ and $W^\pm_e:=\frac{1}{g\sqrt{2}}(\omega^1_e\mp{\rm i}
\omega^2_e)$ get the masses $\frac{1}{2}\sqrt{g^2+g'{}^2}\vert\Phi_v\vert$ and 
$\frac{1}{2}g\vert\Phi_v\vert$, respectively, but $\varpi_e:=\frac{1}{gg'
\sqrt{g^2+g'{}^2}}(g^2\omega^0_e+g'{}^2\omega^3_e)$, identified with the photon 
field, is massless. Thus, the mass of the gauge bosons and the electron is 
given by the corresponding expression in the Weinberg--Salam model except 
that the vacuum value $v=\sqrt{-\mu^2/\lambda}$ is replaced by $\vert\Phi_v
\vert$, whose expansion is $v(1+\Lambda/3\mu^2-\chi^2/9\mu^2-\kappa\mu^2/12
\lambda+...)^{1/2}$. The mass of the real Higgs field $H$ is 

\begin{eqnarray}
m^2_H:\!\!\!\!&=\!\!\!\!&-\bigl(\frac{\partial^2\hat{\cal L}}{\partial H^2}
  \bigr)_v=\vert\Phi_v\vert^{-2}\Bigl(M^2_{\bi\bj}\Phi^{\bi}_v\Phi^{\bj}_v+2M^2
  _{\bi\bj'}\Phi^{\bi}_v\bar\Phi^{\bj'}_v+\bar M^2_{\bi'\bj'}\bar\Phi^{\bi'}_v\bar
  \Phi^{\bj'}_v\Bigr)= \nonumber \\
\!\!\!\!&=\!\!\!\!&\lambda\vert\Phi_v\vert^2+2\vert\Phi_v\vert^{-2}M^2_{\bi\bj'}
  \Phi^{\bi}_v\bar\Phi^{\bj'}_v=2\lambda\vert\Phi_v\vert^2-\frac{1}{9}\chi^2, 
  \label{eq:3.3.4}
\end{eqnarray}
where, to evaluate (\ref{eq:3.3.3a}), we used (\ref{eq:2.3.4dWS}). Because 
of the extra mean curvature term $\chi^2/9$ on the right, the time dependence 
of the Higgs and the other fields is slightly different: e.g. $m^2_e/m^2_H$ 
depends on $\chi$. 

If in a FRW spacetime $\chi^2=\chi^2_c$ (at $4.5\times10^{-43}sec$ Hubble 
time), then the vacuum value of the Higgs field is $\vert\Phi_v\vert^2=6/
\kappa$. Hence, at the instant of genesis, the mass e.g. of the electrons 
was $\simeq6\times10^{26}cm^{-1}$. By (\ref{eq:3.2.6}) this decreased to its 
half by $5.9\times10^{-43}sec$ Hubble time. The electron mass decreased to 
twice its present value (i.e. to $2\times(2.6\times10^{10}cm^{-1})$) by $4.5
\times10^{-27}sec$ Hubble time. The Higgs mass at the moment of genesis was 
$1.3\times10^{32}cm^{-1}$. This decreased to its half by $5.8\times10^{-43}sec$ 
Hubble time, and decreased to twice of its present value, $2\times(6.2\times
10^{15}cm^{-1})$, by $5.5\times10^{-26}sec$ Hubble time. The characteristic 
time scale of the weak interactions, i.e. that the Higgs mass parameter 
defines, is $1/c\vert\mu\vert\simeq5.4\times10^{-27}sec$. Thus, even at this 
time scale the mass of both the Higgs and the other particles were 
approximately twice bigger than their present value. The rest mass of matter 
fields falling into a spherical black hole behaves in the opposite way: The 
(quasi-locally defined) rest mass of the electron, the $W^\pm$ and $Z$ gauge 
bosons and the Higgs field is increasing until the instant corresponding to 
the critical value $\chi_c$ of the mean curvature, and then the masses 
evanesce just before hitting the central singularity. 

Finally, rewriting the Lagrangian ${\cal L}_W$ in terms of the fields 
$W^\pm_e$, $Z_e$ and $\varpi_e$ (as e.g. in \cite{AL73}), for the charge of 
the electron we obtain exactly that in the Weinberg--Salam model. It does 
not depend on the York time. 


\section{Summary and final remarks}
\label{sec:4}

We investigated certain \emph{kinematical} consequences of the conformally 
invariant coupling of the Higgs field of the Standard Model to Einstein's 
theory of gravity. First, we showed that \emph{spacetime} vacuum states, 
i.e. which would have maximal spacetime symmetry, solve the field equations 
and minimize the energy density, do \emph{not} exist. Then, we showed that 
in the $k=1,0$ FRW and the Kantowski--Sachs spacetimes (hence, in particular, 
in spherical black holes) \emph{global instantaneous} vacuum states, i.e. 
field configurations on the transitivity hypersurfaces of the spacetime 
symmetries which would be invariant with respect to the spacetime symmetries, 
solve the \emph{constraint} parts of the field equations and minimize the 
energy functional, do \emph{not} exist. Also, general \emph{quasi-local} 
(i.e. not necessarily global) instantaneous vacuum states do \emph{not} 
exist on hypersurfaces whose mean curvature is greater than a large, but 
finite critical value. If the mean curvature is less than this critical 
value, then instantaneous vacuum states exist, which, as we saw in the FRW 
and Kantowski--Sachs cases, are necessarily gauge symmetry breaking states. 

Using this concept of the global or quasi-local instantaneous (gauge symmetry 
breaking) vacuum states, we determined how the rest mass of the fields of the 
Standard Model depends on the extrinsic York time parameter of the 
hypersurfaces. We found that there are extreme gravitational situations in 
which the notion of rest mass, zero or non-zero, of the Higgs field 
\emph{cannot be introduced at all}, and the BEH mechanism does not work. In 
these situations the fields do not have particle interpretation. If they 
have, then the non-zero rest masses, introduced via the BEH mechanism, are 
time dependent in a non-stationary spacetime. Also, in the absence of the 
vacuum states above, electromagnetism and electric charge do not emerge from 
the Standard Model. 

The time dependence (and the \emph{different} time dependence) of the Higgs 
and the other particles is significant only when the mean curvature is close 
to its critical value. Thus, it may have a potential significance in the 
particle physics processes in the very early Universe, or near the central 
singularity in black holes. Hence it could be interesting to see whether or 
not this time dependence, and in particular the fact that at the 
characteristic time scale of the weak interactions the rest masses were 
twice their present value, could yield observable effect in the particle 
genesis era of the very early Universe. 

If the mean curvature of the hypersurfaces happened to be increasing and 
exceeded the critical value, then the massive fields of the Standard Model 
would lose their rest mass. In our (asymptotically exponentially expanding) 
Universe the mean curvature asymptotically tends to the finite, constant 
value $\sqrt{\Lambda/3}$. Hence the `reverse--BEH' mechanism at the 
cosmological scale cannot provide the way in which the fields lose their 
rest mass (as it would be desirable in the CCC model). The `reverse-BEH' 
mechanism takes place inside black holes, deeply behind the event horizon 
near the central singularity.


\section{Appendix: The total variation of the Lagrangian for 
Weyl spinor fields}
\label{sec:A}

Let $\{E^a_{\ua},\vartheta^{\ua}_a\}$, ${\ua}=0,...,3$, be a $g_{ab}$-orthonormal 
dual frame field and $\{\varepsilon^A_{\uA},\varepsilon^{\uA}_A\}$, ${\uA}=0,1$, 
the corresponding normalized dual spinor basis, i.e. for which $E^a
_{{\uA}{\uA}'}:=E^a_{\ua}\sigma^{\ua}_{{\uA}{\uA}'}=\varepsilon^A_{\uA}\bar\varepsilon
^{A'}_{{\uA}'}$ holds, where $\sigma^{\ua}_{{\uA}{\uA}'}$ are the $SL(2,\mathbb{C})$ 
Pauli matrices (including the factor $1/\sqrt{2}$). If $\Gamma^{\uA}_{e{\uB}}:=
\varepsilon^{\uA}_A\nabla_e\varepsilon^A_{\uB}$, the spacetime connection 1-form 
in the spinor basis above, then the Lagrangian for a multiplet of Weyl spinor 
fields, $\psi^r_A$, is 

\begin{eqnarray}
{\cal L}_W\!\!\!\!&=\!\!\!\!&\frac{\rm i}{2}G_{rr'}g^{ab}\bigl(\bar\psi^{r'}_{A'}
  \nabbla_b\psi^r_A-\psi^r_A\nabbla_b\bar\psi^{r'}_{A'}\bigr)= \nonumber \\
\!\!\!\!&=\!\!\!\!&\frac{\rm i}{2}G_{rr'}\epsilon^{\uA\uB}\epsilon^{{\uA}'{\uB}'}
  E^b_{{\uB}{\uB}'}\Bigl(\bar\psi^{r'}_{{\uA}'}\bigl(\partial_b\psi^r_{\uA}-\psi^r
  _{\uD}\Gamma^{\uD}_{b{\uA}}+\omega^\alpha_b T^r_{\alpha s}\psi^s_{\uA}\bigr)- 
  \nonumber \\
\!\!\!\!&{}\!\!\!\!&-\psi^r_{\uA}
  \bigl(\partial_b\bar\psi^{r'}_{{\uA}'}-\bar\psi^{r'}_{{\uD}'}\bar\Gamma^{{\uD}'}
  _{b{\uA}'}+\omega^\alpha_b\bar T^{r'}_{\alpha s'}\bar\psi^{s'}_{{\uA}'}\bigr)\Bigr), 
\label{eq:A.1}
\end{eqnarray}
where $\epsilon^{\uA\uB}$ is the anti-symmetric Levi-Civita symbol and 
$\partial_b$ denotes the gradient acting on functions. If $E^a_{\ua}(u)$, 
$\psi^r_{\uA}(u)$ are arbitrary smooth 1-parameter families of tetrads and 
spinor components, respectively, such that $E^a_{\ua}(0)=E^a_{\ua}$ and $\psi^r
_{\uA}(0)=\psi^r_{\uA}$, and $\delta$ denotes the derivative with respect to 
$u$ at $u=0$, then by (\ref{eq:A.1}) 

\begin{eqnarray}
\delta{\cal L}_W\!\!\!\!&=\!\!\!\!&\frac{\rm i}{2}G_{rr'}\epsilon^{\uA\uB}
  \epsilon^{{\uA}'{\uB}'}\delta E^b_{{\uB}{\uB}'}\Bigl(\bar\varepsilon^{A'}_{{\uA}'}
  \bar\psi^{r'}_{A'}\varepsilon^A_{\uA}\nabbla_b\psi^r_A-\varepsilon^A_{\uA}\psi^r
  _A\bar\varepsilon^{A'}_{{\uA}'}\nabbla_b\bar\psi^{r'}_{A'}\Bigr)+ \nonumber \\
\!\!\!\!&{}\!\!\!\!&+\frac{\rm i}{2}G_{rr'}\epsilon^{\uA\uB}\epsilon^{{\uA}'{\uB}'} 
  E^b_{{\uB}{\uB}'}\Bigl(\delta\bar\psi^{r'}_{{\uA}'}\varepsilon^A_{\uA}\nabbla_b
  \psi^r_A-\delta\psi^r_{\uA}\varepsilon^{A'}_{{\uA}'}\nabbla_b\bar\psi^{r'}_{A'}
  \Bigr)+ \nonumber \\
\!\!\!\!&{}\!\!\!\!&+\frac{\rm i}{2}G_{rr'}\epsilon^{\uA\uB}\epsilon^{{\uA}'{\uB}'} 
  E^b_{{\uB}{\uB}'}\Bigl(\bar\varepsilon^{A'}_{{\uA}'}\bar\psi^{r'}_{A'}\varepsilon
  ^A_{\uA}\nabbla_b\bigl(\varepsilon^{\uD}_A\delta\psi^r_{\uD}\bigr)-\varepsilon
  ^A_{\uA}\psi^r_{\uA}\bar\varepsilon^{A'}_{{\uA}'}\nabbla_b\bigl(\bar\varepsilon
  ^{{\uD}'}_{A'}\delta\bar\psi^{r'}_{{\uD}'}\bigr)\Bigr)-\nonumber \\
\!\!\!\!&{}\!\!\!\!&-\frac{\rm i}{2}\epsilon^{\uA\uB}\epsilon^{{\uA}'{\uB}'} E^b
  _{{\uB}{\uB}'}\Bigl(\delta^{{\uD}'}_{{\uA}'}\delta\Gamma^{\uD}_{b{\uA}}-\delta^{\uD}
  _{\uA}\delta\bar\Gamma^{{\uD}'}_{b{\uA}'}\Bigr)G_{rr'}\psi^r_{\uD}\bar\psi^{r'}
  _{{\uD}'}= \nonumber \\
\!\!\!\!&=\!\!\!\!&\frac{\rm i}{2}G_{rr'}g^{ab}\nabbla_b\Bigl(\bar\psi^{r'}_{A'}
  \varepsilon_A^{\uA}\delta\psi^r_{\uA}-\psi^r_A\bar\varepsilon_{A'}^{{\uA}'}
  \delta\bar\psi^{r'}_{{\uA}'}\Bigr)- \nonumber \\
\!\!\!\!&{}\!\!\!\!&-{\rm i}G_{rr'}\Bigl(\varepsilon^{A'B'}\bar\varepsilon^{B'}
  _{{\uB}'}\delta\bar\psi^{r'}_{{\uB}'}\bigl(\nabbla_{A'}{}^A\psi^r_A\bigr)-
  \varepsilon^{AB}\varepsilon^{\uB}_B\delta\psi^r_{\uB}\bigl(\nabbla_{A}{}^{A'}
  \bar\psi^{r'}_{A'}\bigr)\Bigr)+ \nonumber \\
\!\!\!\!&{}\!\!\!\!&+\frac{\rm i}{2}\delta E^b_{\ub}E^a_{\ua}\eta^{\ub\ua}G_{rr'}
  \Bigl(\bar\psi^{r'}_{A'}\nabbla_b\psi^r_A-\psi^r_A\nabbla_b\bar\psi^{r'}_{A'}
  \Bigr)- \nonumber \\
\!\!\!\!&{}\!\!\!\!&-\frac{\rm i}{2}\epsilon^{\uA\uB}\epsilon^{{\uA}'{\uB}'} E^b
  _{{\uB}{\uB}'}\Bigl(\delta^{{\uD}'}_{{\uA}'}\delta\Gamma^{\uD}_{b{\uA}}-\delta^{\uD}
  _{\uA}\delta\bar\Gamma^{{\uD}'}_{b{\uA}'}\Bigr)G_{rr'}\psi^r_{\uD}\bar\psi^{r'}
  _{{\uD}'}. \label{eq:A.2}
\end{eqnarray}
Thus, the second term on the right yields the field equations, while the 
variation of the tetrad field in the third term can be written as 

\begin{equation}
\delta E^b_{\ub}E^a_{\ua}\eta^{\ub\ua}=\delta E^{(b}_{\ub}E^{a)}_{\ua}\eta^{\ub\ua}+
\delta E^{[b}_{\ub}E^{a]}_{\ua}\eta^{\ub\ua}=\frac{1}{2}\delta g^{ab}+\varepsilon
^{BA}\bar\lambda^{A'B'}+\varepsilon^{B'A'}\lambda^{AB}, \label{eq:A.3}
\end{equation}
where $\lambda^{AB}=\lambda^{(AB)}$ is defined to be the anti-self-dual part of 
the anti-symmetric part $\delta E^{[b}_{\ub}E^{a]}_{\ua}\eta^{\ub\ua}$ of the 
variation of the tetrad field. To evaluate the last term in (\ref{eq:A.2}), 
we need the explicit form of the variation of the spinor connection 1-form. 
Since 

\begin{equation*}
\Gamma^{\uA}_{e{\uB}}=\frac{1}{2}\sigma^{{\uA}{\uA}'}_{\ua}\sigma^{\ub}_{{\uB}
{\uA}'}\vartheta^{\ua}_a\nabla_eE^a_{\ub}=:\frac{1}{2}\sigma^{{\uA}{\uA}'}
_{\ua}\sigma^{\ub}_{{\uB}{\uA}'}\gamma^{\ua}_{e{\ub}},
\end{equation*}
we should compute the variation of the connection 1-form $\gamma^{\ua}_{e{\ub}}$ 
in the linear frame bundle. It is 

\begin{eqnarray*}
\delta\gamma^{\ua}_{e{\ub}}\!\!\!\!&=\!\!\!\!&\delta\vartheta^{\ua}_a\nabla_e
  E^a_{\ub}+\vartheta^{\ua}_a\frac{1}{2}g^{ac}\bigl(-\nabla_c\delta g_{eb}+
  \nabla_e\delta g_{cb}+\nabla_b\delta g_{ec}\bigr)E^b_{\ub}+\vartheta^{\ua}_a
  \nabla_e\bigl(\delta E^a_{\ub}\bigr)= \\
\!\!\!\!&=\!\!\!\!&\vartheta^{\ua}_a\nabla_e\bigl(\delta E^{[a}_{\um}E^{b]}_{\un}
  \eta^{\um\un}\bigr)\vartheta^{\uc}_b\eta_{\uc\ub}+\frac{1}{2}\vartheta^{\ua}_a
  \Bigl(g^{ac}\bigl(\nabla_c\delta g^{fd}\bigr)g_{fe}g_{db}-\bigl(\nabla_b\delta 
  g^{af}\bigr)g_{fe}\Bigr)E^b_{\ub}. 
\end{eqnarray*}
Substituting this into the last term of (\ref{eq:A.2}) and using 
(\ref{eq:A.3}), by forming total divergences we obtain that 

\begin{eqnarray}
\delta{\cal L}_W\!\!\!\!&=\!\!\!\!&\frac{\rm i}{2}\nabla^a\Bigl(G_{rr'}\bar
  \psi^{r'}_{A'}\varepsilon^{\uA}_A\bigl(\delta\psi^r_{\uA}+\lambda_{\uA}{}^{\uB}
  \psi^r_{\uB}\bigr)-G_{rr'}\psi^r_A\bar\varepsilon^{{\uA}'}_{A'}\bigl(\delta
  \bar\psi^{r'}_{{\uA}'}+\bar\lambda_{{\uA}'}{}^{{\uB}'}\bar\psi^{r'}_{{\uB}'}\bigr)
  \Bigr)- \nonumber \\
\!\!\!\!&{}\!\!\!\!&-{\rm i}G_{rr'}\Bigl(\bigl(\delta\psi^r_{\uA}+\lambda_{\uA}
  {}^{\uB}\psi^r_{\uB}\bigr)\varepsilon^{\uA}_A\bigl(\nabbla^{AA'}\bar\psi^{r'}_{A'}
  \bigr)-\bigl(\delta\bar\psi^{r'}_{{\uA}'}+\bar\lambda_{{\uA}'}{}^{{\uB}'}\bar\psi
  ^{r'}_{{\uB}'}\bigr)\bar\varepsilon^{{\uA}'}_{A'}\bigl(\nabbla^{A'A}\psi^r_A\bigr)
  \Bigr)+ \nonumber \\
\!\!\!\!&{}\!\!\!\!&+\frac{\rm i}{8}G_{rr'}\Bigl(\bar\psi^{r'}_{A'}\nabbla_{BB'}
 \psi^r_A+\bar\psi^{r'}_{B'}\nabbla_{AA'}\psi^r_B-\psi^r_A\nabbla_{BB'}\bar\psi^{r'}
  _{A'}-\psi^r_B\nabbla_{AA'}\bar\psi^{r'}_{B'}\Bigr)\delta g^{ab}, \label{eq:A.4}
\end{eqnarray}
where $\lambda_{\uA}{}^{\uB}:=\varepsilon^A_{\uA}\lambda_A{}^B\varepsilon^{\uB}_B$, 
the components of $\lambda_A{}^B$ in the spinor basis. 

To see the meaning of the combination $\delta\psi^r_{\uA}+\lambda_{\uA}{}^{\uB}
\psi^r_{\uB}$, let us consider a 1-parameter family of spinor dual bases 
$\{\varepsilon^A_{\uA}(u),\varepsilon^{\uA}_A(u)\}$ \emph{in the spinor bundle} 
in which $\{\varepsilon^A_{\uA}(0),\varepsilon^{\uA}_A(0)\}=\{\varepsilon^A_{\uA},
\varepsilon^{\uA}_A\}$ is a normalized dual spinor dyad. Then the variation of 
the spinor basis is 

\begin{eqnarray*}
\delta\varepsilon^{\uA}_A\!\!\!\!&=\!\!\!\!&-\varepsilon^{\uA}_B\bigl(\delta
  \varepsilon^B_{\uB}\bigr)\varepsilon^{\uB}_A=\varepsilon^{\uA}_B\bigl(\delta
  \varepsilon^{(B}_{\uB}\bigr)\varepsilon^{C)}_{\uC}\epsilon^{\uB\uC}
  \varepsilon_{CA}-\frac{1}{2}\varepsilon^{\uA}_A\varepsilon_{BC}\bigl(\delta
  \varepsilon^B_{\uB}\bigr)\varepsilon^C_{\uC}\epsilon^{\uB\uC}= \\
\!\!\!\!&=\!\!\!\!&\varepsilon^{\uA}_B\lambda^B{}_A-\frac{1}{2}\omega\varepsilon
  ^{\uA}_A=\lambda^{\uA}{}_{\uB}\varepsilon^{\uB}_A-\frac{1}{2}\omega\varepsilon
  ^{\uA}_A,
\end{eqnarray*}
where $\lambda^{AB}$ is just the anti-self-dual part of the variation of the 
corresponding tetrad field in (\ref{eq:A.3}), while $\omega$ is defined to be 
$\varepsilon_{BC}\bigl(\delta\varepsilon^B_{\uB}\bigr)\varepsilon^C_{\uC}\epsilon
^{\uB\uC}$. Thus $\lambda^{\uA}{}_{\uB}$ represents the infinitesimal $SL(2,
\mathbb{C})$ transformation of the spinor dyad, while $\omega$ an (in general 
complex) infinitesimal conformal rescaling of the symplectic spinor metric. In 
fact, the variation of the corresponding spinor and spacetime metrics, 
respectively, are $\delta\varepsilon^{AB}=\omega\varepsilon^{AB}$, and hence 
$\delta g^{ab}=(\omega+\bar\omega)g^{ab}$. Nevertheless, as Penrose showed 
\cite{Pe83}, \emph{complex} conformal rescalings of the symplectic spinor 
metric yields \emph{torsion}. Since we keep the variation of the spin frame 
within the framework of Einstein's general relativity, $\omega$ must be real. 
Therefore, $\omega=\frac{1}{2}\varepsilon_{AB}\delta\varepsilon^{AB}=\frac{1}{8}
g_{ab}\delta g^{ab}$ represents an infinitesimal genuine (real conformal) change 
of the metric, while $\lambda^A{}_B$ is only a pure gauge transformation. Thus, 
finally, suppose that $\omega=0$ and consider the transformation of the 
\emph{components} $\psi^r_{\uA}$ of the spinor field $\psi^r_A$ under such a 
variation of the spinor basis. It is $\delta\psi^r_{\uA}=-\lambda_{\uA}{}^{\uB}
\psi^r_{\uB}$. Hence, the combination $\delta\psi^r_{\uA}+\lambda_{\uA}{}^{\uB}
\psi^r_{\uB}$ represents, in fact, the variation of the spinor field components 
\emph{up to pure $SL(2,\mathbb{C})$ basis transformations}, i.e. up to gauge. 
Hence, the second line of (\ref{eq:A.4}) contributes to the field equation 
(\ref{eq:1.3.1c}), while the third line to the energy-momentum tensor 
(\ref{eq:1.3.3}) of the spinor fields.


\bigskip

The author is grateful to \'Arp\'ad Luk\'acs, P\'eter Vecserny\'es and 
Gy\"orgy Wolf for the numerous and enlightening discussions both on the 
structure of the Standard Model and on various aspects of the present 
suggestion. Special thanks to Gy\"orgy Wolf for the careful reading of 
the draft of the paper, his suggestions to improve the text at several 
points and for drawing the figures. 



\begin{thebibliography}{99.}%


\bibitem{PeMc} R. Penrose, M. A. H. MacCallum, Twistor theory: An approach to 
               the quantisation of fields and space-time, Phys. Rep. {\bf 6} 
               241-316 (1972), doi:10.1016/0370-1573(73)90008-2
\bibitem{Pe} R. Penrose, {\it The Road to Reality, A Complete Guide to the 
             Laws of the Universe}, Jonathan Cape, London 2004, 
             ISBN 0-224-04447-8

\bibitem{H} P. W. Higgs, Broken symmetries and the masses of gauge bosons, 
            Phys. Rev. Lett. {\bf 13} 508-509 (1964), 
            doi: 10.1103/PhysRevLett.13.508
\bibitem{EB} F. Englert, R. Brout, Broken symmetry and the mass of gauge 
            vector mesons, Phys. Rev. Lett. {\bf 13} 321-323 (1964), doi: 
            10.1103/PhysRevLett.13.321

\bibitem{AL73} E. S. Abers, B. W. Lee, Gauge theories, Phys. Rep. {\bf 9} 
               1--141 (1973), doi: 10.1016/0370-1573(73)90027-6
\bibitem{ChLi} T.-P. Cheng, L.-F. Li, {\it Gauge Theory of Elementary Particle 
               Physics}, Clarendon Press, Oxford 1984, ISBN 0-19-851961-3

\bibitem{PeCCC} R. Penrose, {\it Cycles of Time}, The Bodley Head, London 
                2010, ISBN 9780224080361

\bibitem{PR1} R. Penrose, W. Rindler, {\it Spinors and Spacetime}, vol 1, 
              Cambridge University Press, Cambridge 1982, ISBN 0-521-24527-3
\bibitem{PR2} R. Penrose, W. Rindler, {\it Spinors and Spacetime}, vol 2, 
              Cambridge University Press, Cambridge 1986, ISBN 0-521-252671-9

\bibitem{SzW} L. B. Szabados, Gy. Wolf, Singularities in Einstein--conformally 
              coupled Higgs cosmological models, arXiv: 1802.00774 [gr-qc]

\bibitem{NP68} E. T. Newman, R. Penrose, New conservation laws for zero 
               rest-mass fields in asymptotically flat spacetimes, Proc. Roy. 
               Soc. A {\bf 305} 175-204 (1968), doi: 10.1098/rspa.1968.0112  

\bibitem{Sz09} L. B. Szabados, Quasi-local energy-momentum and angular 
               momentum in GR, Living Rev. Relativity {\bf 12} (2009) 4, 
               {\tt http://www.livingreviews.org/ lrr-2009-4}, doi: 
               10.12942/lrr-2009-4

\bibitem{ADM} R. Arnowitt, S. Deser, C. W. Misner, The dynamics of general
              relativity, in {\it Gravitation: An Introduction to Current
              Research}, pp. 227--265, Ed.: Witten, L., Wiley, New York,
              London, 1962, ISBN 0124366406, arXiv: gr-qc/0405109
\bibitem{Bondietal} H. Bondi, M. G. J. van der Burg, A. W. K. Metzner,
              Gravitational waves in general relativity. VII. Waves from
              axi-symmetric isolated systems, Proc. R. Soc. London, Ser. A
              {\bf 269} 21-52 (1962), doi: 10.1098/rspa.1962.0161

              R. K. Sachs, Asymptotic symmetries in gravitational theory,
              Phys. Rev. {\bf 128} 2851--2864 (1962), doi: 
              10.1103/PhysRev.128.2851

              A. R. Exton, E. T. Newman, R. Penrose, Conserved quantities in
              the Einstein--Maxwell theory, J. Math. Phys. {\bf 10} 1566--1570
              (1969), doi: 10.1063/1.1665006

\bibitem{AD} G. Abbott, S. Deser, Stability of gravity with a cosmological
             constant, Nucl. Phys. B {\bf 195} 76-96 (1982), doi: 
             10.1016/0550-3213(82)90049-9
\bibitem{SzaTo} L. B. Szabados, P. Tod, A positive Bondi-type mass in 
            asymptotically de Sitter spacetimes, Class. Quantum Grav. {\bf 32} 
            (2015) 205011, doi: 10.1088/0264-9381/32/20/205011, 
            arXiv: 1505.06637 [gr-qc]
\bibitem{Sz12} L. B. Szabados, Mass, gauge conditions and spectral properties 
            of the Sen--Witten and 3-surface twistor operators in closed 
            universes, Class. Quantum Grav. {\bf 29} 095001 (2012), doi: 
            10.1088/0264-9381/29/9/095001, arXiv: 1112.2966 [gr-qc]
\bibitem{Sz13} L. B. Szabados, On the total mass of closed universes with a 
            positive cosmological constant, Class. Quantum Grav. {\bf 30} 
            165013 (2013), doi: 10.1088/0264-9381/30/16/165013, arXiv: 
            1306.3863 [gr-qc]
\bibitem{DM} A. J. Dougan, L. J. Mason, Quasilocal mass constructions with 
             positive energy, Phys. Rev. Lett. {\bf 67} 2119-2122 (1991), doi: 
             10.1103/PhysRevLett.67.2119
\bibitem{Sz93} L. B. Szabados, On the positivity of the quasi-local mass, 
               Class. Quantum Grav. {\bf 10} 1899-1905 (1993), doi: 
               10.1088/0264-9381/10/9/027

\bibitem{HE} S. W. Hawking, G. F. R. Ellis, {\it The Large Scale Structure of 
             Spacetime}, Cambridge University Press, Cambridge 1973, ISBN 
             0-521-09906-4

\bibitem{We73} S. Weinberg, General theory of broken local symmetry, Phys. 
               Rev. D {\bf 7} 1068-1082 (1973), doi: 10.1103/PhysRevD.7.1068

\bibitem{Wa} R. M. Wald, {\it General Relativity}, University of Chicago 
              Press, Chicago 1984, ISBN 0-226-87033-2

\bibitem{Coll} C. B. Collins, Global structure of the ``Kantowski--Sachs'' 
               cosmological models, J. Math. Phys. {\bf 18} 2116-2124 (1977), 
               doi: 10.1063/1.523191
\bibitem{MacCallum} M. A. H. MacCallum, Anisotropic and inhomogeneous 
               relativistic cosmologies, in {\it General Relativity}, An 
               Einstein centenary survey, Eds. W. Israel, S. W. Hawking, 
               Cambridge University Press, Cambridge 1979, ISBN 0-521-22285-0

\bibitem{FoMa} P. Forg\'acs, N. S. Manton, Space-time symmetries in gauge 
               theories, Commun. Math. Phys. {\bf 72} 15-35 (1980), doi: 
               10.1007/BF01200108

\bibitem{HuTo} S. A. Huggett, K. P. Tod, {\it An Introduction to Twistor 
               Theory}, London Mathematical Society Student Texts No 4, 
               Cambridge University Press, Cambridge 1985, ISBN 0-521-45689-4

\bibitem{IsNe} J. Isenberg, J. Nester, Canonical gravity, in {\it General 
             Relativity and Gravitation}, Vol 1, Ed. A. Held, Plenum Press, 
             New York 1980, ISBN 0-306-40365-3(v.1)

\bibitem{Pe83} R. Penrose, Spinors and torsion in general relativity, Found. 
               Phys. {\bf 13} 325--340 (1983), doi: 10.1007/BF01906181

\end{thebibliography}
\end{document}